\RequirePackage{rotating}
\documentclass[fleqn,usenatbib]{mnras}

\usepackage[T1]{fontenc}

\DeclareRobustCommand{\VAN}[3]{#2}
\let\VANthebibliography\thebibliography
\def\thebibliography{\DeclareRobustCommand{\VAN}[3]{##3}\VANthebibliography}

\usepackage{graphicx}	% Including figure files
\usepackage{amsmath}	% Advanced maths commands
\usepackage{amssymb}	% Extra maths symbols
\usepackage{threeparttable}
\usepackage{lscape}
\usepackage{lineno}
\usepackage{eso-pic}% http://ctan.org/pkg/eso-pic
\usepackage{newtxtext,newtxmath}

\AddToShipoutPictureBG*{%
  \AtPageUpperLeft{%
    \hspace{0.75\paperwidth}%
    \raisebox{-3.5\baselineskip}{%
      \makebox[0pt][l]{\textnormal{DES-2021-0671}}
}}}%

\AddToShipoutPictureBG*{%
  \AtPageUpperLeft{%
    \hspace{0.75\paperwidth}%
    \raisebox{-4.5\baselineskip}{%
      \makebox[0pt][l]{\textnormal{FERMILAB-PUB-22-711-PPD}}
}}}%

\newcommand{\allPlanck}{3,130 }

\newcommand{\goodPlanck}{2,913 }
\newcommand{\lfPlanck}{1,092 } % fcont<0.3
\newcommand{\lowfPlanck}{842 }
\newcommand{\madpsz}{853 } % (fcont<0.2&S/N>3) & (fcont<0.3 & S/N>4.5)
\newcommand{\lowerfPlanck}{604 }
\newcommand{\lowestfPlanck}{479 }
\newcommand{\unmatchedPlanck}{329 }
\newcommand{\vect}[1]{\boldsymbol{#1}}

\newcommand{\zspecPlanck}{181 }
\newcommand{\fcont}{$f_{\rm cont}$ }
\newcommand{\Planck}{\textit{Planck} }
\newcommand{\MCMF}{\texttt{MCMF} }

\title[Extended \Planck$\times$ DES cluster catalogue]{The PSZ-MCMF catalogue of \Planck clusters over the DES region}
\author[DES Collaboration]{
\parbox{\textwidth}{
\Large
D.~Hern\'andez-Lang,$^{1,2}$\thanks{E-mail: daniel.hernandez@physik.lmu.de}
M.~Klein,$^{1}$
J.~J.~Mohr,$^{1,3}$
S.~Grandis,$^{1}$
J.-B. Melin,$^{4}$
P. Tarr\'io,$^{4,5,6}$
M. Arnaud,$^{5}$
G.W. Pratt,$^{4,5}$
T.~M.~C.~Abbott,$^{7}$
M.~Aguena,$^{8}$
O.~Alves,$^{9}$
F.~Andrade-Oliveira,$^{9}$
D.~Bacon,$^{10}$
E.~Bertin,$^{11,12}$
D.~Brooks,$^{13}$
D.~L.~Burke,$^{14,15}$
A.~Carnero~Rosell,$^{16,8,17}$
M.~Carrasco~Kind,$^{18,19}$
J.~Carretero,$^{20}$
F.~J.~Castander,$^{21,22}$
M.~Costanzi,$^{23,24,25}$
L.~N.~da Costa,$^{8}$
M.~E.~S.~Pereira,$^{26}$
S.~Desai,$^{27}$
H.~T.~Diehl,$^{28}$
P.~Doel,$^{13}$
S.~Everett,$^{29}$
I.~Ferrero,$^{30}$
B.~Flaugher,$^{28}$
J.~Frieman,$^{28,31}$
J.~Garc\'ia-Bellido,$^{32}$
D.~Gruen,$^{2}$
R.~A.~Gruendl,$^{18,19}$
J.~Gschwend,$^{8,33}$
G.~Gutierrez,$^{28}$
S.~R.~Hinton,$^{34}$
D.~L.~Hollowood,$^{35}$
K.~Honscheid,$^{36,37}$
D.~J.~James,$^{38}$
K.~Kuehn,$^{39,40}$
N.~Kuropatkin,$^{28}$
O.~Lahav,$^{13}$
C.~Lidman,$^{41,42}$
P.~Melchior,$^{43}$
J. Mena-Fern{\'a}ndez,$^{44}$
F.~Menanteau,$^{18,19}$
R.~Miquel,$^{45,20}$
A.~Palmese,$^{46}$
F.~Paz-Chinch\'{o}n,$^{18,47}$
A.~Pieres,$^{8,33}$
A.~A.~Plazas~Malag\'on,$^{43}$
M.~Raveri,$^{48}$
M.~Rodriguez-Monroy,$^{44}$
A.~K.~Romer,$^{49}$
V.~Scarpine,$^{28}$
I.~Sevilla-Noarbe,$^{44}$
M.~Smith,$^{50}$
E.~Suchyta,$^{51}$
G.~Tarle,$^{9}$
D.~Thomas,$^{10}$
and N.~Weaverdyck$^{9,52}$
\begin{center} (DES Collaboration) \end{center}
}
\vspace{0.4cm}
\\
\underline{\normalsize \em Affiliations at the end of the paper.}
}

\date{Accepted XXX. Received YYY; in original form ZZZ}

\pubyear{2022}

\begin{document}
\label{firstpage}
\pagerange{\pageref{firstpage}--\pageref{lastpage}}
%\linenumbers
\maketitle

\begin{abstract}
We present the first systematic follow-up of \Planck Sunyaev-Zeldovich effect (SZE) selected candidates down to signal-to-noise (S/N) of 3 over the 5000 deg$^2$ covered by the Dark Energy Survey.  Using the \MCMF cluster confirmation algorithm, we identify optical counterparts, determine photometric redshifts and richnesses and assign a parameter, $f_{\rm cont}$, {that reflects the probability that each SZE-optical pairing represents a random superposition of physically unassociated systems rather than a real cluster.} 
The new PSZ-MCMF cluster catalogue consists of \madpsz \MCMF confirmed clusters and has a purity of 90\%.  We present the properties of subsamples of the PSZ-MCMF catalogue that have purities ranging from 90\% to 97.5\%, depending on the adopted \fcont threshold. {Halo mass estimates $M_{500}$}, redshifts, richnesses, and optical centers are presented for all PSZ-MCMF clusters.
The PSZ-MCMF catalogue adds 589 previously unknown \Planck identified clusters over the DES footprint and provides redshifts for an additional 50 previously published \Planck selected clusters with S/N$>$4.5. 
 Using the subsample with spectroscopic redshifts, we demonstrate excellent cluster photo-$z$ performance with an RMS scatter in $\Delta z/(1+z)$ of 0.47\%.  
Our \MCMF based analysis allows us to infer the contamination fraction of the initial S/N$>$3 \Planck selected candidate list, which is {$\sim$50\%}.  We present a method of estimating the completeness of the PSZ-MCMF cluster sample. In comparison to the previously published \Planck cluster catalogues, this new S/N$>$3 \MCMF confirmed cluster catalogue populates the lower mass regime at all redshifts and includes clusters up to z$\sim$1.3.

\end{abstract}


%%%%%%%%%%%%%%%%%%%%%%%%%%%%%%%%%%%%%%%%%%%%%%%%%%

%%%%%%%%%%%%%%%%% BODY OF PAPER %%%%%%%%%%%%%%%%%%

\section{Introduction}\label{sec:introduction}

The intracluster medium (ICM) in galaxy clusters can be detected through what are now easily observed ICM signatures, providing a means to select cluster samples based on their ICM properties. At high temperatures of up to T $\sim10^8$~K (for massive clusters), photons are emitted  at X--ray wavelengths via thermal bremsstrahlung. Moreover, the ICM can leave an imprint on the cosmic microwave background (CMB).  At mm-wavelengths, it is possible to study galaxy clusters via the thermal Sunyaev-Zeldovich effect \citep[SZE;][]{SZ1972}, which is produced by inverse Compton scattering of CMB photons by hot electrons in the ICM.

Large X--ray selected galaxy cluster catalogues have been created using X-ray imaging data from the ROSAT All Sky Survey and the XMM-Newton telescope \citep[e.g.][]{Piffaretti2011,Klein2019,CODEX2020, X-class2021} as well as the recently launched eROSITA mission \citep{eFEDS2021, Liu2021, Klein2021}.
The \Planck mission mapped the whole sky between 2009 to 2013 in mm and infrared wavelengths, with the goal of studying CMB anisotropies. The latest cluster catalogue released by the \Planck collaboration is the second \Planck catalogue of Sunyaev-Zeldovich sources \citep[PSZ2;][]{Planck2016}, containing over 1600 cluster candidates down to a signal-to-noise ratio (S/N) of 4.5, detected from the 29-month full-mission data.  Other projects such as the South Pole Telescope \citep[SPT;][]{Carlstrom2011} and the Atacama Cosmology Telescope \citep[ACT;][]{ACT2011} have also been used to create large SZE selected cluster catalogues \citep[e.g.][]{Bleem2015,ACTClusters, Bleem2020}.

Although ICM-based cluster selection from an X-ray or SZE sky survey is efficient, the resulting candidate lists must be optically confirmed to extract galaxy based observables such as precise photometric redshifts \citep[e.g.][]{Staniszewski2009,High2010,Song2012,Liu2015,Klein2019,Klein2021}. The optical followup also allows for a cleaning or removal of the contaminants (falsely identified clusters) from ICM selected samples, because noise fluctuations in the ICM candidate lists do not have physically associated galaxy systems.  It is possible for a noise fluctuation in the ICM candidate list to overlap by chance with a physically unassociated galaxy system.
With the use of the Multi-Component Matched Filter followup technique \citep[\texttt{MCMF};][]{Klein2018, Klein2019}, it is possible to account for this random superposition possibility for each ICM cluster candidate and to deliver
empirically estimated, precise and accurate measurements of the residual contamination in the final cluster catalogue.

To enable efficient optical followup and precise estimates of the purity of the final confirmed cluster catalogue, large and homogeneous photometric datasets are beneficial. The Dark Energy Survey \citep[DES;][]{DES2016} covers $\sim$5000 deg$^2$ with deep, multiband imaging in \textit{g}, \textit{r}, \textit{i}, \textit{z}, \textit{Y} bands with the DECam instrument \citep{Flaugher2015}.  These imaging data are processed and calibrated using the DES data management system \citep{Morganson2018}, and to date two major data releases have taken place \citep{Abbott2018,Abbott_2021}.  

Large, homogeneous multi-band imaging surveys also support the direct galaxy-based selection of cluster catalogues \citep[e.g., ][]{Gladders07,Rykoff2014, Maturi2019, Wen2022}.  However, without a second cluster observable, as in the case of the ICM based selection followed up by optical confirmation, it is more challenging to empirically estimate or control the contamination of the final cluster catalogue.  One can use statistical comparisons to well understood ICM-based samples \citep[see SPTxRM analyses in][]{Grandis20,Grandis21} to estimate the contamination (as well as the mass completeness modeling) or one can attempt to simulate the contamination of the cluster sample directly \citep[e.g.,][]{song2012b,Crocce2015,DeRose2019}, in which case the contamination estimates are impacted by the level of realism of the simulations.

The utility of optically based cluster sample cleaning methods, like that available with the \MCMF algorithm, becomes ever more central to the cluster catalogue creation as one considers lower signal to noise ICM signatures as cluster candidates, because these candidate samples are more contaminated with noise fluctuations.  With an effective optically based cleaning method, it becomes possible to create dramatically larger confirmed cluster samples from a given X-ray or mm-wave survey, while still maintaining low levels of contamination (i.e., high sample purity).  As an example, the X-ray cluster sample MARDY3 selected from ROSAT in combination with DES produced an increase of an order of magnitude in the number of ROSAT selected clusters over the DES area \citep{Klein2019}.  Significant gains are currently being seen in the extraction of cluster samples from lower signal to noise candidate lists from the SPT-SZ 2500d and the SPTpol 500d survey (Klein et al. in prep, Bleem et al. in prep).

Leveraging the rich dataset provided by \textit{Planck}, we have developed a new cluster candidate catalog that extends to lower signal-to-noise levels (S/N$>$3), enhancing the number of candidate clusters identified. However, extending the catalog to lower signal-to-noise levels leads to a higher number of spurious sources or noise fluctuations being classified as \Planck detections, resulting in a decrease in the candidate catalog purity.
To address this reduced purity, we utilize the DES dataset together with the \MCMF cluster confirmation algorithm to confirm \Planck clusters and to reject spurious sources.

In this analysis, we present the PSZ-MCMF\footnote{PSZ stands for the \Planck Sunyaev-Zeldovich cluster candidate list, whereas MCMF comes from the algorithm, which allows us to maximise the number of clusters from any given parent candidate list.} cluster catalog. To construct this catalogue, %In this analysis, 
we extend the \MCMF tool to deal with the larger positional uncertainties that come with \Planck selected cluster candidates and then apply this tool to a \Planck based candidate list down to S/N=3 using DES photometric data. In Section~\ref{sec:data} we give a description of the DES and \Planck data used. In Section~\ref{sec:clusterconfirmation} we describe the enhanced \MCMF cluster confirmation method, while in Section~\ref{sec:results} we report our findings. Finally, in Section~\ref{sec:summary} we summarise our findings and report our conclusions. Throughout this paper we adopt a flat $\Lambda$CDM cosmology with $\Omega_M$ = 0.3 and H$_0$ = 70~km~s$^{-1}$~Mpc$^{-1}$.

\section{Data}\label{sec:data}

\subsection{DES multi-band photometric data}\label{sec:data_des}

In this work we use the DES Y3A2 GOLD photometric data, which is based on DES imaging data obtained from the first three years of the survey \citep{Abbott2018}. We employ $g,r,i,z$ band photometry, which has 95\% completeness limits of 23.72, 23.34, 22.78 and 22.25~mag, respectively. The YA32 GOLD catalogue has been optimized for cosmological studies with DES, similar to the Y1A1 GOLD catalogue \citep{Drlica2018}. Because we build upon the same \MCMF cluster confirmation method applied in a ROSAT$\times$DES analysis \citep{Klein2019}, we refer the reader to that source for further details of the filtering and handling of the optical multi-band data.

In summary, we make use of the single-object fitting photometry (SOF), which is based on the ngmix code \citep{Sheldon2014}. The photometry is performed by fitting a galaxy model for each source in each single epoch %and band 
image of a given band at the same time, interpolating the point-spread functions (PSFs) at the location of each source. This fitting is done masking neighbouring sources. We make use of the star-galaxy separator included in the GOLD catalogs \citep{Drlica2018} and exclude unresolved objects with 
$i<$22.2~mag. We also make use of the 
masking provided by Y3A2 GOLD \citep[similar to that described in Y1A1 GOLD,][]{Drlica2018} to exclude regions around bright stars. 

\begin{figure}
    \centering
    \includegraphics[width=\linewidth]{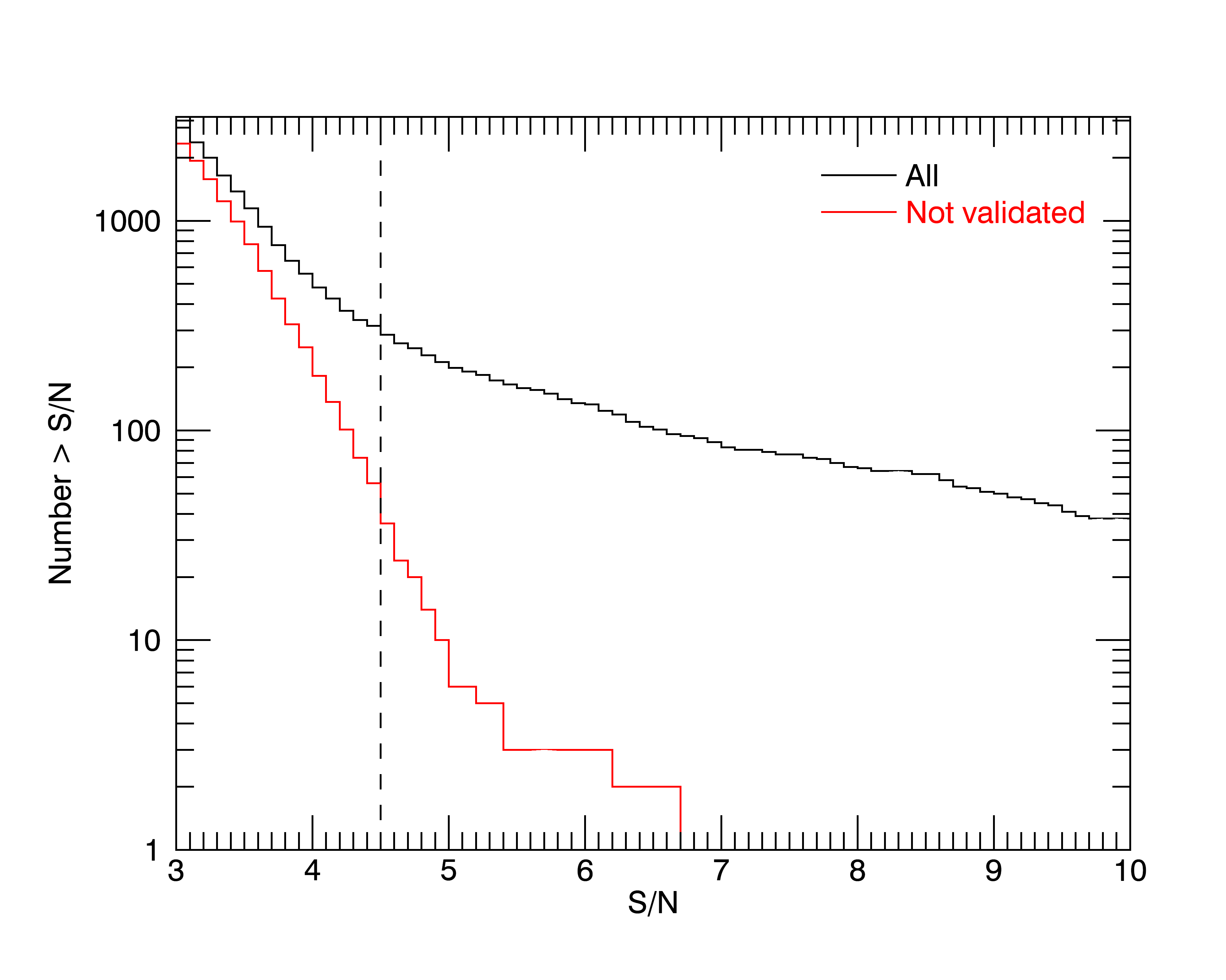}
    \vskip-0.15in
    \caption{Cumulative number of \Planck cluster candidates down to S/N=3. The solid black line represents the full sample of 3130 candidates. The solid red line represents the 2670 candidates that have not been validated in previous works via a simple cross-identification with known SZE and X-ray clusters (see text). The dashed line corresponds to the S/N=4.5 limit for the PSZ2 catalogue \citep{Planck2016}.}
    \label{fig:snr_validated}
\end{figure}

\subsection{Planck SZE candidate list}\label{sec:data_sz}

We build a catalogue of \Planck SZE sources with S/N$>$3 located within the DES footprint. The SZE catalogue is created using a matched multi-filter (MMF) approach \citep[see for example][]{Herranz2002,Melin2006}, namely the MMF3 algorithm used and described in \cite{PSZ1} and improved for the PSZ2 catalogue. The cluster detection is done using a combination of the \Planck maps and assuming prior knowledge on the cluster profile. In this application of MMF3, we divide the sky into patches of $10^\circ \times 10^\circ$, generating 504 overlapping patches, and run the detection algorithm with two iterations; the first iteration detects the SZE signal and the second refines the SZE candidate position to allow for improved estimation of the S/N and other properties.

\begin{figure}
    \centering
    \includegraphics[width=\linewidth]{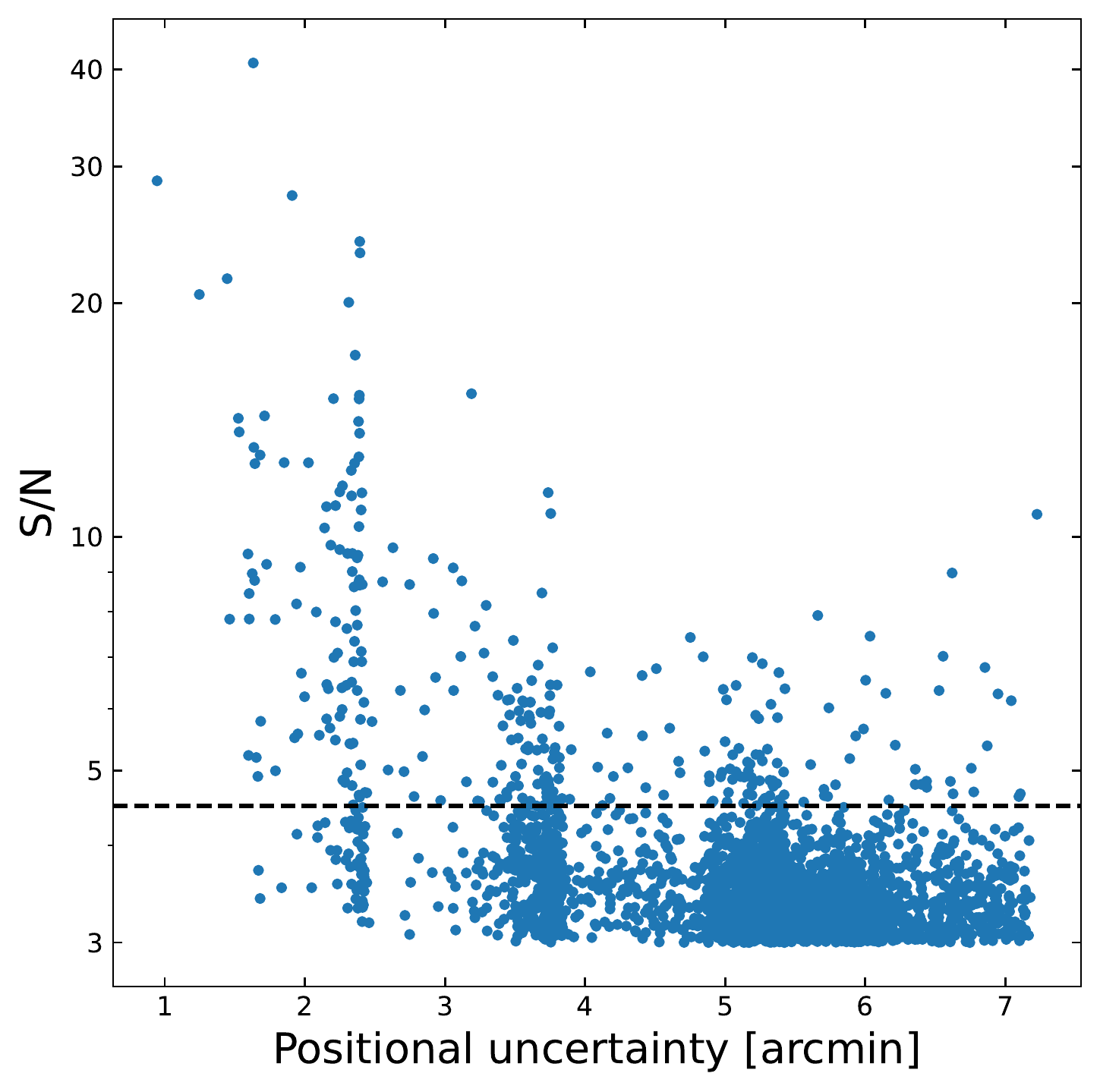}
    \vskip-0.15in
    \caption{Positional uncertainty distribution in units of arcminutes versus \Planck candidate S/N. The mean and median of the sources with S/N$<$4.5 are 5.26~arcmin and 5.36~arcmin, respectively. The black dashed line represents the threshold at S/N=4.5.}
    \label{fig:pos_uncertainty}
\end{figure}

The filter works by combining the frequency maps from the \Planck survey into a vector $\vect{M(x)}$, where each component corresponds to a map at frequency $\nu_i$ with $i=1,\ ...,N$ with $N$ being the total number of maps. For \textit{Planck}, we use the channel maps from 100 to 857 GHz, which correspond to the six
highest-frequency maps.

For each cluster candidate at a given central position $\vect{x_0}$, the algorithm fits:
\begin{equation}
    \vect{M_\nu (x)} = y_0\ \vect{j_\nu}\ T_{\theta_c} (\vect{x} - \vect{x_0}) + \vect{n_\nu(x)}
\end{equation}
\noindent
where $y_0$ is the central value at position $\vect{x_0}$ and $\vect{n_\nu(x)}$ corresponds to the noise vector, which is the sum of the other emission components in the map that do not correspond to the cluster SZE (such as, e.g., primordial CMB anisotropies and diffuse galactic emission). %The cluster signal due to the SZE, in each frequency map, $\nu$, is represented by
The frequency dependence of the SZE is represented by $\vect{j_\nu}$. The spatial profile is defined as $T_{\theta_c}$, with $\theta_c$ as the core radius. The assumed profile is chosen to be the universal pressure profile \citep{Arnaud2010}.

The filter is then employed to minimize the total variance estimate $\sigma_{\theta_C}^2$ on $y_0$ for each detected candidate, which yields an estimate $\hat{y}_0$. The S/N is then defined as $\hat{y}_0 / \sigma_{\theta_C}$.

From this analysis we get the positions and associated uncertainties of the SZE sources plus the S/N and the SZE flux. At S/N$>$3, we get a total of \allPlanck \Planck SZE sources (i.e. cluster candidates). Fig.~\ref{fig:snr_validated} shows the cumulative number of cluster candidates (black) and unvalidated cluster candidates (red) for each S/N bin within the DES footprint. A candidate is considered to be validated if 1) it is less than 5~arcmin from a confirmed cluster (with known redshift) of the Meta-Catalog of SZ detected clusters (MCSZ) of the M2C database\footnote{\url{https://www.galaxyclusterdb.eu/m2c/}}, or 2) it is less than 10~arcmin and less than $\theta_{500}$ from a confirmed cluster in the Meta-Catalog of X-Ray Detected Clusters of Galaxies \citep[MCXC,][]{Piffaretti2011}. From the full sample of 3,130 candidates, 460 have been validated in this way (with 414 matching MCSZ clusters, and 46 matching MCXC only), while the remaining 2,670 are non-validated candidates but may nevertheless be real galaxy clusters. 

\begin{figure}
    \centering
    \includegraphics[width=\hsize]{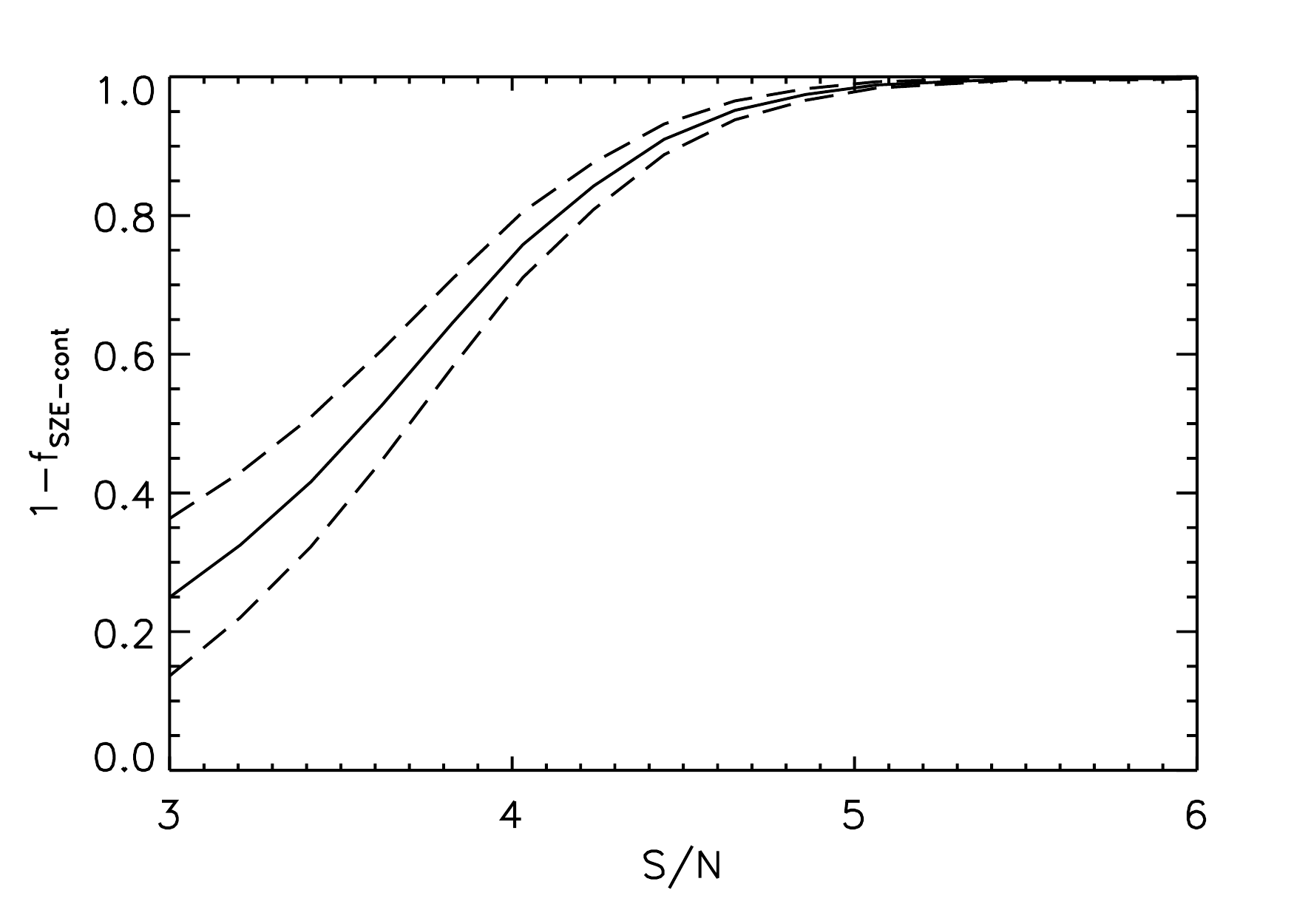}
    \vskip-0.15in
    \caption{Purity as a function of signal-to-noise threshold of the cluster candidate list, estimated on \Planck\ simulations. Dashed lines show the uncertainty of the estimated purity. The purity decreases from $\sim$1 at S/N$>$6 to $\sim$0.25 at S/N$>$3.}
    \label{fig:purity_vs_snr_planck}
\end{figure}

Fig.~\ref{fig:pos_uncertainty} contains the S/N versus the positional uncertainties of the \Planck sources, where the black dashed line represents a S/N=4.5. The apparent structure of the positional uncertainty is due to the pixelization of the \Planck maps. The detection algorithm filters the maps and finds the pixel which maximizes the S/N. The position assigned for a detection corresponds to the pixel center. % with no subpixel positioning. 
The positional uncertainty is also computed on a pixelized grid.

We estimate the contamination of the \Planck\ SZE candidate list using simulations. We use the \Planck Sky Model \citep[version 1.6.3;][]{Delabrouille2013}, to produce realistic all-sky mock observations. The simulations contain primary cosmic microwave background anisotropies, galactic components (synchrotron, thermal dust, free-free, spinning dust), extra-galactic radio and infrared point sources, and kinetic and thermal SZE. Each frequency map is convolved with the corresponding beam, and the instrumental noise consistent with the full mission is added. We run the thermal SZE detection algorithm down to S/N=3, and we match the candidate list with the input cluster catalogue adopting a 5~arcmin matching radius. We perform the matching after removing regions of the sky with high dust emission, leaving 75\% of the sky available, and we only use input clusters with a measured Compton parameter $Y$ in a circle of radius $5 \times R_{500}$\footnote{ $R_{500}$ is defined as the radius within which the density is 500 times the critical density}, $Y_{5R500}$, above $2 \times 10^{-4} \, {\rm arcmin^2}$. We adopt the SZE flux-mass relation
\begin{equation}\label{eq:y5r500}
E^{-2/3}(z)D_{A}^2(z)Y_{5R500} = A  \left [ \frac{M_{500}}{3 \times 10^{14} h_{70}^{-1} M_\odot} \right ]^{5/3}
\end{equation}
with $A=2.59 \times 10^{-5} \, h_{70}^{-1} \, {\rm Mpc}^2$ \citep[see equation~B.3 in][]{Arnaud2010}. $E(z)=H(z)/H_0$ is the Hubble parameter normalized to its present value and $D_A(z)$ is the angular diameter distance. The Compton parameter $Y_{5R500}$ is given in steradians. We estimate the purity of the sample as the number of real clusters divided by the number of detected clusters. This ratio is computed for various S/N thresholds. The result is shown in Fig.~\ref{fig:purity_vs_snr_planck}. The uncertainty in purity is considered to be the difference between the best estimate and the lower limit of the purity \citep[Fig.~11 and Fig.~12 in][respectively]{Planck2016} of the PSZ2 catalog, for the union 65\% case. We fit this difference as a function of the contamination with a power law in the range S/N=4.5-20. We extrapolate this down to S/N=3. From here on, we refer to this contamination as the initial contamination: $f_{\rm SZE-cont}$.
At high S/N threshold (S/N$>$6), the purity, $1-f_{\rm SZE-cont}$, is close to unity. Reducing the S/N threshold to 4.5 leads to a purity close to 0.9, which is consistent with previous estimates \citep{Planck2016}. When reducing the threshold to S/N=3, we measure a purity of $\sim$25$\%$ corresponding to a contamination $f_{\rm SZE-cont}=0.75$ in the simulations.

\section{Cluster confirmation method}\label{sec:clusterconfirmation}

To identify optical counterparts and estimate photometric redshifts we use a modified version of the \MCMF cluster confirmation algorithm on the \Planck candidate list and DES-Y3 photometric catalogues. For each potential cluster, the radial position and the galaxy color weightings are summed over all cluster galaxy candidates to estimate the excess number of galaxies, or richness ($\lambda$), with respect to the background. \citet{Klein2019} contains further details of \MCMF weights and the counterpart identification method. 

We expect only a fraction 1-$f_\mathrm{SZE-cont}$ of the \Planck candidates to be real clusters, with a large fraction ($f_\mathrm{SZE-cont}=0.75$) corresponding to contaminants (we return to the value of $f_\mathrm{SZE-cont}$ in Section~\ref{sec:incompleteness}). Most of these contaminants have no associated optical system, but some will happen to lie on the sky near a physically unassociated optical system or a projection of unassociated galaxies along the line of sight.  We refer to these contaminants as "random superpositions".  The \MCMF method has been designed to enable us to remove these contaminants from the \Planck candidate list. To estimate the likelihood of a ``random superposition'' (e.g, a spurious \Planck candidate being associated with one of the two cases above), we run \MCMF at random positions in the portion of the sky survey that lies away from the candidates. With this information we can reconstruct the frequency and redshift distribution of optical systems, and this allows us to estimate the probability that each candidate is a contaminant (see details in Section~\ref{sec:fcont}).

\subsection{Cluster confirmation with \MCMF}

In the \MCMF method the sky coordinates of the cluster candidates are used to search the multi-band photometric catalogues with an associated galaxy red sequence (RS) model, to estimate galaxy richness $\lambda$ as a function of redshift along the line of sight to each candidate. The weighted richnesses are estimated within a default aperture of $R_{500}$ centered at the candidate sky position \citep{Klein2018, Klein2019}. The weights include both a radial and a color component, with the radial filter following a projected Navarro, Frenk and White profile \citep[NFW;][]{NFW0, NFW1}, giving higher weights to galaxies closer to the center. The color filter uses the RS models and is tuned to give higher weights to cluster red sequence galaxies. These RS models are calibrated using over 2,500 clusters and groups with spectroscopic redshifts from the literature, including: the SPT-SZ cluster catalogue \citep[][]{Bleem2015}, the redMaPPer Y1 catalogue \citep[only for clusters with spectroscopic redshifts,][]{McClintock2019}, and the 2RXS X-ray sources cross-matched with the MCXC cluster catalogue \citep{Piffaretti2011}. These richnesses are estimated for each redshift bin with steps of $\Delta z=0.005$. The richness as a function of redshift is then searched to find richness peaks; the three strongest $\lambda$ peaks, each with a different photometric redshift, are recorded for each candidate. 

The mean positional uncertainty of the \Planck sources is $\sim$5.3~arcmin, which, adopting the cosmology from Section~\ref{sec:introduction}, translates into an uncertainty of $\sim$0.6~Mpc and $\sim$1.9~Mpc at $z=0.1$ and $z=0.5$, respectively. Given the large positional uncertainty of the \Planck candidates, the SZE position of a cluster could in some cases be offset by several times $R_{500}$. These large positional uncertainties enhance the probability of a spurious \Planck candidate being paired to a physically unassociated optical system. To address this large positional uncertainty, we run the \MCMF algorithm twice. The first run adopts the positions from the \Planck candidate catalogue, and carries out a search for possible optical counterparts within an aperture that is 3 times the positional uncertainty of the candidate, corresponding to a mean aperture of $\sim$15.9~arcmin.

\begin{figure*}
    \centering
    \includegraphics[width=0.49\linewidth]{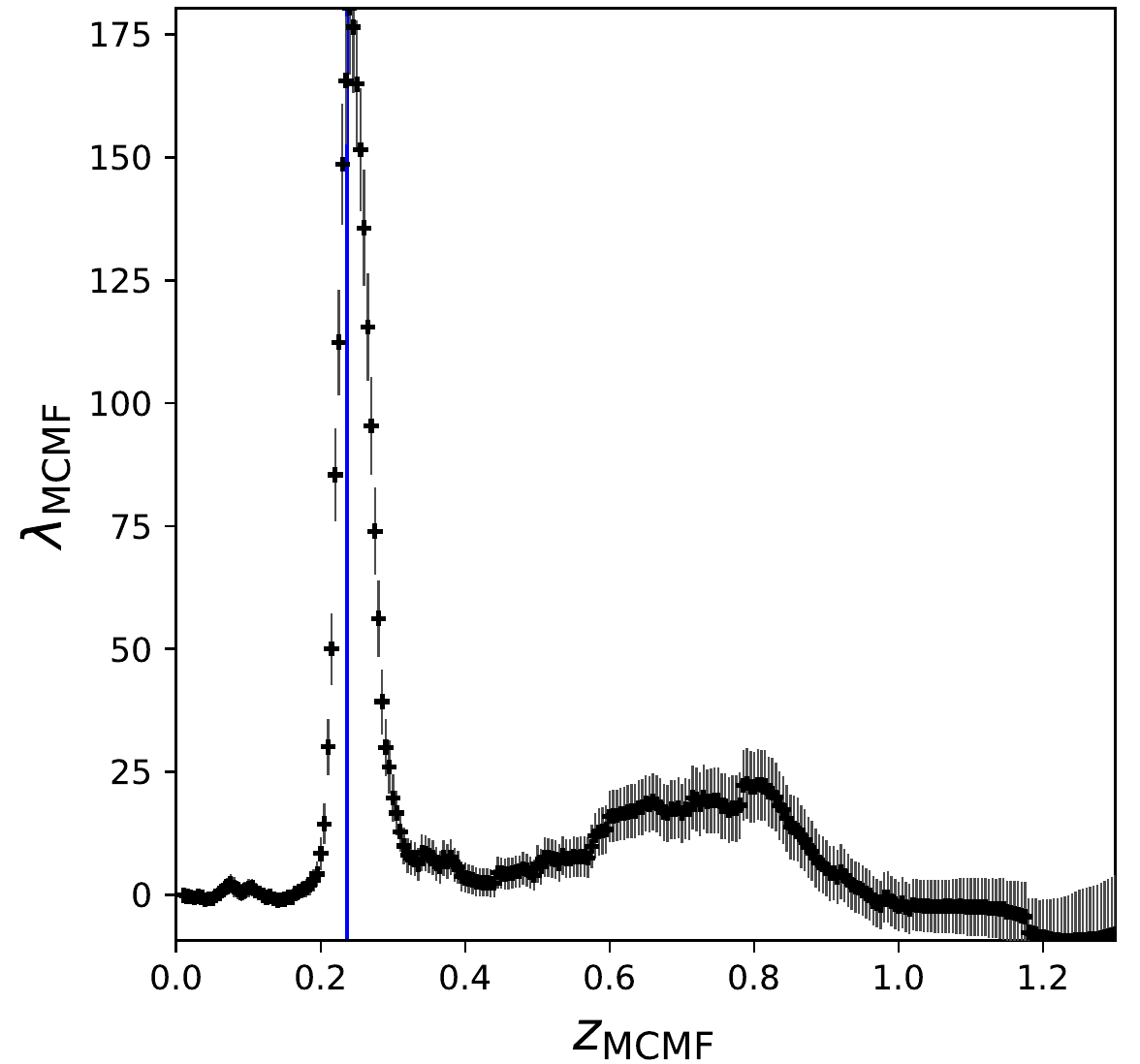}
    \includegraphics[width=0.4815\linewidth]{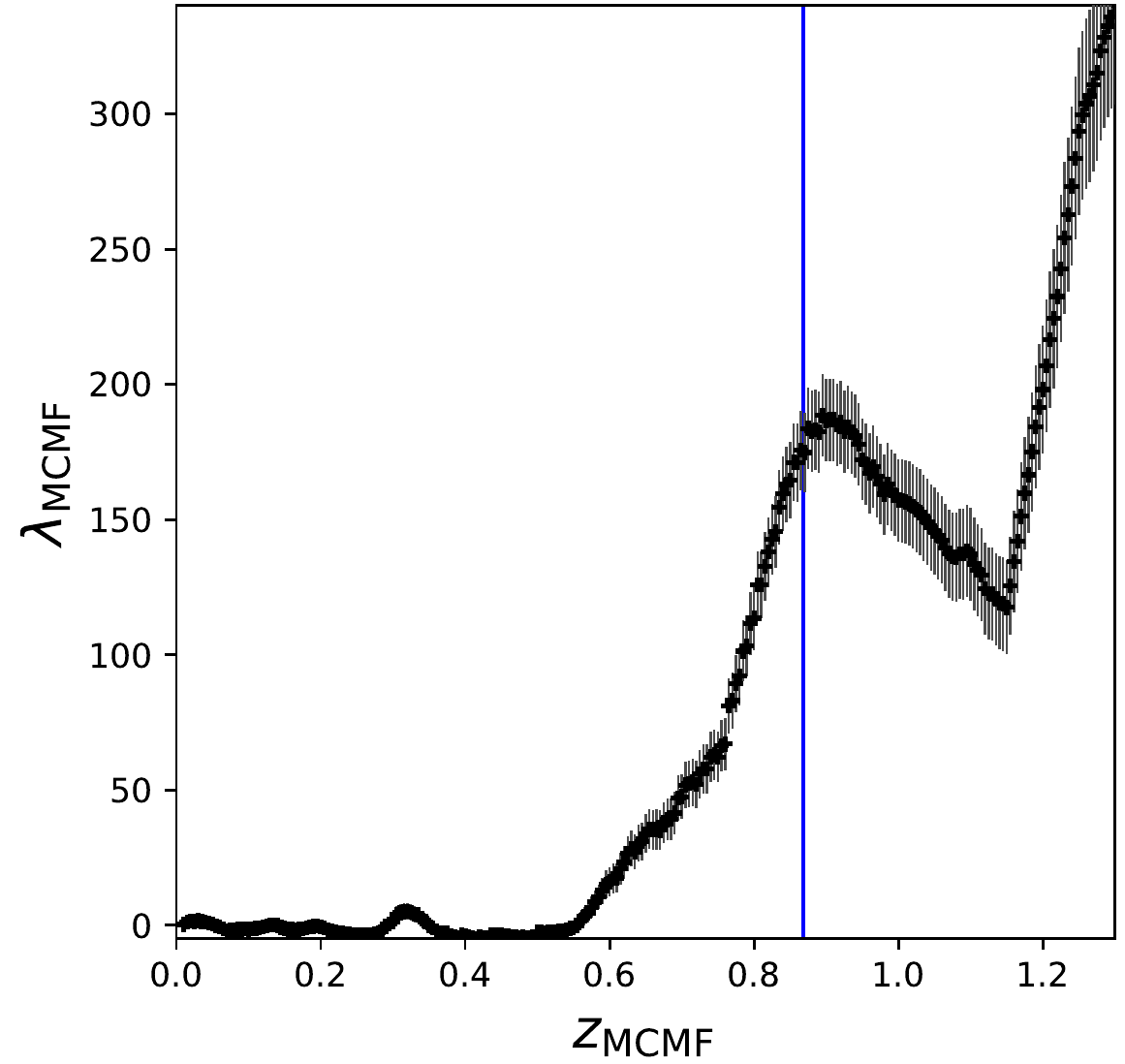}
    \includegraphics[width=0.49\linewidth]{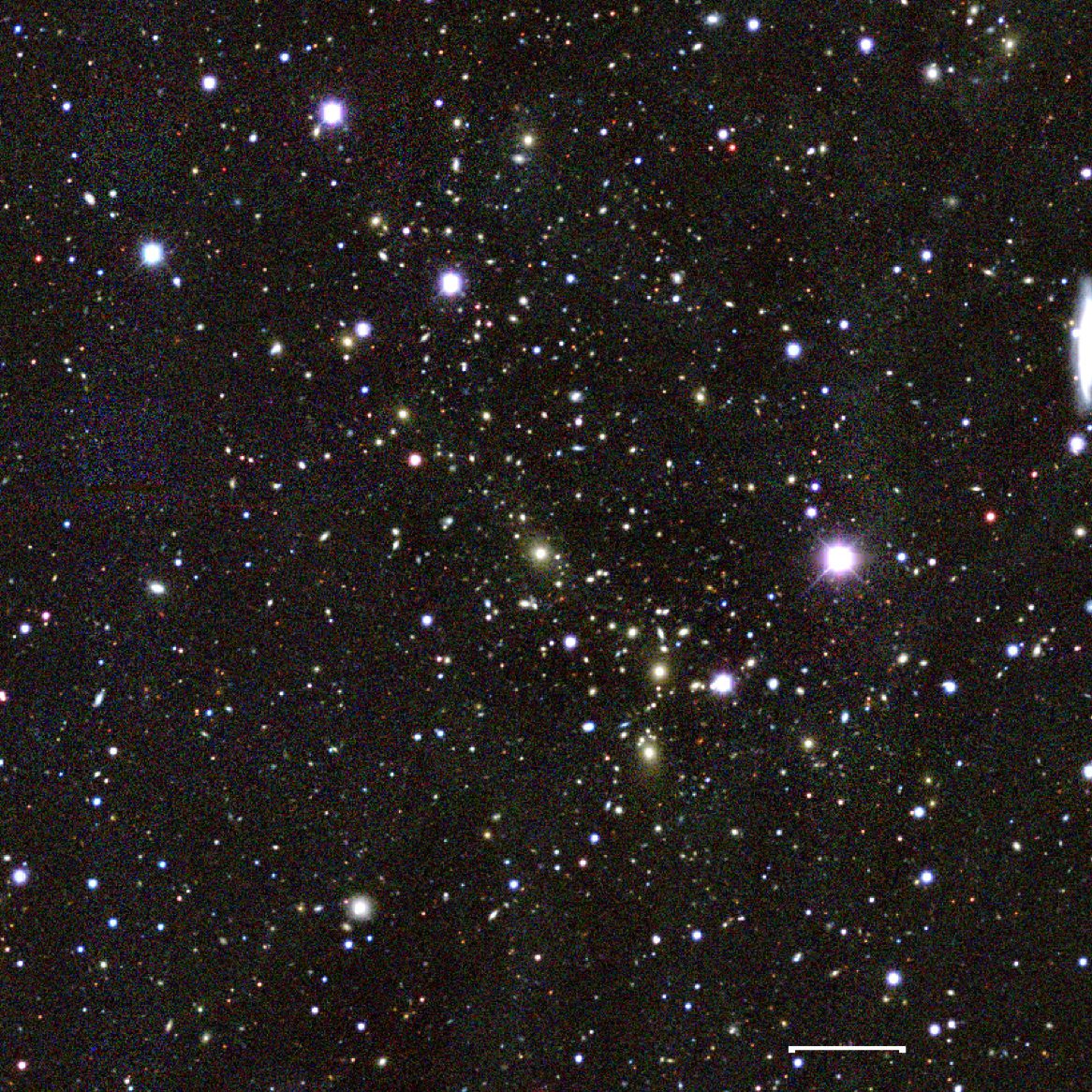}
    \includegraphics[width=0.49\linewidth]{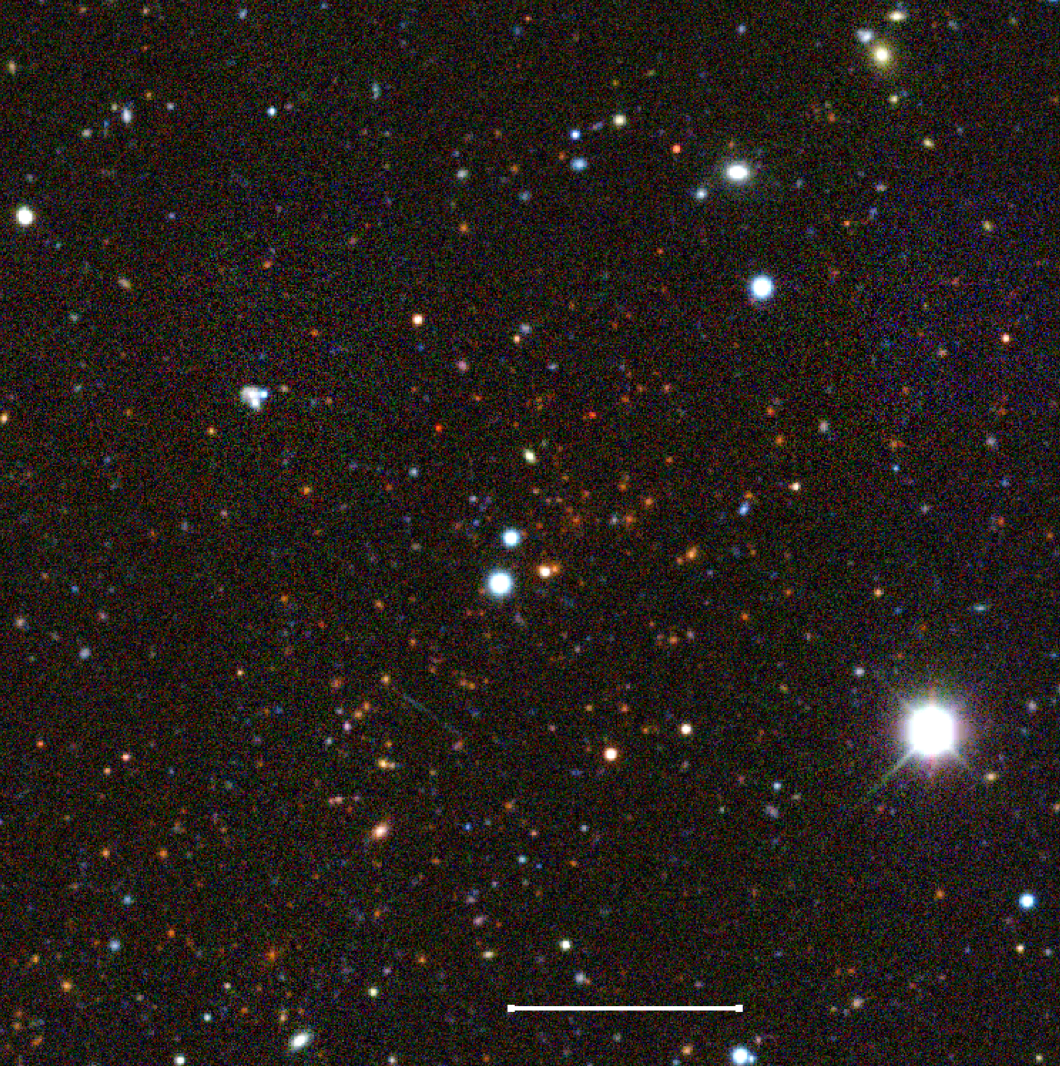}    
    \caption{Example \Planck cluster candidates with IDs PSZ-SN3~J2135+0124 ($z_{\rm MCMF}= 0.24$, left) and PSZ-SN3~J0102-4915 ($z_{\rm MCMF}= 0.87$, right). \textit{Above:} Richness as a function of redshift for each candidate. The blue line marks the most likely redshift of the candidate. \textit{Below:} DES pseudo-color images at the cluster positions. The white bar at the bottom denotes a scale of 1~arcmin. North is up and east is to the left.}
    \label{fig:mcmftest}
\end{figure*}

\begin{figure}
    \centering
    \includegraphics[width=\linewidth]{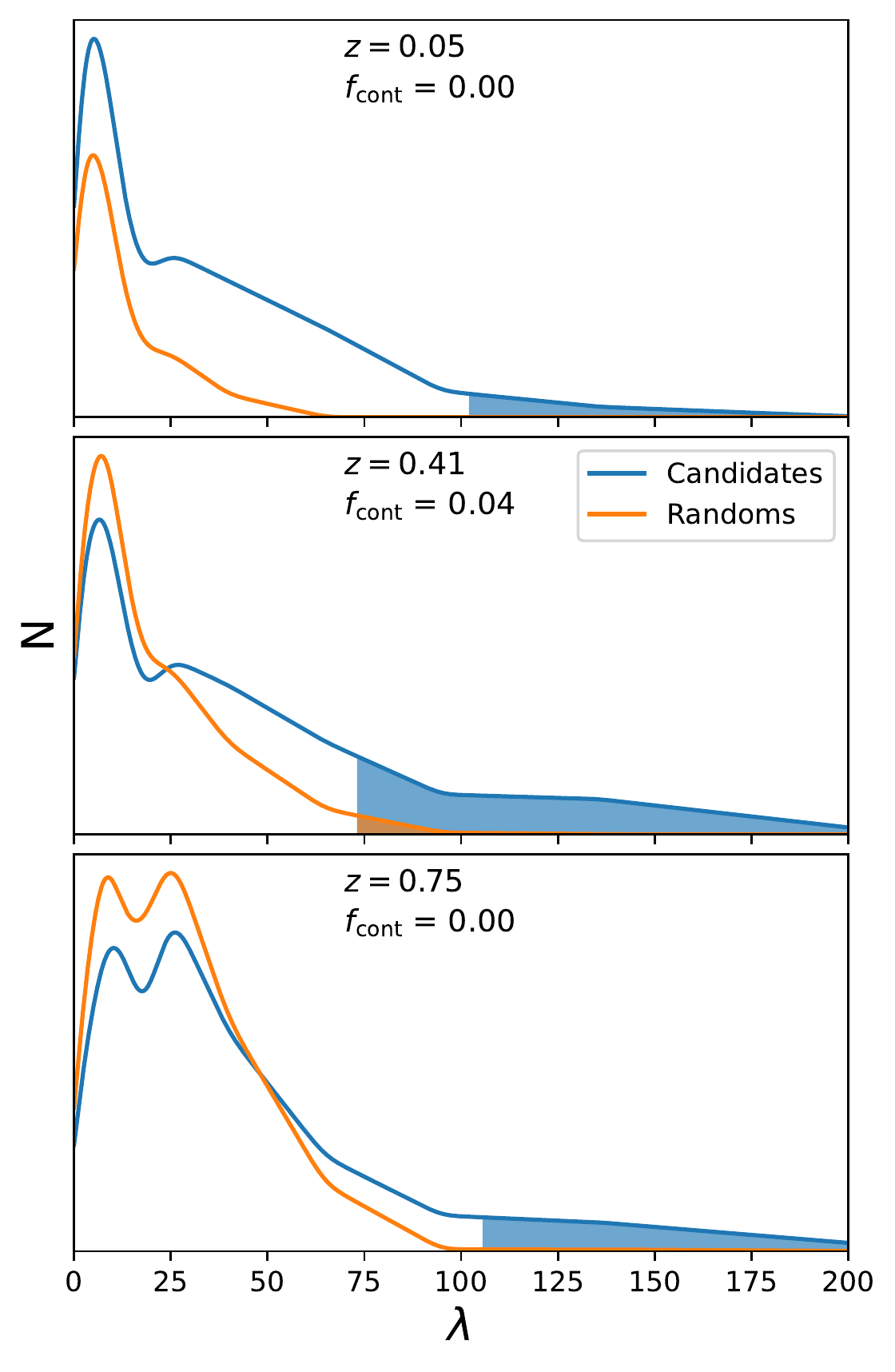}
    \vskip-0.15in
    \caption{Examples of normalized richness distributions for random lines of sight (orange) and for \Planck cluster candidates (blue) for all sources within an estimated $\delta z<0.05$ of 3 \Planck candidates shown from top to bottom at $z=0.05, 0.41, 0.75$. For each of the sources, the area under the curves where the richness is equal to or greater than that of the \Planck candidate is shaded.  These shaded regions correspond to the numerator (orange) and denominator (blue) of  equation~(\ref{eq:fcont}).}
    \label{fig:f_cont_examples}
\end{figure}

This first run gives us up to three possible optical counterparts for each \Planck candidate, with the corresponding photometric redshift, optical center and $\lambda$ for each. For all potential counterparts, the RS galaxy density maps are used to identify the peak richness, which is adopted as the optical center. In the top row of Fig.~\ref{fig:mcmftest} we show the richness distribution in redshift (estimated in this first run) of two different \Planck candidates, at $z_{\rm MCMF} \approx 0.24$ (left) and $z_{\rm MCMF}\approx0.88$ (right), with their corresponding pseudo-color images shown on the bottom row. 

All potential counterparts identified in the first run are then used for a second \MCMF run with the goal of identifying the most likely optical counterpart for each \Planck candidate and refining the estimation of the photometric redshift and richness. We proceed with the second run of \MCMF using the optical counterpart positions as the input, but now using $R_{500}$ as the aperture within which to search for counterparts. $R_{500}$ %is the radius within which the density is 500 times the critical density, and it
is derived using a NFW profile and the \Planck candidate mass estimation, $M_{500}$, at the redshift of each potential counterpart. For each candidate, redshift-dependant masses are estimated using the SZE mass proxy \citep[for details see Section~7.2.2 of][]{PSZ1}. The \Planck flux measured with the matched filter is degenerate with the assumed size. We break this size-flux degeneracy using the flux-mass relation given by \citep[see also  equation 5 in][]{PSZ1}
\begin{equation}\label{eq:mass_y500}
    E^{-2/3} (z) \left[ \dfrac{D^2_{\rm{A}} (z) Y_{500}}{10^{-4} \rm{Mpc}^2} \right] = 10^{-0.19} h_{70}^{-0.21} \left[ \dfrac{M_{500} }{6 \times 10^{14} M_\odot }  \right] ^ {1.79},
\end{equation}
where $E=H(z)/H_0$ and $H(z)$ is the Hubble parameter, and $D_{\rm{A}}$ is the angular diameter distance.
This second run also gives us up to three redshift peaks for each source, but we select the richness peak whose redshift 
is the closest to the output redshift from the first run.  

In summary, we obtain the positions and the redshifts of up to three potential optical counterparts with the first \MCMF run, and in the second run we obtain the final redshifts and richnesses of each of these optical counterparts.  The information from the second run allows us to select the most probable counterpart in most cases, with some candidates having more than one probable counterpart, as discussed below.

\subsection{Quantifying probability of random superpositions}
\label{sec:contamination}

As already noted, with \MCMF we leverage the richness distributions along random lines of sight in the survey as a basis for assigning a probability that each potential optical counterpart of a \Planck selected candidate is a random superposition (e.g., it is not physically associated with the \Planck candidate).  We describe this process below.

\subsubsection{Richness distributions from random lines of sight}

A catalogue along random lines of sight is generated from the original \Planck catalogue, where for each candidate position we generate a random position on the sky, with a minimum radius of approximately 3 times the mean positional uncertainty (5.5~arcmin). We also impose the condition that the random position has to be at least $\sim3\times5.5$~arcmin away from any of the \Planck candidates. We analyze the catalogue of random positions using \MCMF in the same manner as for the data, except that, for the NFW profile used in the second run, the mass information needed to estimate the $R_{500}$ is randomly selected from any of the \Planck candidates (removing the candidate from which the random was generated).

To have sufficient statistics we select two random positions for each \Planck candidate, so we have approximately two times as many random lines of sight as \Planck candidates. Given the large positional uncertainties in the \Planck candidate catalogue, optical counterparts of random lines of sight might be assigned to an optical counterpart of a \Planck candidate. To account for this, we remove from our random lines of sight catalogue those positions that 1) have $\lambda \geq 30$ (e.g., lines of sight with massive clusters), and 2) are within 3~Mpc of any \Planck source from our final, confirmed catalogue and have $|z_{\rm Planck} - z_{\rm random}| < 0.1$. Also, once the second set of random lines of sight has been analysed, we remove those positions that lie within 3~arcmin from any random source position from the first set to avoid double counting  the  same optical structures.

\subsubsection{Estimating the random superposition probability $f_{\rm cont}$}\label{sec:fcont}

With the random lines of sight we can use the $f_{\rm cont}$ estimator presented in \cite{Klein2019}, which is proportional to the probability of individual \Planck candidates being random superpositions of physically unassociated structures \citep[][]{Klein2021}.  By imposing an \fcont\ threshold on our final cluster catalog, we are able to quantify (and therefore also control) the contamination fraction. To estimate $f_{\rm cont}$ for each \Planck candidate, we integrate the normalized richness distributions along random lines of sight $f_{\rm rand}(\lambda, z)$, within multiple redshift bins, 
that have $\lambda \geq \lambda_{\rm src}$, where $\lambda_{\rm src}$ is
the richness of the \Planck candidate. We do the same for the richness distribution of the \Planck candidates $f_{\rm obs}(\lambda,z)$ and then we estimate $f_{\rm cont}$ as the ratio
\begin{equation}\label{eq:fcont}
    f_{\rm cont} (\lambda_{\rm src}, z) = \dfrac{\int^\infty_{\lambda_{\rm src}}d\lambda\,f_{\rm rand}(\lambda, z) }{\int^\infty_{\lambda_{\rm src}} d\lambda\,f_{\rm obs}(\lambda, z)}.
\end{equation}
In Fig.~\ref{fig:f_cont_examples} we show three examples of \Planck candidates with the estimated $f_{\rm cont}$. The blue and orange lines are the interpolated richness distributions of \Planck candidates and of random lines of sight, respectively, at the redshift of the best optical counterpart.  The orange (blue) shaded area shows the integral in the numerator (denominator) in  equation~(\ref{eq:fcont}), starting at the richness $\lambda_{\rm{src}}$ of the \Planck candidate. 

In simple terms, a constant value of $f_{\rm cont}$ can be translated to a redshift-varying richness value $\lambda(z)$. Thus, selecting candidates with a value of $f_{\rm cont}$ lower than some threshold, is similar to requiring the final cluster sample to have a minimum richness that can vary with redshift ($\lambda_\mathrm{min}(z)$), above which the catalogue has a fixed level of contamination. We refer to this threshold as $f_{\rm cont}^{\rm max}$, which yields a catalogue contamination estimated as $f_{\rm cont}^{\rm max} \times (\mathrm{initial\ contamination})$, independent of redshift. Because the initial contamination of the \Planck selected sample is $f_{\rm{SZE-cont}}$ and the final contamination of the cluster sample selected to have $f_{\rm{SZE-cont}}<f_{\rm cont}^{\rm max}$ is $f_{\rm cont}^{\rm max} \times f_{\rm{SZE-cont}}$, one can think of the $f_{\rm cont}^{\rm max}$ selection threshold as the fraction of the contamination in the original candidate sample that ends up being included in the final confirmed cluster sample.  Thus, through selecting an $f_{\rm cont}$ threshold one can control the level of contamination in the final confirmed cluster catalogue.

\section{Results}\label{sec:results}

In Section~\ref{sec:catalogue} we present PSZ-MCMF, the confirmed cluster catalogue extracted from the \Planck candidate list after an analysis of the DES optical followup information using the \MCMF algorithm.  
We then discuss in more detail the mass estimates (Section~\ref{sec:masses}), the cross-comparison with other ICM selected cluster catalogues (Section~\ref{sec:comparisonICM}) and the catalogue contamination and incompleteness (Section~\ref{sec:contaminationandincompleteness}).

\subsection{Creating the PSZ-MCMF cluster catalogue}
\label{sec:catalogue}

As mentioned above, the \MCMF algorithm allows us to identify up to three different richness peaks, corresponding to different possible optical counterparts, for each of the \allPlanck \Planck candidates. To generate a final cluster catalogue, we select the most likely optical counterpart for each of the \allPlanck \Planck candidates by choosing the counterpart that has the lowest probability $f_{\rm cont}$ of being a random superposition (i.e., of being a contaminant rather than a real cluster). 

With \MCMF we identify optical counterparts for 2,938 of the \allPlanck \Planck candidates, whereas for the remaining 192 \Planck candidates no counterpart is found (see Section~\ref{sec:mcmf_nooptical_counterpart} for details). Of the 2,938 candidates with optical counterparts, 2,913 have unique counterparts, while the remaining 25 share their counterpart with another candidate that is closer to that counterpart (see Section~\ref{sec:mcmf_sameopt_counter} for details). Finally, we consider a candidate to be confirmed when its optical counterpart has \fcont below the threshold value $f_{\rm cont}^{\rm max}=0.3$. This results in \lfPlanck confirmed \Planck clusters. Of these confirmed clusters, 120 have two prominent redshift peaks with \fcont below the threshold value $f_{\rm cont}^{\rm max}$, and are considered to be candidates with multiple optical counterparts.

\begin{figure}
    \centering
    \includegraphics[width=\linewidth]{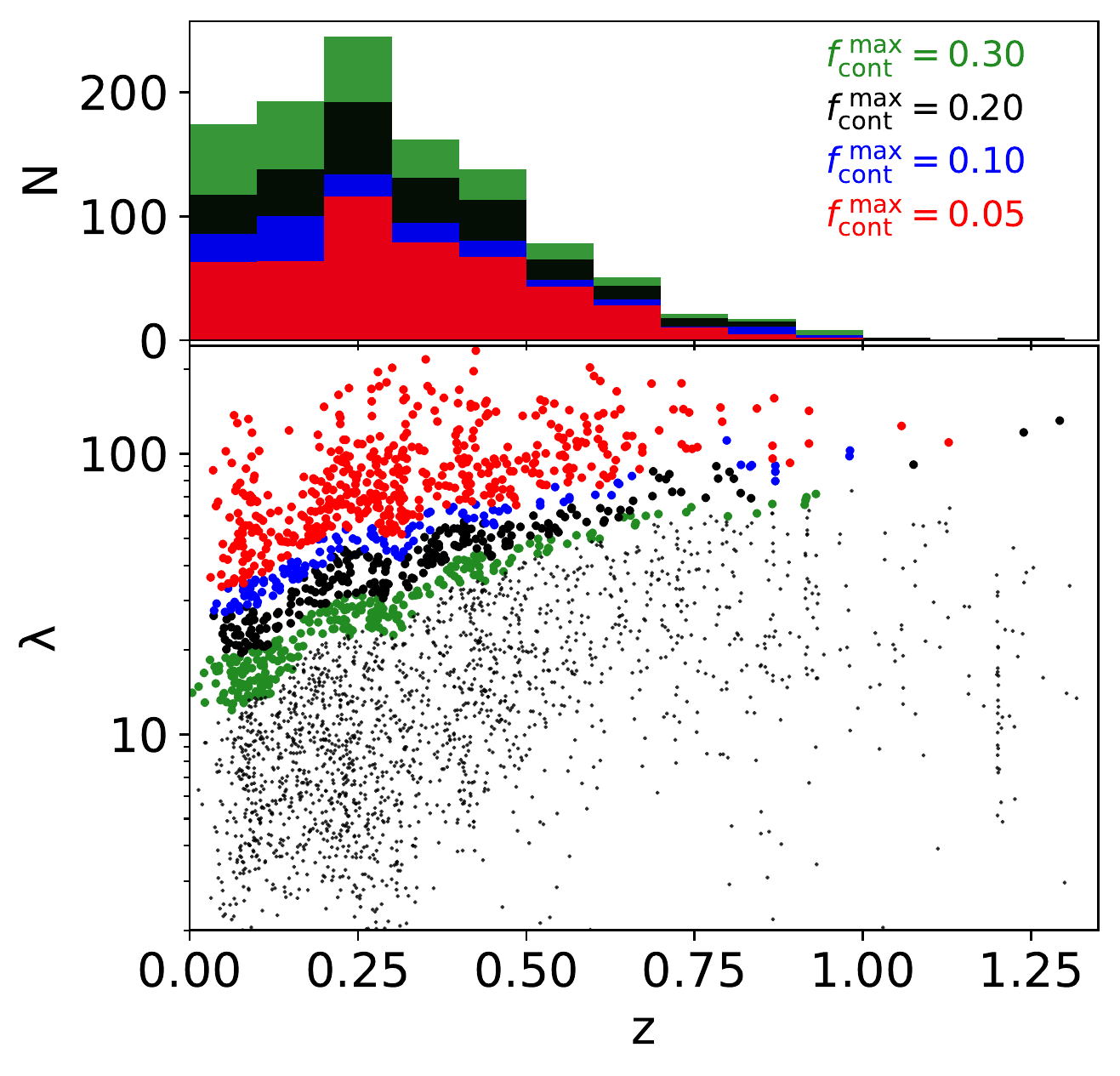}
    \vskip-0.15in
    \caption{\textit{top:} Redshift distribution of the \goodPlanck \Planck candidates. The green, black, blue and red histograms show the distributions of candidates with \fcont below $f_{\rm cont}^{\rm max} = 0.3, 0.2, 0.1$ and $0.05$, respectively. \textit{bottom:}  Richness versus redshift for the best optical counterpart for each \Planck candidate.  Pairs with a probability of being random superpositions (contamination)  $f_{\rm cont} > 0.3$ are shown as small black dots. Bigger green, black, blue and red dots represent counterparts with $0.2\le f_{\rm cont} < 0.3$, $0.1\le f_{\rm cont} < 0.2$, $0.05\le f_{\rm cont} < 0.1$ and $f_{\rm cont} < 0.05$, respectively, corresponding to subsamples with decreasing contamination.
   }
    \label{fig:f_cont}
\end{figure}

The top panel of Fig.~\ref{fig:f_cont} shows the redshift distribution for different values of the threshold $f_{\rm cont}^{\rm max}$, while the bottom panel shows the richness as a function of the redshift for the best optical counterpart of the \Planck candidates in this final catalogue. Small dots represent sources with an estimated $f_{\rm cont}$$\geq$0.3, while bigger dots are color coded as green, black, blue or red according to whether 0.2$\le$$f_{\rm cont}$$<$0.3, 0.1$\le$$f_{\rm cont}$$<$ 0.2, 0.05$\le$$f_{\rm cont}$$<$0.1 or $f_{\rm cont}$$<$0.05, respectively. 

In Table~\ref{tab:f_cont_cluster} we show the number of cluster candidates with \fcont below different values of the threshold $f_{\rm cont}^{\rm max}$, and different \Planck candidate S/N thresholds. With this analysis we are adding 589 (828) clusters to the \Planck cluster sample at $f_{\rm cont}^{\rm max} = 0.2$ (0.3) when going from the \Planck S/N$>$4.5 to S/N$>$3.

\begin{table}\caption{Number of confirmed \Planck clusters with $f_{\rm cont}^{\rm max} = 0.3, 0.2, 0.1$ and $0.05$ presented by row. Results are split by S/N. The second and third columns, for each S/N subsample, show the purity of the sample (Section~\ref{sec:realcontamination}) and the completeness (Section~\ref{sec:incompleteness}). The PSZ-MCMF sample presented in this paper corresponds to clusters with more restrictive \MCMF cleaning in the case of the low signal to noise sample than in the higher signal to noise sample.  These subsamples (see discussion in Section~\ref{sec:madpsz_sample}) are listed in bold face.}
    \label{tab:f_cont_cluster}
    \centering
    \begin{tabular}{ccccccc}
        \hline
        \hline
        $f_{\rm cont}^{\rm max}$ & \multicolumn{3}{c}{S/N$>$3}& \multicolumn{3}{c}{S/N$>$4.5} \\
        & N$_{\rm cl}$ &  Purity & Comp. & N$_{\rm cl}$ & Purity & Comp. \\
        \hline
        0.3 & \lfPlanck & 0.847 & 0.648 & {\bf 264} & {\bf 0.974} & {\bf 0.990}\\
        {\bf 0.2} & {\bf \lowfPlanck} & {\bf 0.898} & {\bf 0.530} & 253 & 0.983 & 0.957\\
        0.1 & \lowerfPlanck & 0.949 & 0.402 & 236 & 0.992 & 0.900\\
        0.05 & \lowestfPlanck & 0.975 & 0.327 & 213 &0.996 & 0.816\\
         \hline
    \end{tabular}
\end{table}

\subsubsection{Candidates with a second optical counterpart}\label{sec:mcmf_secodoptical_counterpart}

If the cluster candidate has two prominent redshift peaks with $f_{\rm cont}$$<$$f_{\rm cont}^{\rm max}$=0.3, where either (1) the redshift offset ($\delta z$ = ($z_1-z_2$)/($1+z_1$)) is greater than 2\% or (2) the on-sky separation is greater than 10~arcmin, then we classify this candidate as a one with multiple optical systems, because a second optical counterpart with $f_{\rm cont}$$<$0.3 is an indication that the probability of being a chance superposition is lower than $f_{\rm cont}^{\rm max} \times f_{\rm{SZE-cont}}$. We give the redshifts, sky-positions, richnesses and other values for this second optical counterpart in the full cluster catalogue. %(i.e., a multi-cluster system that we would expect to merge in the future). 
In the case that both counterparts have the same $f_{\rm cont}$, we select the one that is closer to the \Planck candidate position. In Appendix~\ref{sec:multiple_systems} we discuss a specific example.

\subsubsection{Candidates with no optical counterpart}\label{sec:mcmf_nooptical_counterpart}

Out of the \allPlanck \Planck candidates, there are 192 for which the \MCMF analysis delivers no optical counterpart-- not even with a high $f_{\rm cont}$. Most of these candidates (all but 26) are located near the edges of the DES footprint, suggesting that with more complete optical data many of these candidates could be associated with an optical counterpart. The 26 candidates that lie away from the DES survey edge show either a bright star or bright low-z galaxy near the \Planck position or a lack of photometric information in one or more DES bands. Regions of the sky with these characteristics are masked by \MCMF and this is the likely reason that no optical counterpart is identified for those candidates.

\subsubsection{Candidates sharing the same optical counterpart}\label{sec:mcmf_sameopt_counter}

Given the rather generous search aperture used in the first run of \MCMF\hspace{-1 ex}, it is possible that some \Planck candidates lying near one another on the sky share the same optical counterpart. There are 41 candidates, at $f_{\rm cont}$$<$0.3, that share 20 optical counterparts. The criteria we use to identify these 41 candidates is similar to the one used above to identify candidates with more than one possible optical counterpart. If the distance between the optical counterparts for the two \Planck candidates is less than 10~arcmin and the redshift offset satisfies $|\delta z| \leq 0.02$, then we consider the two candidates to be sharing the same optical counterpart. In Appendix~\ref{sec:same_counterpart} we discuss a specific example.

To account for such cases, we add a column to our catalogue that refers to which \Planck candidate is the most likely SZE counterpart by using the distance between the SZE and the optical centers. The \Planck candidate with the smallest projected distance from the optical center normalized by the positional uncertainty of the \Planck candidate is considered to be the most likely SZE source.

\subsubsection{Final PSZ-MCMF sample}\label{sec:madpsz_sample}

With considerations of this last class we end up with \goodPlanck \Planck candidates, which are the closest to their respective optical counterparts. 
Table~\ref{tab:f_cont_cluster} contains the numbers of confirmed clusters, the purity (Section~\ref{sec:realcontamination}) and the completeness (Section~\ref{sec:incompleteness}) for different selection thresholds in $f_{\rm cont}$ and S/N. Given how the catalogue contamination of \Planck candidates depends strongly on the S/N threshold (see Fig.~\ref{fig:purity_vs_snr_planck}), we decide to use two different values of $f_{\rm{cont}}^{\rm max}$ for the low S/N (S/N$>$3) and high S/N (S/N$>$4.5) samples. The low S/N sample will be defined as clusters with S/N$>$3 that meet the $f_{\rm{cont}}^{\rm max}$=0.2 threshold (second row of the S/N$>$3 sample in Table~\ref{tab:f_cont_cluster}), whereas the high S/N sample will be defined as clusters with S/N$>$4.5 that meet the $f_{\rm{cont}}^{\rm max}$=0.3 threshold (first row of the S/N$>$4.5 sample in Table~\ref{tab:f_cont_cluster}). The combination of these two samples corresponds to the PSZ-MCMF cluster sample, with a total of \madpsz clusters. 

As previously noted in Section~\ref{sec:fcont}, the contamination fraction of the confirmed cluster sample is $f_{\rm{cont}}^{\rm max} \times f_{\rm SZE-cont}$ and depends on the $f_{\rm cont}$ selection threshold applied. The full PSZ-MCMF cluster catalogue %($f_{\rm cont}<0.3$, 85\% purity at S/N$>$3)
will be made available online at the VizieR archive\footnote{http://vizier.u-strasbg.fr/}.  Table~\ref{table:sample_catalog} contains a random subsample of the PSZ-MCMF catalogue with a subset of the columns.

In much of the discussion that follows we focus on the PSZ-MCMF cluster catalog; however, we will define two subsamples that will be used in specific cases: the low S/N sample and the high S/N sample. The low S/N sample ($f_{\rm{cont}}^{\rm max}=0.2$ and S/N$>$3), consists of \lowfPlanck clusters with a $\sim$90\% purity and 53\% completeness. The high S/N sample ($f_{\rm{cont}}^{\rm max}$=0.3 and S/N$>$4.5) consists of 264 clusters with a $\sim$97\% purity and 99\% completeness.  Other sample selections could be made, and the basic properties of twelve samples are presented in Table~\ref{tab:f_cont_cluster}.

%%%%%%%%%%%%%%%%%%%%%%%%%%%%%%%%%%%%%%%%%%%%%%%%%%%%%%%%%%%
%%%%%%%%%%%%%%%%%%%%%%%%%%%%%%%%%%%%%%%%%%%%%%%%%%%%%%%%%%%
%%%%%%%%%%%%%%%%%%%%%%%%%%%%%%%%%%%%%%%%%%%%%%%%%%%%%%%%%%%

\subsubsection{Comparison with spectroscopic redshifts}
\label{sec:spec-z_comparison4}

Starting with the $\sim$2,500 clusters and groups with spectroscopic redshifts used to calibrate the RS models of \texttt{MCMF}, we cross-match the cluster positions with the optical coordinates of each of our \Planck candidates, selecting as matches those that lie within an angular distance of 3~arcmin. We choose to match with the optical counterpart positions, because they provide a more accurate sky position than the \Planck SZE positions, which have a typical uncertainty of 5~arcmin.  We use this cross-matched sample of clusters with spectroscopic redshifts to refine the red-sequence models of the \MCMF algorithm \citep[][]{Klein2019}. 

We find \zspecPlanck clusters in common with the PSZ-MCMF cluster catalogue, including a $z=1.1$ cluster (SPT-CL~J2106-5844).  Of this sample, 18 clusters have another \MCMF richness peak with $f_{\rm cont}$ below the threshold value $f_{\rm cont}^{\rm max}$=0.2. Of these 18 candidates, the primary richness peak (lowest $f_{\rm cont}$) in 16 shows good agreement with the corrsponding spectroscopic redshift $z_{\rm spec}$, while for the remaining two the secondary peak lies at the $z_{\rm spec}$. Of the full cross-matched sample, there are two sources that have no secondary peak and exhibit a large redshift offset in the primary richness peak. We discuss these two cases in Appendix~\ref{sec:specz_comp}.

To characterise the redshift offset, we fit a Gaussian to the distribution of $\Delta z$ = ($z_{\rm spec} - z_{\rm MCMF}$) / (1+z$_{\rm spec}$) of the \zspecPlanck clusters, finding that the standard deviation is $\sigma$=0.00468 (indicating a typical \MCMF redshift uncertainty of 0.47\%), with a mean offset $\mu = -0.00005$ (indicating no \MCMF redshift bias). This is consistent with the previously reported results from applications of the \MCMF algorithm \citep{Klein2019}.

%%%%%%%%%%%%%%%%%%%%%%%%%%%%%%%%%%%%%%%%%%%%%%%%%%%%%%%%%%%
%%%%%%%%%%%%%%%%%%%%%%%%%%%%%%%%%%%%%%%%%%%%%%%%%%%%%%%%%%%
%%%%%%%%%%%%%%%%%%%%%%%%%%%%%%%%%%%%%%%%%%%%%%%%%%%%%%%%%%%

\subsection{Estimating PSZ-MCMF cluster masses}
\label{sec:masses}

Each \Planck candidate comes with a function $M_{500}^{\rm Pl} (Y_{500}, z)$ that allows an initial mass estimate using the redshift and the SZE signal $Y_{500}$ of the candidate (see equation~\ref{eq:mass_y500}).  Therefore, for each of the \madpsz PSZ-MCMF clusters, we use the final photometric redshift from our \MCMF analysis to estimate a mass.   

It is important to note that candidates with multiple optical counterparts may have a biased SZE signature $Y_{500}$ due to contributions from both physical systems, which would impact the estimated $M_{500}^{\rm Pl}$. However, because we do not have enough information to be able to separate the SZE emission coming for each component of the multiple counterparts, we adopt masses that are derived from the redshift of the first ranked richness peak. These masses are biased as discussed further below, and we therefore present a different mass esimate $M_{500}$ in the final PSZ-MCMF catalogue (see the example Table~\ref{table:sample_catalog}).

\begin{figure}
    \centering
    \includegraphics[width=\linewidth]{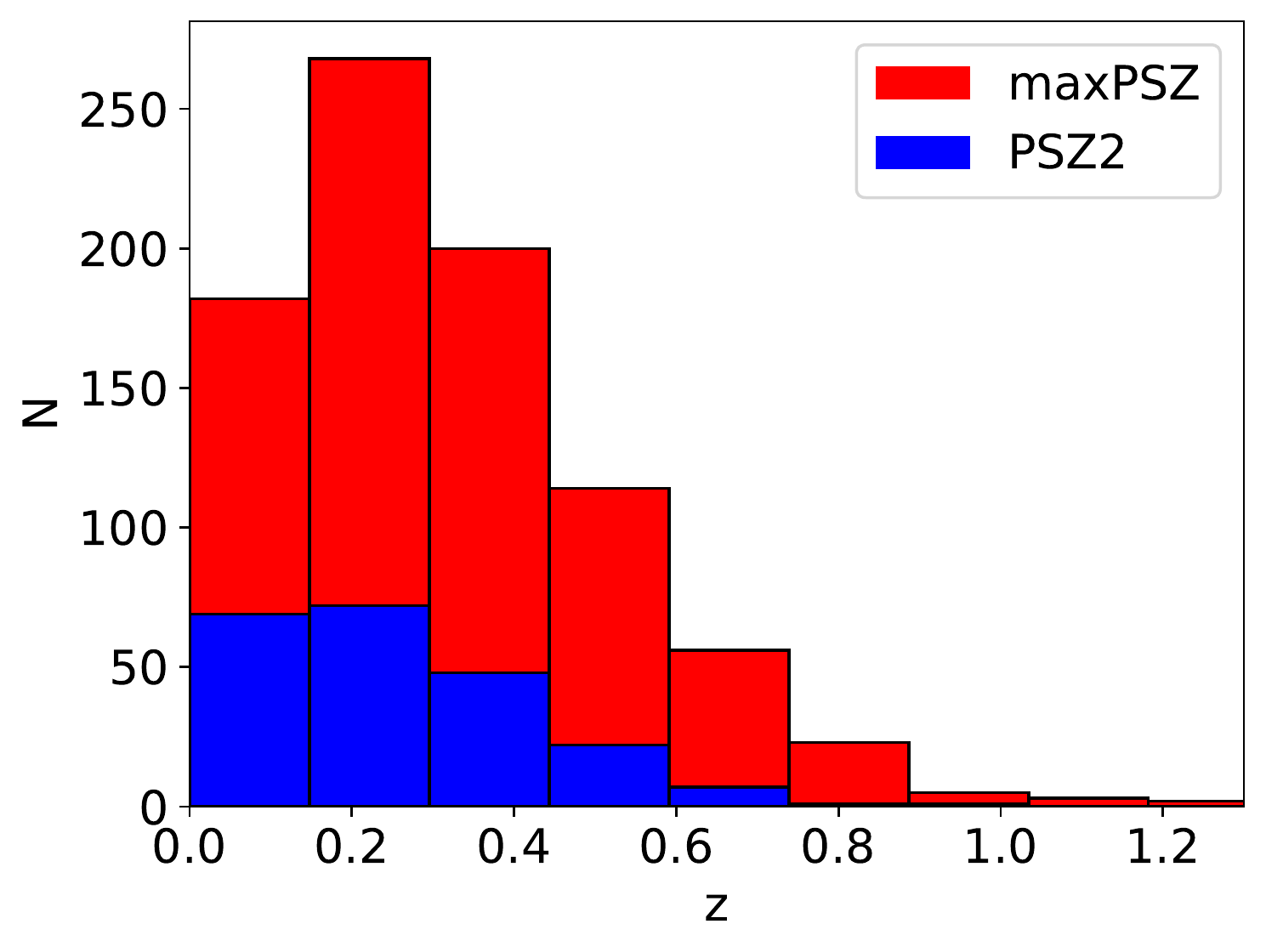}
    \vskip-0.15in
    \caption{Redshift distribution of the PSZ-MCMF clusters (red) and the PSZ2 clusters within the DES region (blue).  The new PSZ-MCMF catalogue presented here is significantly larger and extends to higher redshift.
    }
    \label{fig:redshift_dist_psz2}
\end{figure}

\begin{figure*}
    \centering
    \includegraphics[width=\linewidth]{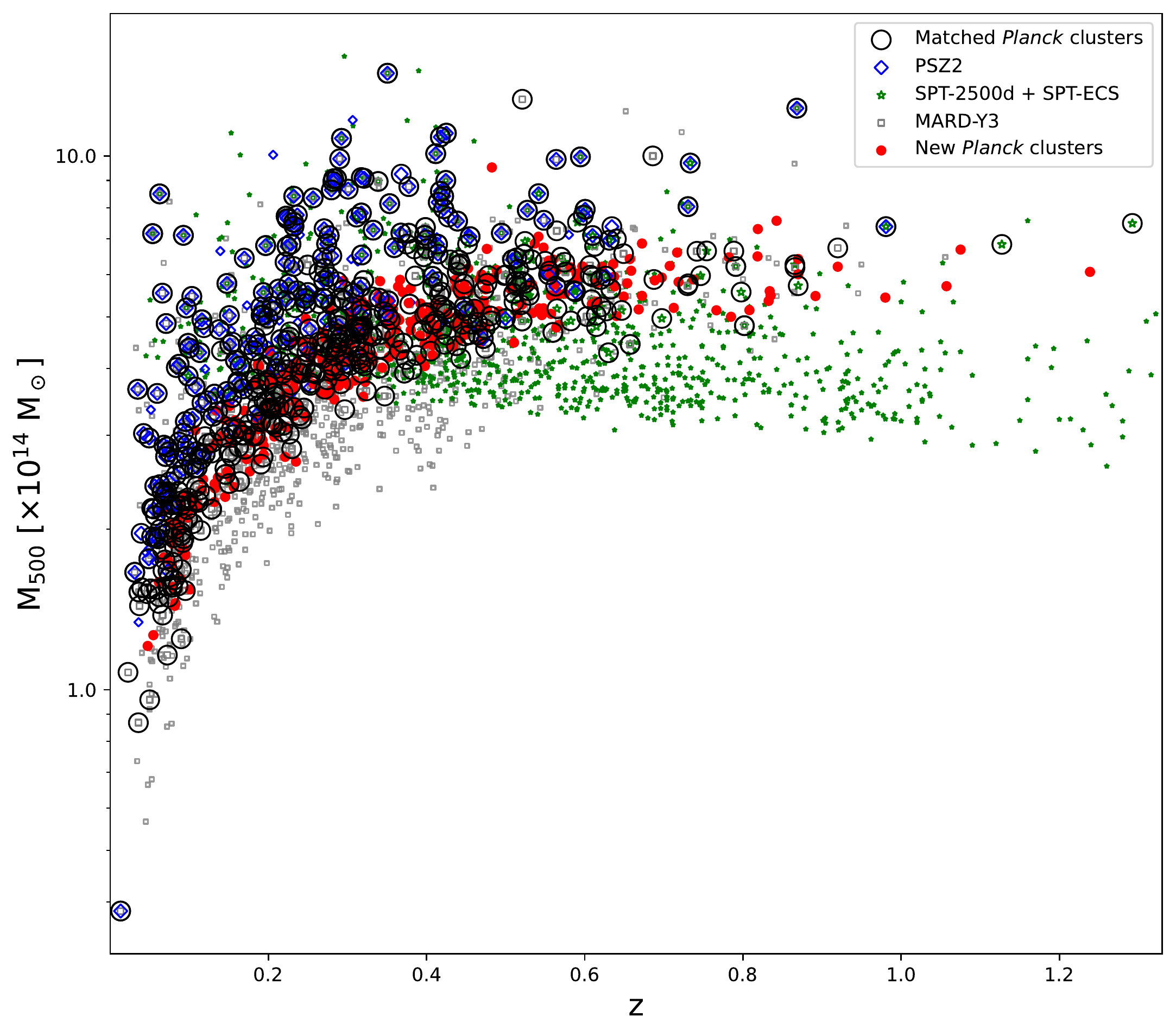}
    \vskip-0.15in
    \caption{Mass versus redshift for the different cluster samples MARD-Y3, PSZ2, SPT and the PSZ-MCMF cluster catalog. SPT, PSZ2 and MARD-Y3 clusters are shown as green stars, blue diamonds or gray squares, respectively.  New PSZ-MCMF clusters identified in this analysis (no match to PSZ2, SPT or MARD-Y3) are shown with red dots whereas clusters that match with at least one of the other catalogues appear as black circles. In the case of matches, masses and redshifts are those of our PSZ-MCMF catalog.}
    \label{fig:mass_vs_redshift}
\end{figure*}

We expect a mass shift between the PSZ-MCMF cluster sample and both SPT and MARD-Y3, that is largely due to the hydrostatic mass bias that has not been accounted for in the \Planck estimated masses \citep[see, e.g.,][]{vonderlinden2014,Hoekstra2015,Planck21CMBlensing,Melin2021}.  In contrast, the SPT and MARD-Y3 masses are calibrated to weak lensing mass measurements \citep{Bocquet2019}, and should not be impacted by hydrostatic mass bias. We therefore apply a systematic bias correction to the \Planck masses to bring all samples onto a common mass baseline represented by $M_{500}$.  

To be able to compare our masses with different surveys accurately, we use cross-matched clusters and estimate the median mass ratio between the SPT/MARD-Y3 and the \Planck mass estimates (see Section \ref{sec:cross-matchcomparison} for details), finding a median of $M_{500}^{\rm Pl}$/$M_{500} \approx 0.8$. This value is in agreement with both weak lensing \citep{vonderlinden2014,Hoekstra2015} and CMB lensing \citep{Planck21CMBlensing} analyses of \Planck clusters. Therefore, we correct the masses of the PSZ-MCMF clusters identified in our current analysis by this factor. Because the previously published PSZ2 catalogue has masses that are calculated in a manner similar to the $M_{500}^{\rm Pl}$ described above, 
we correct PSZ2 masses also using a correction of $(1-b)=0.8$. However, we note a further shift of $M_{500}^{\rm Pl}/M_{500}^{\rm PSZ2} \approx 0.95$ with respect to our corrected masses, and so we further correct the PSZ2 masses for the final comparison.

It should be noted that the mass bias of \Planck clusters is still an ongoing topic. In summary, the masses we present in the following sections and the final cluster catalogue Table~\ref{table:sample_catalog} are denoted as $M_{500}$ and are rescaled to be consistent with results from a range of weak lensing calibration analyses.  These masses are larger than the \Planck masses $M_{500}^{\rm Pl}$ by a factor $1/0.8=1.25$.  

\subsection{Comparison to other ICM selected cluster catalogues}\label{sec:comparisonICM}

%To compare our PSZ-MCMF redshift and mass estimations, we cross-match 
To check how the PSZ-MCMF cluster sample compares to others, we select three cluster catalogues that have been selected using ICM signatures and that lie within the DES footprint: MARD-Y3 \citep{Klein2019}, SPT-2500d \citep{Bocquet2019} along with SPT-ECS \citep[][]{Bleem2020} and PSZ2. MARD-Y3 is an X--ray selected cluster catalogue confirmed with DES Y3 photometric data, using the same tools as for the \Planck analysis presented here. This MARD-Y3 catalogue has 2,900 clusters with $f_{\rm cont} < 0.2$. On the other hand, both the SPT and PSZ2 cluster catalogues are based on SZE selection. For SPT we select sources with a redshift measurement (photometric or spectroscopic), giving a total of 964 clusters. It is worth noting that PSZ2 is an all sky survey, and for the comparison we select sources that lie within the DES survey region and have a redshift measurement (226 clusters).

\subsubsection{Comparison to PSZ2 catalogue}
\label{sec:mcmf_vs_psz2}

We compare the estimated redshifts of our 2,938 candidates with optical counterparts (no $f_{\rm{cont}}^{\rm max}$ applied)  %\Planck sample 
with those from the PSZ2 catalogue \citep{Planck2016}, because the two catalogues should contain a similar number of clusters at S/N$>$4.5, with small variations expected due to the different algorithms used to detect clusters. There are 1,094 PSZ2 clusters with a measured redshift, and, out of those, 226 lie within the DES footprint. We match these 226 clusters with sources from our catalogue that have good photometric redshift estimations and S/N$\geq$4.5, using a matching radius of 3~arcmin. In this case we do the matching using both the \Planck SZE position and the optical positions. 

We find 217 matching sources, but one of those matches does not correspond to the closest cluster in our catalogue so we exclude it and use the 216 remaining sources. Of the 9 PSZ2 sources for which we find no match, 7 have missing photometric information in one or more DES bands. The remaining two clusters with IDs PSZ2 G074.08-54.68 and PSZ2 G280.76-52.30, are further discussed in Appendix~\ref{sec:PSZ2_comp_example}. 

Of this matched sample of 216 systems, 207 (214) systems have \fcont$<0.2$ (0.3) and redshifts that are in good agreement with ours. The cases of disagreement are discussed in detail in Appendix~\ref{sec:PSZ2_comp}. By comparing the 214 matching clusters with $f_{\rm cont} < 0.3$ to the numbers shown on the Table~\ref{tab:f_cont_cluster} (264 at S/N$>$4.5), it becomes apparent that the analysis we describe here has led to photometric redshifts and optical counterparts for 50 PSZ2 clusters that previously had no redshift information. Fig.~\ref{fig:redshift_dist_psz2} shows the redshift distribution of our cluster catalogue (red histogram) and of the PSZ2 catalogue within DES (blue histogram).

\subsubsection{PSZ-MCMF mass-redshift distribution}\label{sec:cross-matchcomparison}
We compare the mass-redshift distribution of PSZ-MCMF clusters with that of MARD-Y3, SPT and PSZ2. Our first step in cross-matching is to select clusters that are the closest to their respective optical counterpart (\madpsz clusters). Then the cross-match comparison is done by using both a positional match within 3~arcmin from the \Planck positions or from the optical positions.  We also add a redshift constraint, where only candidates with a redshift offset $\delta z < 0.02$ (using only the first peak) are considered. This gives a total of 500, 187 and 233 matches with MARD-Y3, PSZ2 and SPT (2500d + ECS), respectively. 
In total, then, \unmatchedPlanck PSZ-MCMF clusters are not matched to any of the three published catalogues.

In Fig.~\ref{fig:mass_vs_redshift} we show the mass versus redshift distribution for the different cluster samples. The SPT, PSZ2 and MARD-Y3 samples are shown as green stars, blue diamonds or gray squares, respectively. PSZ-MCMF clusters are shown with red dots if they are unmatched to clusters in SPT, PSZ2 or MARD-Y3 and as black circles if they are matched.  The red systems are the previously unknown SZE selected clusters in the DES region.  In the case of matches to previously published samples, we adopt the mass and redshift estimates from the PSZ-MCMF sample to ensure the points lie on top of one another.  
Fig.~\ref{fig:mass_vs_redshift} contains more than 10 massive clusters ($M_{500} \gtrsim 10^{15}$ M$_\odot$ and $z < 0.5$) with no matches to the PSZ-MCMF cluster sample. Visual inspection shows that those systems were slightly outside the DES footprint or within masked regions within the general DES footprint.

For MARD-Y3, we clean the unmatched sources by selecting those without multiple X--ray sources to avoid double counting clusters, and also exclude clusters with strong AGN contamination as indicated by their AGN exclusion filter \citep[see section 4.2.1 in][]{Klein2019}. Also, following their mass versus redshift distribution, we use a threshold of $f_{\rm cont} < 0.05$ and also remove sources with a second counterpart with $f_{\rm cont} < 0.05$.

The mass-redshift distribution of our \Planck sample is similar to that of the MARD-Y3 X--ray selected sample, which finds more lower mass systems at lower redshifts.  In contrast, the SPT sample mass-redshift distribution exhibits only a slight redshift trend \citep{Bleem2015}, but it lacks the lower mass systems seen at low redshift in the \Planck and MARD-Y3 samples.  For the \Planck selection, it is the multi-frequency mapping that enables the separation of the thermal SZE from the contaminating CMB primary temperature anisotropy, and this enables the detection of low redshift and low mass systems in a way that resembles the flux limited selection in the MARD-Y3 catalogue.  SPT, on the other hand, has coverage over a narrow range of frequency and cannot as effectively separate the thermal SZE and the primary CMB anisotropies.  The SPT cluster extraction is therefore restricted to a smaller range of angular scales, which is well matched to cluster virial regions at $z\gtrsim0.3$, but at lower redshifts an ever smaller fraction of the SZE signature is obtained, making it ineffective at detecting the low mass and low redshift systems seen in the \Planck and MARD-Y3 samples. At $z \lesssim 0.6$, MARD-Y3 selects lower mass clusters than we are able to with our \Planck sample, but at higher redshifts both catalogues follow similar distributions. When comparing with PSZ2, our new \Planck catalogue contains lower mass clusters  at all redshifts, which is expected given that we are pushing to lower S/N with our \Planck catalogue. Our \Planck sample also contains the first $z>1$ \Planck selected clusters.

%%%%%%%%%%%%%%%%%%%%%%%%%%%%%%%%%%%%%%%%%%%%%%%%%%%%%%%%%%%
%%%%%%%%%%%%%%%%%%%%%%%%%%%%%%%%%%%%%%%%%%%%%%%%%%%%%%%%%%%
%%%%%%%%%%%%%%%%%%%%%%%%%%%%%%%%%%%%%%%%%%%%%%%%%%%%%%%%%%%
\begin{figure}
    \centering
    \includegraphics[width=\linewidth]{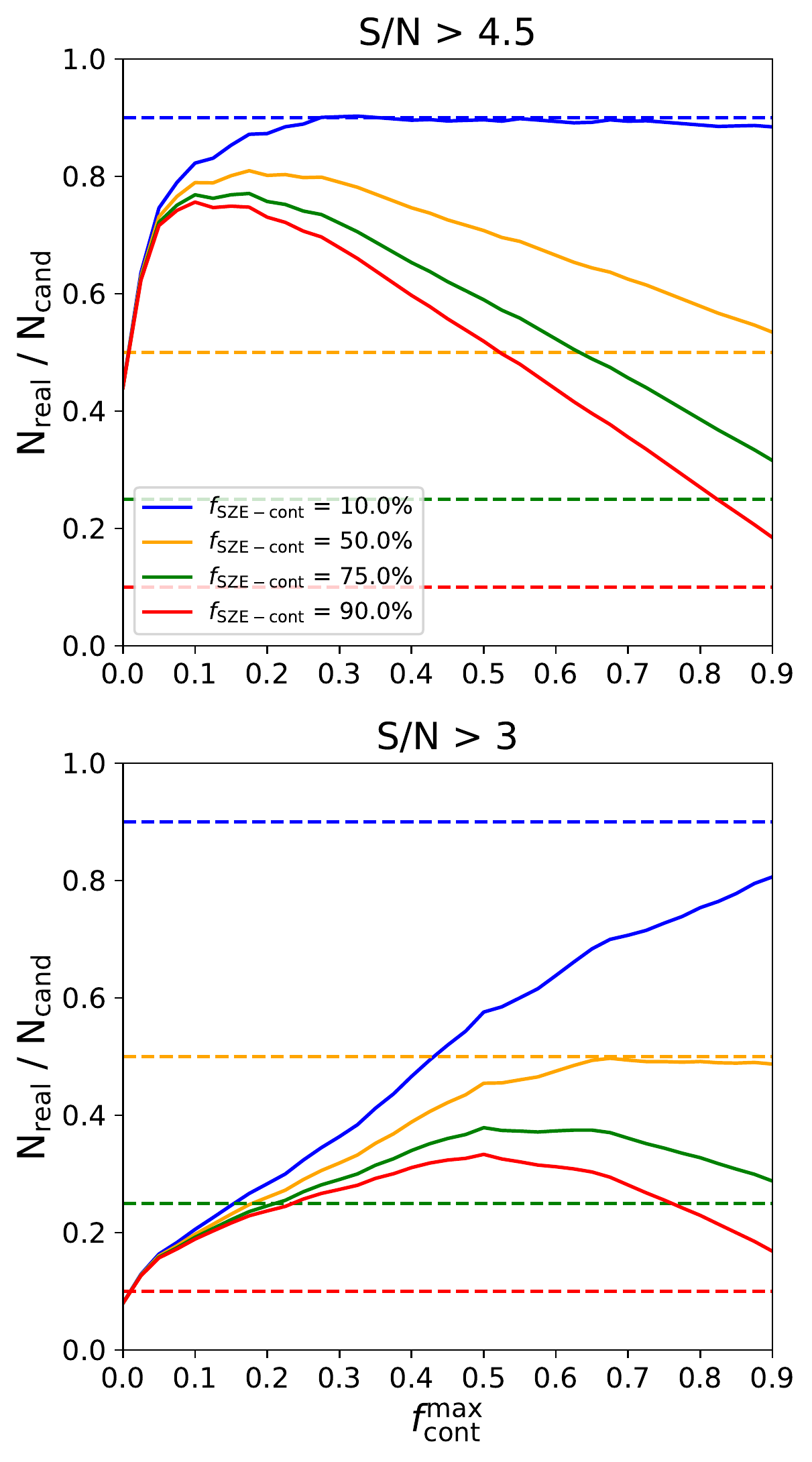}
    \vskip-0.15in
    \caption{Ratio of the estimated number of real clusters $N_{\rm real}$ 
    %out of the \MCMF confirmed cluster candidates $N_{\rm cl}$ 
    to the total number of candidate clusters $N_{\rm cand}$ in the \Planck sample as a function of the \fcont threshold value applied. 
    The solid lines show different curves from  equation~(\ref{eq:nreal_nclus}) with four different values of the contamination $f_{\rm{SZE-cont}}$ of the initial \Planck candidate list. The dashed lines show $1-f_{\rm{SZE-cont}}$, with colors encoding different initial contamination levels.  The analysis indicates an initial contamination of 10\% in the S/N>4.5 (upper) and 50\% in the S/N>3 (lower) \Planck candidate samples.}
    \label{fig:ntrue_ncand}
\end{figure}

\subsection{PSZ-MCMF contamination and incompleteness}
\label{sec:contaminationandincompleteness}

An application of the \Planck based cluster finding algorithm to mock data suggests that at S/N$>$3 we should expect about 75\% of the candidates to be contamination (noise fluctuations; see Section~\ref{sec:data_sz}). In this section we explore that expectation using information from the \MCMF followup. Moreover, as one subjects the confirmed PSZ-MCMF sample to more restrictive \fcont selection thresholds (i.e., smaller values), one is removing not only chance superpositions (contaminants) from the sample, but also some real clusters.  In the following subsections we also explore the incompleteness introduced by the \fcont selection.

\subsubsection{Estimating contamination}
\label{sec:realcontamination}
 
 With the \MCMF analysis results in hand, we can now estimate the true contamination fraction of the initial candidate list by analysing the number of real cluster candidates $N_{\rm real}$ from the number of selected clusters $N_{\rm cl}$ as a function of the $f_{\rm cont}$ threshold $f^{\rm max}_{\rm cont}$ and input \Planck candidate catalogue contamination $f_{\rm SZE-cont}$. The number of real clusters is estimated as
\begin{equation}\label{eq:nreal_nclus}
    N_{\rm{real}} (f^{\rm max}_{\rm cont} ) = N_{\rm{cl}}(f_{\rm cont} < f^{\rm max}_{\rm cont} ) \left[1 - f^{\rm max}_{\rm cont}f_{\rm{SZE-cont}}\right]
\end{equation}
\noindent
where $N_{\rm cl}(f_{\rm cont} < f^{\rm max}_{\rm cont})$ is the total number of confirmed \Planck candidates with $f_{\rm cont} < f^{\rm max}_{\rm cont}$ and $\left[1 - f^{\rm max}_{\rm cont} f_{\rm SZE-cont}\right]$ represents the fraction of real clusters in a sample of \MCMF confirmed clusters.  As discussed in Section~\ref{sec:fcont}, $f_{\rm cont}$ is defined in a cumulative manner and the final contamination of an $f_{\rm cont}<f_{\rm cont}^{\rm max}$ selected sample is the product $f_{\rm cont}^{\rm max}f_{\rm SZE-cont}$ where $f_{\rm SZE-cont}$ is the contamination fraction of the original \Planck candidate list, and $f_{\rm cont}^{\rm max}$ is the fraction of this contamination that makes it into the final confirmed cluster sample.

In this way, we can estimate $N_{\rm real}$ for a number of values of $f^{\rm max}_{\rm cont}$ and $f_{\rm SZE-cont}$. Under the assumption that the \fcont selection restricts contamination as expected, we can then solve for the input candidate list contamination $f_{\rm SZE-cont}$, which again was estimated through \Planck sky simulations to be $\sim$0.75. The catalogue contamination should give a constant ratio of $N_{\rm real}/N_{\rm cand} = 1 - f_{\rm SZE-cont}$ at higher $f^{\rm max}_{\rm cont}$ where this $f_{\rm cont}$ selection becomes unimportant.

It is instructive to start with a less contaminated sample similar to PSZ2 by taking into account only \Planck candidates with S/N$>$4.5 (284 candidates). In Fig.~\ref{fig:ntrue_ncand} we plot the ratio of the number of estimated real clusters $N_{\rm real}$ to the total number of \Planck candidates as a function of the \fcont threshold value $f^{\rm max}_{\rm cont}$ used to select the sample. Each solid curve represents the estimated number of real clusters $N_{\rm real}$, color coded according to the assumed \Planck candidate sample contamination $f_{\rm SZE-cont}$. The horizontal dashed lines show $1-f_{\rm SZE-cont}$, which is showing the fraction of \Planck candidates that are expected to be real clusters and therefore could be confirmed using \MCMF.  We would expect that for threshold values $f^{\rm max}_{\rm cont}$ approaching 1, where the \MCMF selection is having no impact, that the fraction plotted in the figure would reach the value $1-f_{\rm SZE-cont}$.

The input contamination that best describes the high S/N sample is $f_{\rm SZE-cont} = 8.5$\%, where at $f_{\rm cont} < 0.3$ the fraction of confirmed candidates has reached the maximum possible within the \Planck candidate list.  A further relaxing of the \fcont threshold has essentially no impact on the number of real clusters $N_{\rm real}$; it just adds contaminants to the list of \MCMF confirmed clusters $N_{\rm cl}$ at just the rate that matches the expected increase in contamination described in equation (\ref{eq:nreal_nclus}).  
This contamination is in line with the $\sim 91\%$ reliability estimated for the PSZ2 cluster cosmology sample \citep[see Fig.~11 in][]{Planck2016}.

Note the behavior of the blue line at \fcont values $<0.3$.  The confirmed ratio falls away from 90\%, indicating the onset of significant incompleteness in the \MCMF selected sample.  This is an indication that as one uses \fcont to produce cluster samples with lower and lower contamination fractions, one is also losing real systems and thereby increasing incompleteness.  We discuss this further in the next subsection (Section~\ref{sec:incompleteness}).

For the more contaminated S/N$>$3 \Planck candidate sample (\goodPlanck candidates) the results are shown in the bottom panel of Fig.~\ref{fig:ntrue_ncand}.
When the $f_{\rm cont}$ threshold is $0.2$, the estimated number of real clusters $N_{\rm real}$ is roughly 25\% of the total number of \Planck candidates, which implies a 75\% contamination. However, unlike the S/N$>$4.5, the curve does not flatten until $f_{\rm cont} \geq 0.65$, and only for initial contamination values $f_{\rm SZE-cont}=50$\%.  This later flattening reflects the low mass range (and therefore lower richness range) of the S/N$>$3 candidate list.
Additionally, our analysis indicates that the initial contamination of the \Planck S/N$>$3 candidate list is 
$\approx$51\% rather than the estimated 75\% from \Planck mock sky experiments. We explore these differences further in Appendix~\ref{sec:Planckcontamination}.  

Finally, using this 51\% initial contamination (yellow lines), we expect to lose 286 clusters when going from an $f_{\rm cont}$ threshold of $<$0.2 ($\sim$90\% purity) to $<$0.05 ($\sim$97.5\% purity).  Indeed, any $f_{\rm cont}$ threshold below 0.6 will remove real \Planck selected clusters from the \MCMF confirmed sample, but including these systems comes at the cost of higher contamination (purity drops to $\sim$70\%). The purity for different thresholds of $f_{\rm cont}^{\rm max}$ is listed in Table~\ref{tab:f_cont_cluster} for the two \Planck candidate S/N ranges.

Given how the PSZ-MCMF cluster catalog is constructed (the combination of the low and high S/N subsamples), the final purity is estimated to be $\sim$90\%.

\begin{figure}
    \centering
    \includegraphics[width=\linewidth]{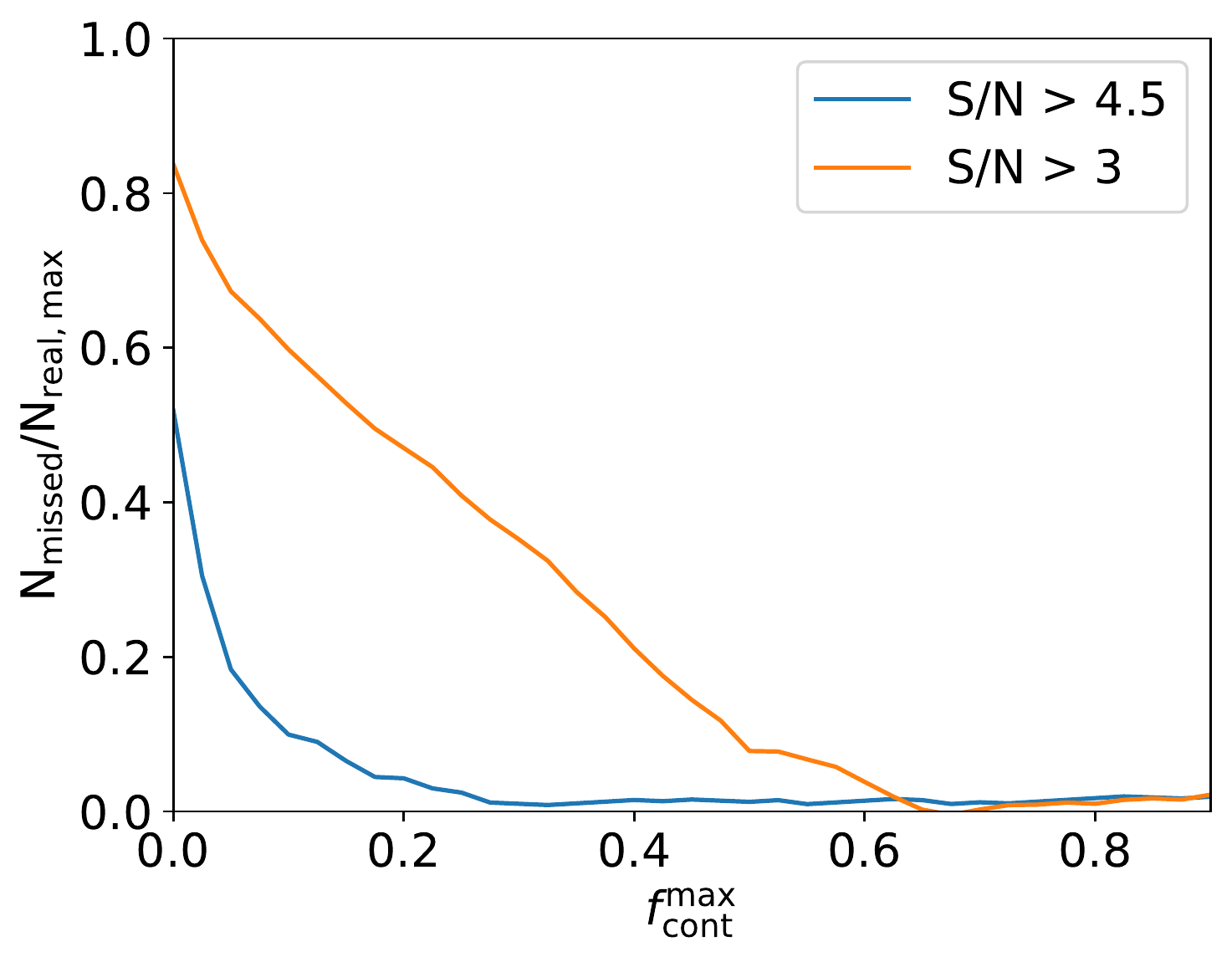}
    \vskip-0.15in
    \caption{Estimate of the fractional incompleteness versus the \fcont threshold $f^{\rm max}_{\rm cont}$ for the S/N$>$4.5 ($f_{\rm{SZE-cont}}=8.5$\%) and S/N$>$3 ($f_{\rm{SZE-cont}}=51$\%) confirmed cluster samples, shown with blue and orange lines, respectively. The contamination of the resulting cluster catalogue is given by $f_{\rm{SZE-cont}}f_{\rm cont}^{\rm max}$ (see Section~\ref{sec:realcontamination} and Table~\ref{tab:f_cont_cluster}).}
    \label{fig:nmissed_ntrueall}
\end{figure}

\subsubsection{Estimating incompleteness}
\label{sec:incompleteness}
From this analysis, we can estimate the number of missed clusters N$_{\rm missed}$ or equivalently the fractional incompleteness for a given \fcont threshold. First, we estimate the maximum number of real clusters in the full sample as $N_{\rm real,max} = N_{\rm{cand}}( 1 - f_{\rm{SZE-cont}}) = 1,427 $ (for the S/N$>$3 sample), where in this case $N_{\rm{cand}}$ is the full \Planck candidate list.  Then, we estimate the number of missed clusters using the total number of expected real clusters minus the number of real SZE selected clusters at a particular \fcont threshold value: 
\begin{equation}
N_{\rm{missed}}(f_{\rm{cont}}<f^{\rm max}_{\rm cont}) = N_{\rm{real,max}} - N_{\rm{real}}(f^{\rm max}_{\rm cont}) 
\end{equation}
where $N_{\rm{real}}(f^{\rm max}_{\rm cont})$ is defined as in equation~(\ref{eq:nreal_nclus}).
In Fig.~\ref{fig:nmissed_ntrueall} we show the ratio of missed clusters over the expected maximum number of real clusters for the samples at S/N$>$3 (orange line) and S/N$>$4.5 (blue line). An $f_{\rm cont}$ threshold of  0.2 in the S/N$>$4.5 \Planck sample would be missing slightly over $3\%$ of the real clusters, while at S/N$>$3 and the same threshold 0.2, we expect to miss $\sim47\%$ of the real clusters.  With an \fcont threshold of 0.05 we miss $\sim70\%$ of the real clusters.  The completeness for different selection thresholds $f_{\rm cont}^{\rm max}$ is shown in Table~\ref{tab:f_cont_cluster}, for the two \Planck candidate S/N ranges. We estimate the completeness of the PSZ-MCMF cluster catalog to be $\sim$54\%.

The higher incompleteness for the lower S/N sample is expected, because as discussed in Section~\ref{sec:masses} this sample pushes to lower masses and therefore lower richnesses than the S/N$>$4.5 sample.  At lower richness, real clusters cannot be as effectively differentiated from the typical background richness distribution (see random line of sight discussion in Section~\ref{sec:contamination}).  In this low mass regime, along with the large positional uncertainties, the cost of creating a higher purity \Planck sample is the introduction of high incompleteness. 

\section{Summary \& Conclusions}\label{sec:summary}

In this analysis we create the PSZ-MCMF cluster catalogue by applying the \MCMF cluster confirmation algorithm to DES photometric data and an SZE selected cluster candidate list extracted down to S/N=3 from \Planck sky maps.
%to create the PSZ-MCMF cluster catalogue.
In contrast to previous analyses employing the \MCMF algorithm, the low angular resolution of \Planck together with the low S/N threshold result in much larger positional uncertainties of the SZE selected candidates.
To overcome this challenge we apply the \MCMF algorithm twice, first using the \Planck candidate coordinates to define a search region with an aperture that is 3 times the \Planck candidate positional uncertainty, and then second using the positions of the optical counterparts found in the first run, with an aperture based on an estimate of the halo radius $R_{500}(z)$ that employs the mass constraints from the \Planck dataset.

We control the contamination of the final, confirmed sample by measuring the parameter \fcont for each \Planck candidate.  As discussed in Section~\ref{sec:fcont}, the value of this parameter is proportional to the probability that the \Planck candidate and its optical counterpart are a chance superposition of physically unassociated systems rather than a real cluster of galaxies.   About 10\% of the \Planck candidates exhibit multiple potential optical counterparts.  In such cases we select the most likely optical counterpart by choosing the one with the lowest \fcont value (lowest chance of being contamination). 

Our analysis of the PSZ-MCMF sample indicates that the initial contamination fraction of the \Planck S/N$>$4.5 candidate list is $f_{\rm SZE-cont}\sim$9\% and the S/N$>$3 candidate list is $f_{\rm SZE-cont}\sim$50\%.  The optical followup with \MCMF allows us to reduce this contamination substantially to the product $f_{\rm cont}^{\rm max}\times f_{\rm SZE-cont}$, where $f_{\rm cont}^{\rm max}$ is the maximum allowed \fcont value in a particular subsample.

Table~\ref{table:sample_catalog} contains the full PSZ-MCMF sample of \madpsz confirmed clusters, defined using an \fcont threshold of 0.3 for S/N$>$4.5 candidates and an \fcont threshold of 0.2 for S/N$>$3 candidates.  Table~\ref{tab:f_cont_cluster} contains the number of clusters, the purity and the completeness of this {cl} catalogue (in bold face) together with other subsamples constructed using smaller \fcont thresholds of 0.2, 0.1 and 0.05 for both \Planck S/N ranges.  Whereas the full catalogue contains \madpsz clusters with a purity of 90\% and completeness of 54\%, the subsample with \fcont$<$0.2 ($<$0.1) contains 842 (604) clusters with purity and completeness of 90\% (95\%) and 53\% (40\%), respectively.

Furthermore, the {cl} cluster sample at S/N$>$3 excludes 47\% of the real clusters when applying a limiting value at $f_{\rm cont} < 0.2$, while the same threshold on the S/N$>$4.5 sample excludes around 4\%.  We attribute the higher incompleteness of the confirmed low S/N sample to the fact that these systems have lower masses and richnesses.  The lower richnesses for the real clusters in this regime are simply more difficult to separate from the characteristic richness variations along random lines of sight in the DES survey.  The relatively large positional uncertainties of the \Planck candidates makes this effect even stronger.

Users are encouraged to select subsamples of the {cl} sample with lower contamination, depending on their particular scientific application.
The PSZ-MCMF catalogue adds 828 previously unknown \Planck identified clusters at S/N$>$3, and it delivers redshifts for 50 previously published S/N$>$4.5 \Planck clusters.

For each of the confirmed clusters we derive photometric redshifts. By comparing the PSZ-MCMF cluster sample with spectroscopic redshifts from the literature, we find a mean redshift offset $<10^{-4}$ and an RMS scatter of 0.47\%.  With these redshifts together with the \Planck mass constraints, we estimate halo masses for all confirmed clusters.  These original \Planck based mass estimates contain no correction for hydrostatic mass bias, and so these are rescaled by the factor $1/0.8=1.25$ to make them consistent with the weak lensing derived SPT cluster masses \citep{Bocquet2019}.  Optical positions, redshifts and halo masses $M_{500}$ are provided for each confirmed cluster in Table~\ref{table:sample_catalog}.

We crossmatch the PSZ-MCMF cluster catalogue to different SZE and X-ray selected cluster catalogues within the DES footprint. 
We find that the PSZ-MCMF mass distribution with redshift is similar to that of the X--ray selected MARD-Y3 cluster catalogue. However, at redshifts lower than $z<0.5$ the PSZ-MCMF catalogue does not contain the lower mass systems that the X-ray selected MARD-Y3 catalogue contains. 
When comparing with the previous \Planck SZE source catalogue PSZ2, we have optical counterparts for most of the systems that lie within the DES footprint, finding in general good agreement with their previously reported redshifts. Compared to the higher S/N PSZ2 sample, we find that most of our new lower S/N PSZ-MCMF systems lie at lower masses at all redshifts and extend to higher redshift, as expected.  Probing to lower masses allows for the confirmation of the first $z>1$ \Planck identified galaxy clusters.  Crudely scaling these results to the full extragalactic sky ($\sim30000$~deg$^2$) implies that the \Planck full sky candidate list confirmed using \MCMF applied to DES like multi-band optical data would yield a sample of $\sim$6000 clusters, which is $\sim6$ times the number of clusters in the PSZ2 all-sky cluster catalogue with redshift information.

\section*{Acknowledgements}

We would like to thank Guillaume Hurier for providing a quality assessment \citep{Aghanim2015} for the Planck candidates in a first version of this work.
We acknowledge financial support from the MPG Faculty Fellowship program, the ORIGINS cluster funded by the Deutsche Forschungsgemeinschaft (DFG, German Research Foundation) under Germany's Excellence Strategy - EXC-2094 - 390783311, and the Ludwig-Maximilians-Universit\"at Munich.
Part of the research leading to these results has received funding from the European Research Council under the European Union's Seventh Framework Programme (FP7/2007–2013)/ERC grant agreement n$^\circ$ 340519.

Funding for the DES Projects has been provided by the U.S. Department of Energy, the U.S. National Science Foundation, the Ministry of Science and Education of Spain, 
the Science and Technology Facilities Council of the United Kingdom, the Higher Education Funding Council for England, the National Center for Supercomputing 
Applications at the University of Illinois at Urbana-Champaign, the Kavli Institute of Cosmological Physics at the University of Chicago, 
the Center for Cosmology and Astro-Particle Physics at the Ohio State University,
the Mitchell Institute for Fundamental Physics and Astronomy at Texas A\&M University, Financiadora de Estudos e Projetos, 
Funda{\c c}{\~a}o Carlos Chagas Filho de Amparo {\`a} Pesquisa do Estado do Rio de Janeiro, Conselho Nacional de Desenvolvimento Cient{\'i}fico e Tecnol{\'o}gico and 
the Minist{\'e}rio da Ci{\^e}ncia, Tecnologia e Inova{\c c}{\~a}o, the Deutsche Forschungsgemeinschaft and the Collaborating Institutions in the Dark Energy Survey. 

The Collaborating Institutions are Argonne National Laboratory, the University of California at Santa Cruz, the University of Cambridge, Centro de Investigaciones Energ{\'e}ticas, 
Medioambientales y Tecnol{\'o}gicas-Madrid, the University of Chicago, University College London, the DES-Brazil Consortium, the University of Edinburgh, 
the Eidgen{\"o}ssische Technische Hochschule (ETH) Z{\"u}rich, 
Fermi National Accelerator Laboratory, the University of Illinois at Urbana-Champaign, the Institut de Ci{\`e}ncies de l'Espai (IEEC/CSIC), 
the Institut de F{\'i}sica d'Altes Energies, Lawrence Berkeley National Laboratory, the Ludwig-Maximilians Universit{\"a}t M{\"u}nchen and the associated Excellence Cluster Universe, 
the University of Michigan, NSF's NOIRLab, the University of Nottingham, The Ohio State University, the University of Pennsylvania, the University of Portsmouth, 
SLAC National Accelerator Laboratory, Stanford University, the University of Sussex, Texas A\&M University, and the OzDES Membership Consortium.

Based in part on observations at Cerro Tololo Inter-American Observatory at NSF's NOIRLab (NOIRLab Prop. ID 2012B-0001; PI: J. Frieman), which is managed by the Association of Universities for Research in Astronomy (AURA) under a cooperative agreement with the National Science Foundation.

The DES data management system is supported by the National Science Foundation under Grant Numbers AST-1138766 and AST-1536171.
The DES participants from Spanish institutions are partially supported by MICINN under grants ESP2017-89838, PGC2018-094773, PGC2018-102021, SEV-2016-0588, SEV-2016-0597, and MDM-2015-0509, some of which include ERDF funds from the European Union. IFAE is partially funded by the CERCA program of the Generalitat de Catalunya.
Research leading to these results has received funding from the European Research
Council under the European Union's Seventh Framework Program (FP7/2007-2013) including ERC grant agreements 240672, 291329, and 306478.
We  acknowledge support from the Brazilian Instituto Nacional de Ci\^encia
e Tecnologia (INCT) do e-Universo (CNPq grant 465376/2014-2).

This manuscript has been authored by Fermi Research Alliance, LLC under Contract No. DE-AC02-07CH11359 with the U.S. Department of Energy, Office of Science, Office of High Energy Physics.

\section*{Data Availability}

The optical data underlying this article corresponds to the DES Y3A2 GOLD photometric data, which is available through the DES data management website\footnote{\url{https://des.ncsa.illinois.edu/}}. The full PSZ-MCMF cluster catalogue (\lfPlanck clusters) will be made available online at the VizieR archive. 

%%%%%%%%%%%%%%%%%%%%%%%%%%%%%%%%%%%%%%%%%%%%%%%%%%



\bibliographystyle{mnras}
\bibliography{mnras_template} % if your bibtex file is called example.bib

\begin{thebibliography}{}
\makeatletter
\relax
\def\mn@urlcharsother{\let\do\@makeother \do\$\do\&\do\#\do\^\do\_\do\%\do\~}
\def\mn@doi{\begingroup\mn@urlcharsother \@ifnextchar [ {\mn@doi@}
  {\mn@doi@[]}}
\def\mn@doi@[#1]#2{\def\@tempa{#1}\ifx\@tempa\@empty \href
  {http://dx.doi.org/#2} {doi:#2}\else \href {http://dx.doi.org/#2} {#1}\fi
  \endgroup}
\def\mn@eprint#1#2{\mn@eprint@#1:#2::\@nil}
\def\mn@eprint@arXiv#1{\href {http://arxiv.org/abs/#1} {{\tt arXiv:#1}}}
\def\mn@eprint@dblp#1{\href {http://dblp.uni-trier.de/rec/bibtex/#1.xml}
  {dblp:#1}}
\def\mn@eprint@#1:#2:#3:#4\@nil{\def\@tempa {#1}\def\@tempb {#2}\def\@tempc
  {#3}\ifx \@tempc \@empty \let \@tempc \@tempb \let \@tempb \@tempa \fi \ifx
  \@tempb \@empty \def\@tempb {arXiv}\fi \@ifundefined
  {mn@eprint@\@tempb}{\@tempb:\@tempc}{\expandafter \expandafter \csname
  mn@eprint@\@tempb\endcsname \expandafter{\@tempc}}}

\bibitem[\protect\citeauthoryear{{Abbott} et~al.,}{{Abbott}
  et~al.}{2016}]{DES2016}
{Abbott} T.,  et~al., 2016, \mn@doi [\mnras] {10.1093/mnras/stw641}, \href
  {https://ui.adsabs.harvard.edu/abs/2016MNRAS.460.1270D} {460, 1270}

\bibitem[\protect\citeauthoryear{{Abbott} et~al.,}{{Abbott}
  et~al.}{2018}]{Abbott2018}
{Abbott} T.~M.~C.,  et~al., 2018, \mn@doi [\apjs] {10.3847/1538-4365/aae9f0},
  \href {https://ui.adsabs.harvard.edu/abs/2018ApJS..239...18A} {239, 18}

\bibitem[\protect\citeauthoryear{{Abbott} et~al.,}{{Abbott}
  et~al.}{2021}]{Abbott_2021}
{Abbott} T.~M.~C.,  et~al., 2021, \mn@doi [\apjs] {10.3847/1538-4365/ac00b3},
  \href {https://ui.adsabs.harvard.edu/abs/2021ApJS..255...20A} {255, 20}

\bibitem[\protect\citeauthoryear{{Aghanim} et~al.,}{{Aghanim}
  et~al.}{2015}]{Aghanim2015}
{Aghanim} N.,  et~al., 2015, \mn@doi [\aap] {10.1051/0004-6361/201424963},
  \href {https://ui.adsabs.harvard.edu/abs/2015A&A...580A.138A} {580, A138}

\bibitem[\protect\citeauthoryear{{Arnaud}, {Pratt}, {Piffaretti},
  {B{\"o}hringer}, {Croston}  \& {Pointecouteau}}{{Arnaud}
  et~al.}{2010}]{Arnaud2010}
{Arnaud} M.,  {Pratt} G.~W.,  {Piffaretti} R.,  {B{\"o}hringer} H.,  {Croston}
  J.~H.,   {Pointecouteau} E.,  2010, \mn@doi [\aap]
  {10.1051/0004-6361/200913416}, \href
  {https://ui.adsabs.harvard.edu/abs/2010A&A...517A..92A} {517, A92}

\bibitem[\protect\citeauthoryear{{Bleem} et~al.,}{{Bleem}
  et~al.}{2015}]{Bleem2015}
{Bleem} L.~E.,  et~al., 2015, \mn@doi [\apjs] {10.1088/0067-0049/216/2/27},
  \href {https://ui.adsabs.harvard.edu/abs/2015ApJS..216...27B} {216, 27}

\bibitem[\protect\citeauthoryear{{Bleem} et~al.,}{{Bleem}
  et~al.}{2020}]{Bleem2020}
{Bleem} L.~E.,  et~al., 2020, \mn@doi [\apjs] {10.3847/1538-4365/ab6993}, \href
  {https://ui.adsabs.harvard.edu/abs/2020ApJS..247...25B} {247, 25}

\bibitem[\protect\citeauthoryear{{Bocquet} et~al.,}{{Bocquet}
  et~al.}{2019}]{Bocquet2019}
{Bocquet} S.,  et~al., 2019, \mn@doi [\apj] {10.3847/1538-4357/ab1f10}, \href
  {https://ui.adsabs.harvard.edu/abs/2019ApJ...878...55B} {878, 55}

\bibitem[\protect\citeauthoryear{{B{\"o}hringer} et~al.,}{{B{\"o}hringer}
  et~al.}{2004}]{RXC}
{B{\"o}hringer} H.,  et~al., 2004, \mn@doi [\aap] {10.1051/0004-6361:20034484},
  \href {https://ui.adsabs.harvard.edu/abs/2004A&A...425..367B} {425, 367}

\bibitem[\protect\citeauthoryear{{Brunner} et~al.,}{{Brunner}
  et~al.}{2022}]{eFEDS2021}
{Brunner} H.,  et~al., 2022, \mn@doi [\aap] {10.1051/0004-6361/202141266},
  \href {https://ui.adsabs.harvard.edu/abs/2022A&A...661A...1B} {661, A1}

\bibitem[\protect\citeauthoryear{{Carlstrom} et~al.,}{{Carlstrom}
  et~al.}{2011}]{Carlstrom2011}
{Carlstrom} J.~E.,  et~al., 2011, \mn@doi [\pasp] {10.1086/659879}, \href
  {https://ui.adsabs.harvard.edu/abs/2011PASP..123..568C} {123, 568}

\bibitem[\protect\citeauthoryear{{Crocce}, {Castander}, {Gazta{\~n}aga},
  {Fosalba}  \& {Carretero}}{{Crocce} et~al.}{2015}]{Crocce2015}
{Crocce} M.,  {Castander} F.~J.,  {Gazta{\~n}aga} E.,  {Fosalba} P.,
  {Carretero} J.,  2015, \mn@doi [\mnras] {10.1093/mnras/stv1708}, \href
  {https://ui.adsabs.harvard.edu/abs/2015MNRAS.453.1513C} {453, 1513}

\bibitem[\protect\citeauthoryear{{DeRose} et~al.,}{{DeRose}
  et~al.}{2019}]{DeRose2019}
{DeRose} J.,  et~al., 2019, arXiv e-prints, \href
  {https://ui.adsabs.harvard.edu/abs/2019arXiv190102401D} {p. arXiv:1901.02401}

\bibitem[\protect\citeauthoryear{{Delabrouille} et~al.,}{{Delabrouille}
  et~al.}{2013}]{Delabrouille2013}
{Delabrouille} J.,  et~al., 2013, \mn@doi [\aap] {10.1051/0004-6361/201220019},
  \href {https://ui.adsabs.harvard.edu/abs/2013A&A...553A..96D} {553, A96}

\bibitem[\protect\citeauthoryear{{Drlica-Wagner} et~al.,}{{Drlica-Wagner}
  et~al.}{2018}]{Drlica2018}
{Drlica-Wagner} A.,  et~al., 2018, \mn@doi [\apjs] {10.3847/1538-4365/aab4f5},
  \href {https://ui.adsabs.harvard.edu/abs/2018ApJS..235...33D} {235, 33}

\bibitem[\protect\citeauthoryear{{Finoguenov} et~al.,}{{Finoguenov}
  et~al.}{2020}]{CODEX2020}
{Finoguenov} A.,  et~al., 2020, \mn@doi [\aap] {10.1051/0004-6361/201937283},
  \href {https://ui.adsabs.harvard.edu/abs/2020A&A...638A.114F} {638, A114}

\bibitem[\protect\citeauthoryear{{Flaugher} et~al.,}{{Flaugher}
  et~al.}{2015}]{Flaugher2015}
{Flaugher} B.,  et~al., 2015, \mn@doi [\aj] {10.1088/0004-6256/150/5/150},
  \href {https://ui.adsabs.harvard.edu/abs/2015AJ....150..150F} {150, 150}

\bibitem[\protect\citeauthoryear{{Gladders}, {Yee}, {Majumdar}, {Barrientos},
  {Hoekstra}, {Hall}  \& {Infante}}{{Gladders} et~al.}{2007}]{Gladders07}
{Gladders} M.~D.,  {Yee} H.~K.~C.,  {Majumdar} S.,  {Barrientos} L.~F.,
  {Hoekstra} H.,  {Hall} P.~B.,   {Infante} L.,  2007, \mn@doi [\apj]
  {10.1086/509909}, \href
  {https://ui.adsabs.harvard.edu/abs/2007ApJ...655..128G} {655, 128}

\bibitem[\protect\citeauthoryear{{Grandis} et~al.,}{{Grandis}
  et~al.}{2020}]{Grandis20}
{Grandis} S.,  et~al., 2020, \mn@doi [\mnras] {10.1093/mnras/staa2333}, \href
  {https://ui.adsabs.harvard.edu/abs/2020MNRAS.498..771G} {498, 771}

\bibitem[\protect\citeauthoryear{{Grandis} et~al.,}{{Grandis}
  et~al.}{2021}]{Grandis21}
{Grandis} S.,  et~al., 2021, \mn@doi [\mnras] {10.1093/mnras/stab869}, \href
  {https://ui.adsabs.harvard.edu/abs/2021MNRAS.504.1253G} {504, 1253}

\bibitem[\protect\citeauthoryear{{Herranz}, {Sanz}, {Hobson}, {Barreiro},
  {Diego}, {Mart{\'\i}nez-Gonz{\'a}lez}  \& {Lasenby}}{{Herranz}
  et~al.}{2002}]{Herranz2002}
{Herranz} D.,  {Sanz} J.~L.,  {Hobson} M.~P.,  {Barreiro} R.~B.,  {Diego}
  J.~M.,  {Mart{\'\i}nez-Gonz{\'a}lez} E.,   {Lasenby} A.~N.,  2002, \mn@doi
  [\mnras] {10.1046/j.1365-8711.2002.05704.x}, \href
  {https://ui.adsabs.harvard.edu/abs/2002MNRAS.336.1057H} {336, 1057}

\bibitem[\protect\citeauthoryear{{High} et~al.,}{{High}
  et~al.}{2010}]{High2010}
{High} F.~W.,  et~al., 2010, \mn@doi [\apj] {10.1088/0004-637X/723/2/1736},
  \href {https://ui.adsabs.harvard.edu/abs/2010ApJ...723.1736H} {723, 1736}

\bibitem[\protect\citeauthoryear{{Hilton} et~al.,}{{Hilton}
  et~al.}{2021}]{ACTClusters}
{Hilton} M.,  et~al., 2021, \mn@doi [\apjs] {10.3847/1538-4365/abd023}, \href
  {https://ui.adsabs.harvard.edu/abs/2021ApJS..253....3H} {253, 3}

\bibitem[\protect\citeauthoryear{{Hoekstra}, {Herbonnet}, {Muzzin}, {Babul},
  {Mahdavi}, {Viola}  \& {Cacciato}}{{Hoekstra} et~al.}{2015}]{Hoekstra2015}
{Hoekstra} H.,  {Herbonnet} R.,  {Muzzin} A.,  {Babul} A.,  {Mahdavi} A.,
  {Viola} M.,   {Cacciato} M.,  2015, \mn@doi [\mnras] {10.1093/mnras/stv275},
  \href {https://ui.adsabs.harvard.edu/abs/2015MNRAS.449..685H} {449, 685}

\bibitem[\protect\citeauthoryear{{Klein} et~al.,}{{Klein}
  et~al.}{2018}]{Klein2018}
{Klein} M.,  et~al., 2018, \mn@doi [\mnras] {10.1093/mnras/stx2929}, \href
  {https://ui.adsabs.harvard.edu/abs/2018MNRAS.474.3324K} {474, 3324}

\bibitem[\protect\citeauthoryear{{Klein} et~al.,}{{Klein}
  et~al.}{2019}]{Klein2019}
{Klein} M.,  et~al., 2019, \mn@doi [\mnras] {10.1093/mnras/stz1463}, \href
  {https://ui.adsabs.harvard.edu/abs/2019MNRAS.488..739K} {488, 739}

\bibitem[\protect\citeauthoryear{{Klein} et~al.,}{{Klein}
  et~al.}{2022}]{Klein2021}
{Klein} M.,  et~al., 2022, \mn@doi [\aap] {10.1051/0004-6361/202141123}, \href
  {https://ui.adsabs.harvard.edu/abs/2022A&A...661A...4K} {661, A4}

\bibitem[\protect\citeauthoryear{{Koulouridis} et~al.,}{{Koulouridis}
  et~al.}{2021}]{X-class2021}
{Koulouridis} E.,  et~al., 2021, \mn@doi [\aap] {10.1051/0004-6361/202140566},
  \href {https://ui.adsabs.harvard.edu/abs/2021A&A...652A..12K} {652, A12}

\bibitem[\protect\citeauthoryear{{Lin}, {Mohr}  \& {Stanford}}{{Lin}
  et~al.}{2004}]{Lin2004}
{Lin} Y.,  {Mohr} J.~J.,   {Stanford} S.~A.,  2004, \apj, \href
  {http://adsabs.harvard.edu/cgi-bin/nph-bib_query?bibcode=2004ApJ...610..745L&amp;db_key=AST}
  {610, 745}

\bibitem[\protect\citeauthoryear{{Liu} et~al.,}{{Liu} et~al.}{2015}]{Liu2015}
{Liu} J.,  et~al., 2015, \mn@doi [\mnras] {10.1093/mnras/stv458}, \href
  {https://ui.adsabs.harvard.edu/abs/2015MNRAS.449.3370L} {449, 3370}

\bibitem[\protect\citeauthoryear{{Liu} et~al.,}{{Liu} et~al.}{2022}]{Liu2021}
{Liu} A.,  et~al., 2022, \mn@doi [\aap] {10.1051/0004-6361/202141120}, \href
  {https://ui.adsabs.harvard.edu/abs/2022A&A...661A...2L} {661, A2}

\bibitem[\protect\citeauthoryear{{Marriage} et~al.,}{{Marriage}
  et~al.}{2011}]{ACT2011}
{Marriage} T.~A.,  et~al., 2011, \mn@doi [\apj] {10.1088/0004-637X/737/2/61},
  \href {https://ui.adsabs.harvard.edu/abs/2011ApJ...737...61M} {737, 61}

\bibitem[\protect\citeauthoryear{{Maturi}, {Bellagamba}, {Radovich},
  {Roncarelli}, {Sereno}, {Moscardini}, {Bardelli}  \& {Puddu}}{{Maturi}
  et~al.}{2019}]{Maturi2019}
{Maturi} M.,  {Bellagamba} F.,  {Radovich} M.,  {Roncarelli} M.,  {Sereno} M.,
  {Moscardini} L.,  {Bardelli} S.,   {Puddu} E.,  2019, \mn@doi [\mnras]
  {10.1093/mnras/stz294}, \href
  {https://ui.adsabs.harvard.edu/abs/2019MNRAS.485..498M} {485, 498}

\bibitem[\protect\citeauthoryear{{McClintock} et~al.,}{{McClintock}
  et~al.}{2019}]{McClintock2019}
{McClintock} T.,  et~al., 2019, \mn@doi [\mnras] {10.1093/mnras/sty2711}, \href
  {https://ui.adsabs.harvard.edu/abs/2019MNRAS.482.1352M} {482, 1352}

\bibitem[\protect\citeauthoryear{{Melin}, {Bartlett}  \&
  {Delabrouille}}{{Melin} et~al.}{2006}]{Melin2006}
{Melin} J.~B.,  {Bartlett} J.~G.,   {Delabrouille} J.,  2006, \mn@doi [\aap]
  {10.1051/0004-6361:20065034}, \href
  {https://ui.adsabs.harvard.edu/abs/2006A&A...459..341M} {459, 341}

\bibitem[\protect\citeauthoryear{{Melin}, {Bartlett}, {Tarr{\'\i}o}  \&
  {Pratt}}{{Melin} et~al.}{2021}]{Melin2021}
{Melin} J.~B.,  {Bartlett} J.~G.,  {Tarr{\'\i}o} P.,   {Pratt} G.~W.,  2021,
  \mn@doi [\aap] {10.1051/0004-6361/202039471}, \href
  {https://ui.adsabs.harvard.edu/abs/2021A&A...647A.106M} {647, A106}

\bibitem[\protect\citeauthoryear{{Morganson} et~al.,}{{Morganson}
  et~al.}{2018}]{Morganson2018}
{Morganson} E.,  et~al., 2018, \mn@doi [\pasp] {10.1088/1538-3873/aab4ef},
  \href {https://ui.adsabs.harvard.edu/abs/2018PASP..130g4501M} {130, 074501}

\bibitem[\protect\citeauthoryear{{Navarro}, {Frenk}  \& {White}}{{Navarro}
  et~al.}{1996}]{NFW0}
{Navarro} J.~F.,  {Frenk} C.~S.,   {White} S. D.~M.,  1996, \mn@doi [\apj]
  {10.1086/177173}, \href
  {https://ui.adsabs.harvard.edu/abs/1996ApJ...462..563N} {462, 563}

\bibitem[\protect\citeauthoryear{{Navarro}, {Frenk}  \& {White}}{{Navarro}
  et~al.}{1997}]{NFW1}
{Navarro} J.~F.,  {Frenk} C.~S.,   {White} S. D.~M.,  1997, \mn@doi [\apj]
  {10.1086/304888}, \href
  {https://ui.adsabs.harvard.edu/abs/1997ApJ...490..493N} {490, 493}

\bibitem[\protect\citeauthoryear{{Piffaretti}, {Arnaud}, {Pratt},
  {Pointecouteau}  \& {Melin}}{{Piffaretti} et~al.}{2011}]{Piffaretti2011}
{Piffaretti} R.,  {Arnaud} M.,  {Pratt} G.~W.,  {Pointecouteau} E.,   {Melin}
  J.~B.,  2011, \mn@doi [\aap] {10.1051/0004-6361/201015377}, \href
  {https://ui.adsabs.harvard.edu/abs/2011A&A...534A.109P} {534, A109}

\bibitem[\protect\citeauthoryear{{Planck Collaboration} et~al.,}{{Planck
  Collaboration} et~al.}{2014}]{PSZ1}
{Planck Collaboration} et~al., 2014, \mn@doi [\aap]
  {10.1051/0004-6361/201321523}, \href
  {https://ui.adsabs.harvard.edu/abs/2014A&A...571A..29P} {571, A29}

\bibitem[\protect\citeauthoryear{{Planck Collaboration} et~al.,}{{Planck
  Collaboration} et~al.}{2016}]{Planck2016}
{Planck Collaboration} et~al., 2016, \mn@doi [\aap]
  {10.1051/0004-6361/201525823}, \href
  {https://ui.adsabs.harvard.edu/abs/2016A&A...594A..27P} {594, A27}

\bibitem[\protect\citeauthoryear{{Planck Collaboration} et~al.,}{{Planck
  Collaboration} et~al.}{2020}]{Planck21CMBlensing}
{Planck Collaboration} et~al., 2020, \mn@doi [\aap]
  {10.1051/0004-6361/201833880}, \href
  {https://ui.adsabs.harvard.edu/abs/2020A&A...641A...1P} {641, A1}

\bibitem[\protect\citeauthoryear{{Rozo} \& {Rykoff}}{{Rozo} \&
  {Rykoff}}{2014}]{Rozo2014}
{Rozo} E.,  {Rykoff} E.~S.,  2014, \mn@doi [\apj] {10.1088/0004-637X/783/2/80},
  \href {https://ui.adsabs.harvard.edu/abs/2014ApJ...783...80R} {783, 80}

\bibitem[\protect\citeauthoryear{{Rozo} et~al.,}{{Rozo}
  et~al.}{2009a}]{Rozo2009}
{Rozo} E.,  et~al., 2009a, \mn@doi [\apj] {10.1088/0004-637X/699/1/768}, \href
  {https://ui.adsabs.harvard.edu/abs/2009ApJ...699..768R} {699, 768}

\bibitem[\protect\citeauthoryear{{Rozo} et~al.,}{{Rozo}
  et~al.}{2009b}]{Rozo2009b}
{Rozo} E.,  et~al., 2009b, \mn@doi [\apj] {10.1088/0004-637X/703/1/601}, \href
  {https://ui.adsabs.harvard.edu/abs/2009ApJ...703..601R} {703, 601}

\bibitem[\protect\citeauthoryear{{Rykoff} et~al.,}{{Rykoff}
  et~al.}{2014}]{Rykoff2014}
{Rykoff} E.~S.,  et~al., 2014, \mn@doi [\apj] {10.1088/0004-637X/785/2/104},
  \href {https://ui.adsabs.harvard.edu/abs/2014ApJ...785..104R} {785, 104}

\bibitem[\protect\citeauthoryear{{Sheldon}}{{Sheldon}}{2014}]{Sheldon2014}
{Sheldon} E.~S.,  2014, \mn@doi [\mnras] {10.1093/mnrasl/slu104}, \href
  {https://ui.adsabs.harvard.edu/abs/2014MNRAS.444L..25S} {444, L25}

\bibitem[\protect\citeauthoryear{{Song}, {Mohr}, {Barkhouse}, {Warren}, {Dolag}
   \& {Rude}}{{Song} et~al.}{2012a}]{song2012b}
{Song} J.,  {Mohr} J.~J.,  {Barkhouse} W.~A.,  {Warren} M.~S.,  {Dolag} K.,
  {Rude} C.,  2012a, \mn@doi [\apj] {10.1088/0004-637X/747/1/58}, \href
  {https://ui.adsabs.harvard.edu/abs/2012ApJ...747...58S} {747, 58}

\bibitem[\protect\citeauthoryear{{Song} et~al.,}{{Song}
  et~al.}{2012b}]{Song2012}
{Song} J.,  et~al., 2012b, \mn@doi [\apj] {10.1088/0004-637X/761/1/22}, \href
  {https://ui.adsabs.harvard.edu/abs/2012ApJ...761...22S} {761, 22}

\bibitem[\protect\citeauthoryear{{Staniszewski} et~al.,}{{Staniszewski}
  et~al.}{2009}]{Staniszewski2009}
{Staniszewski} Z.,  et~al., 2009, \mn@doi [\apj] {10.1088/0004-637X/701/1/32},
  \href {https://ui.adsabs.harvard.edu/abs/2009ApJ...701...32S} {701, 32}

\bibitem[\protect\citeauthoryear{{Sunyaev} \& {Zeldovich}}{{Sunyaev} \&
  {Zeldovich}}{1972}]{SZ1972}
{Sunyaev} R.~A.,  {Zeldovich} Y.~B.,  1972, Comments on Astrophysics and Space
  Physics, \href {https://ui.adsabs.harvard.edu/abs/1972CoASP...4..173S} {4,
  173}

\bibitem[\protect\citeauthoryear{{Wen} \& {Han}}{{Wen} \&
  {Han}}{2022}]{Wen2022}
{Wen} Z.~L.,  {Han} J.~L.,  2022, \mn@doi [\mnras] {10.1093/mnras/stac1149},
  \href {https://ui.adsabs.harvard.edu/abs/2022MNRAS.513.3946W} {513, 3946}

\bibitem[\protect\citeauthoryear{{von der Linden} et~al.,}{{von der Linden}
  et~al.}{2014}]{vonderlinden2014}
{von der Linden} A.,  et~al., 2014, \mn@doi [\mnras] {10.1093/mnras/stu1423},
  \href {https://ui.adsabs.harvard.edu/abs/2014MNRAS.443.1973V} {443, 1973}

\makeatother
\end{thebibliography}

%%%%%%%%%%%%%%%%%%%%%%%%%%%%%%%%%%%%%%%%%%%%%%%%%%

%%%%%%%%%%%%%%%%% APPENDICES %%%%%%%%%%%%%%%%%%%%%

\appendix

\section{Multiple optical counterparts}

\label{sec:multiple_systems}

\begin{figure}
    \centering
    \includegraphics[width=\linewidth]{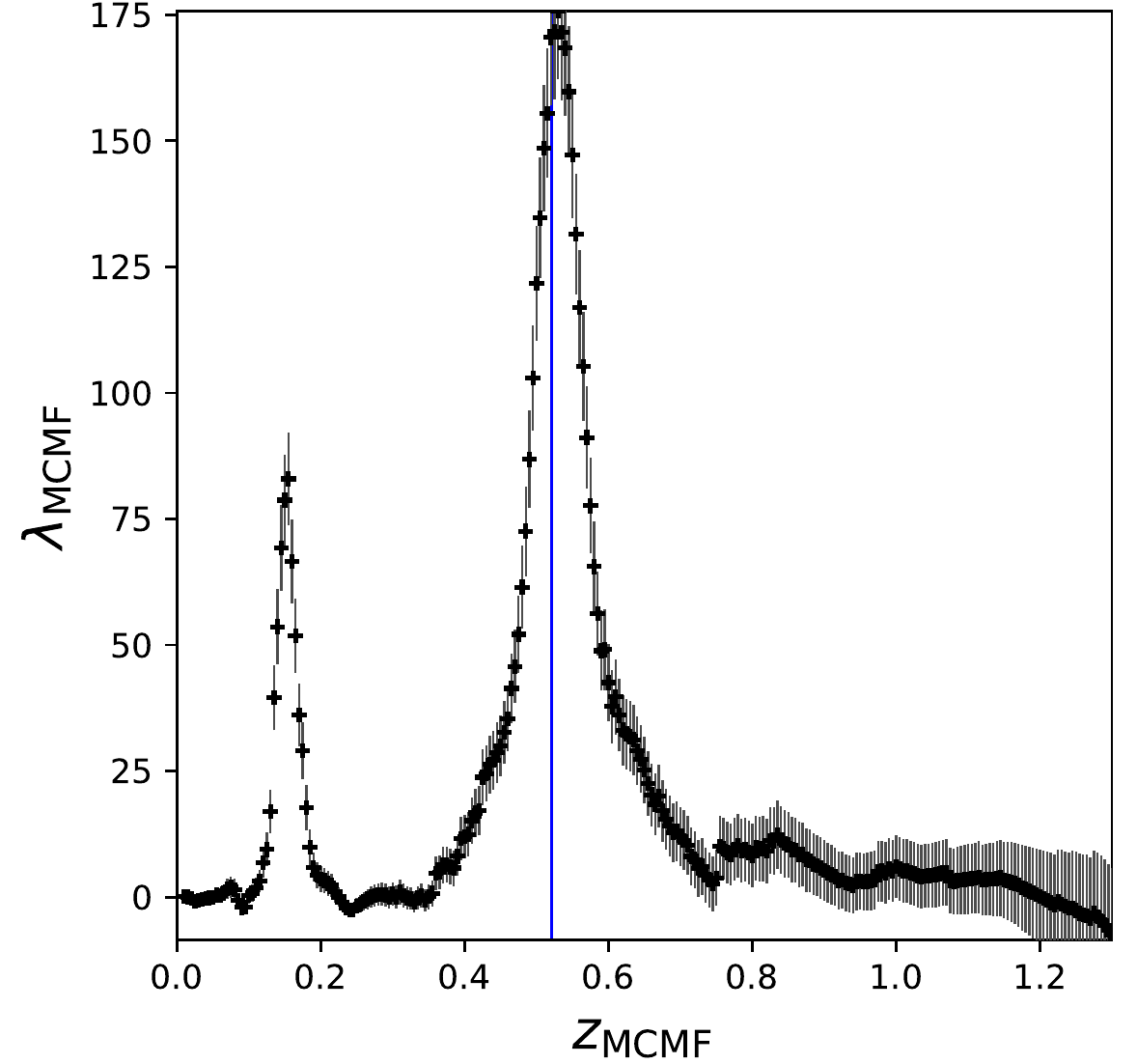}
    \includegraphics[width=\linewidth]{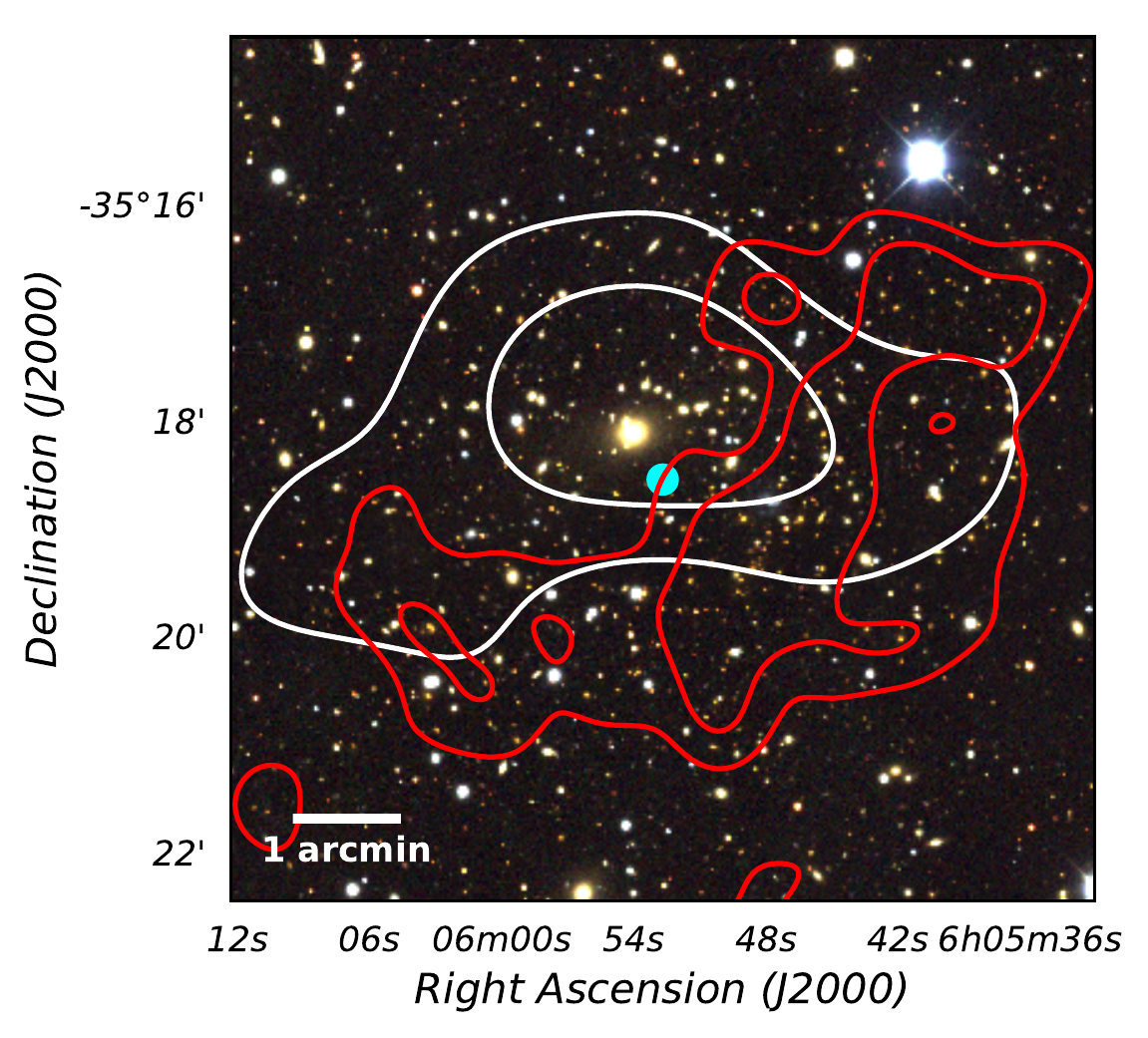}
    \vskip-0.15in
    \caption{\textit{top}: Richness as a function of redshift for the \Planck candidate PSZ-SN3~J0605-3519. Two richness peaks at $z_{\rm MCMF} = 0.15$ and $z_{\rm MCMF} = 0.52$ are evident. \textit{bottom}: pseudo-color image from DES \textit{g}, \textit{r}, \textit{i} cutouts around the \Planck candidate coordinates, marked by a cyan dot. Contours are from the galaxy density maps of the counterpart at $z_{\rm MCMF} = 0.15$ (white) and at $z_{\rm MCMF} = 0.52$ (red).}
    \label{fig:multi_counterpart}
\end{figure}

In Fig.~\ref{fig:multi_counterpart} we show an example of the \Planck candidate PSZ-SN3~J0605-3519, which is classified as a candidate with multiple optical counterparts. The upper figure shows the richness as a function of redshift, which shows two prominent peaks at $z_{\rm MCMF} =0.15$ and $z_{\rm MCMF} =0.52$. The lower image contains the pseudo-color image from $gri$ DES cutouts. White and red contours are derived from the RS galaxy density map for galaxies at $z_{\rm MCMF} = 0.15$ and $z_{\rm MCMF} = 0.52$, respectively. The richness for these two counterparts are $\lambda_{\rm MCMF} = 84$ and $\lambda_{\rm MCMF} = 156$ for the white and red contours, for the two optical candidates at $z_{\rm MCMF} = 0.15$ and $z_{\rm MCMF}=0.52$ respectively. For this candidate, the estimated \fcont of both redshift peaks is 0, indicating a vanishing small probability that either one is a random superposition. We choose the one at $z_{\rm MCMF} = 0.15$ as the ``preferred'' counterpart because it lies nearer to the \Planck candidate position. The reported spectroscopic redshift for this cluster comes from the REFLEX cluster  catalogue, with $z_{\rm spec} = 0.1392$ for cluster RXCJ0605.8-3518 \citep[][]{RXC}.

\section{Shared optical counterpart}

\label{sec:same_counterpart}

\begin{figure}
    \centering
    \includegraphics[width=\linewidth]{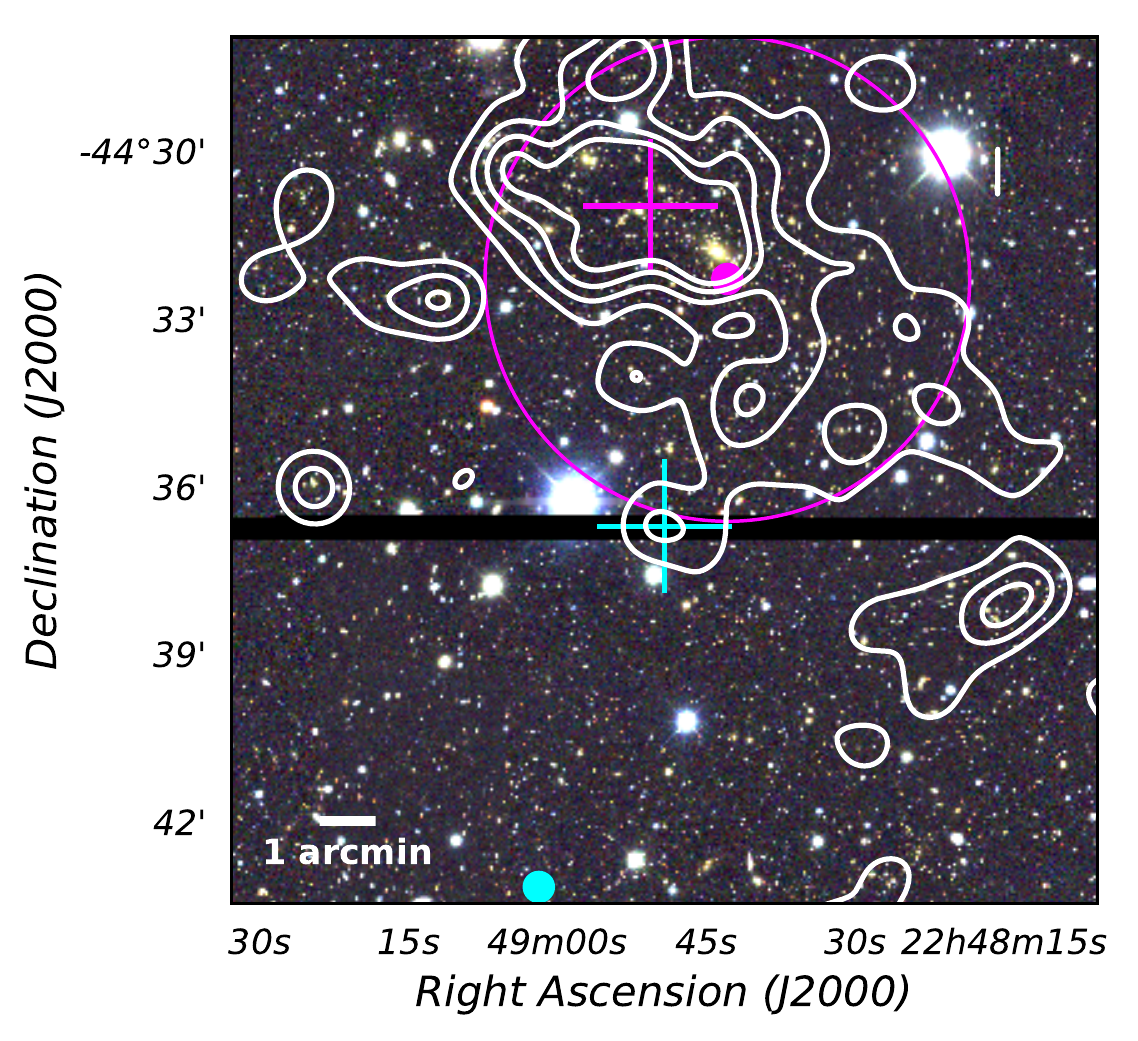}
    \vskip-0.15in
    \caption{Pseudo-color image from DES \textit{g}, \textit{r}, \textit{i} cutouts around the coordinates of \Planck candidates PSZ-SN3~J2248-4430 and PSZ-SN3~J2248-4436, marked by magenta and cyan dots, respectively. White contours are from the galaxy density maps of the counterpart at $z_{\rm MCMF} = 0.35$. Crosses mark the position of the optical counterparts associated with each of the \Planck sources, color coded according to the \Planck source.}
    \label{fig:multiple_candidates}
\end{figure}

In Fig.~\ref{fig:multiple_candidates} we show an example of two \Planck candidates (PSZ-SN3~J2248-4430 with $f_{\rm cont} = 0.00$ and PSZ-SN3~J2248-4436 with $f_{\rm cont} = 0.18$) sharing the same optical counterpart, where the \Planck positions are marked with dots. The optical center of the preferred counterpart for each candidate is marked with a cross of the same color. White contours are the RS galaxy density map from the first \MCMF run, where the optical centers are determined. Both redshifts point toward a cluster at $z_{\rm MCMF} = 0.35$, but it is pretty clear that the two \Planck candidates have resolved to the same optical counterpart. Interestingly, this optical system also corresponds to a South Pole Telescope (SPT) cluster, namely SPT-CL~J2248-4431, with a spectroscopic redshift of $z_{\rm spec}= 0.351$ \citep{Bocquet2019}.

To resolve such cases, we select the \Planck candidate with the smallest projected distance from the optical center normalized by 
the positional uncertainty of the \Planck candidate.  We add a column to the catalogue that identifies which \Planck candidate is the most likely SZE counterpart, flag$_{\rm closest}$, with a value of 0 for candidates pointing to a unique optical counterpart and 1 for candidates which share the optical counterpart with another candidate but are selected as the most likely SZE counterpart. We visually inspected each of the 41 ($f_{\rm cont} < 0.3$) cases, looking not only at the separation, but also at the S/N of the candidates, and the estimated \fcont and $\lambda$. The method described above correctly identifies the most likely candidate for a counterpart in 18 out of 20 cases for candidates at $f_{\rm cont} < 0.3$. For the remaining two, we manually select the most likely SZE source. The final PSZ-MCMF cluster catalogue contains \madpsz clusters, which are the most likely SZE counterparts of their respective optical counterpart.

\begin{landscape}

\begin{table}
\centering
\caption{Thirty entries of the new PSZ-MCMF cluster catalogue, limited to the most important columns of the catalogue. The full cluster catalogue containing \madpsz clusters will be available online via VizieR server at CDS (http://vizier.u-strasbg.fr) and as supplementary material, including additional columns. The first five columns show the cluster name, the \Planck position, the S/N and the positional uncertainty of the candidates. The next columns are \MCMF derived quantities for the most likely optical counterpart (lowest $f_{\rm cont}$). These are: (6 and 7) the optical position of the cluster, (8) the distance between the optical and the \Planck positions normalised by the positional uncertainty, (9) the redshift and (10) the richness. Column (11) is the estimated probability that the source is a random superposition and column (12) contains the mass estimate $M_{500}$. Columns (13--15) contain flags that are described in the text.}
\label{table:sample_catalog}
\begin{threeparttable}
\begin{tabular}{l r r r c r r c c r c r c c c}
    
    \hline
    \hline
    \multicolumn{1}{c}{Name} & 
    \multicolumn{1}{c}{R.A.$_{\rm{\Planck}}$} & 
    \multicolumn{1}{c}{Dec.$_{\rm \Planck}$} & 
    \multicolumn{1}{c}{S/N} & 
    \multicolumn{1}{c}{$\sigma_{\rm pos}$} & 
    \multicolumn{1}{c}{R.A.$^*$} & 
    \multicolumn{1}{c}{Dec.$^*$} & 
    \multicolumn{1}{c}{$\sigma_{\rm SZ-Op}^*$} & 
    \multicolumn{1}{c}{$z_{\rm \MCMF}^*$} & 
    \multicolumn{1}{c}{$\lambda_{\rm \MCMF}^*$} & 
    \multicolumn{1}{c}{$f_{\rm cont}^*$} & 
    \multicolumn{1}{c}{$M_{500}$} & 
    \multicolumn{1}{c}{FLAG} & 
    \multicolumn{1}{c}{FLAG} & 
    \multicolumn{1}{c}{FLAG}\\
    & 
    \multicolumn{1}{c}{deg.} & 
    \multicolumn{1}{c}{deg.} &  
    & 
    \multicolumn{1}{c}{arcmin} & 
    \multicolumn{1}{c}{deg.} & 
    \multicolumn{1}{c}{deg.} & 
    & 
    & 
    & 
    & 
    \multicolumn{1}{c}{$\times 10^{14}$ M$_\odot$} & 
    \multicolumn{1}{c}{COSMO} & 
    \multicolumn{1}{c}{CLEAN} & 
    \multicolumn{1}{c}{QNEURAL}\\
\hline
 PSZ-SN3~J0543-1857 & 85.9328 & -18.9048 & 3.8526 & 5.5398 & 85.9314 & -18.9595 & 0.5922 & 0.6568 & 83.1978 & 0.0530 & 6.4115 & 1 & 1 & 1\\
 PSZ-SN3~J0550-2233 & 87.6461 & -22.6607 & 3.4805 & 6.9070 & 87.6388 & -22.5545 & 0.9247 & 0.0801 & 34.5207 & 0.0476 & 1.9277 & 1 & 1 & 1\\
 PSZ-SN3~J0548-2152 & 87.1619 & -21.9176 & 4.0710 & 6.7710 & 87.0182 & -21.8678 & 1.2612 & 0.0862 & 42.4760 & 0.0197 & 2.3530 & 1 & 1 & 1\\
 PSZ-SN3~J0558-2629 & 89.7651 & -26.5085 & 4.2204 & 6.1337 & 89.7495 & -26.4901 & 0.2257 & 0.2681 & 84.8770 & 0.0000 & 4.4636 & 1 & 1 & 1\\
 PSZ-SN3~J0555-2637 & 88.8111 & -26.6496 & 3.3028 & 5.2495 & 88.8133 & -26.6195 & 0.3446 & 0.2766 & 71.8997 & 0.0000 & 3.9173 & 1 & 1 & 1\\
 PSZ-SN3~J0554-2556 & 88.7011 & -25.9919 & 3.0569 & 5.5877 & 88.6952 & -25.9386 & 0.5750 & 0.2714 & 54.1370 & 0.0505 & 3.6558 & 1 & 1 & 1\\
 PSZ-SN3~J0616-3948 & 94.1341 & -39.8328 & 8.7490 & 2.5549 & 94.1227 & -39.8033 & 0.7235 & 0.1506 & 41.0926 & 0.0545 & 5.0250 & 1 & 1 & 1\\
 PSZ-SN3~J0609-3700 & 92.4479 & -36.9542 & 3.9032 & 4.9092 & 92.4358 & -37.0106 & 0.6994 & 0.2685 & 72.3089 & 0.0000 & 4.1736 & 1 & 1 & 1\\
 PSZ-SN3~J0638-5358 & 99.7055 & -53.9805 & 13.5941 & 2.3905 & 99.6977 & -53.9745 & 0.1891 & 0.2320 & 92.2980 & 0.0000 & 8.4066 & 1 & 1 & 1\\
 PSZ-SN3~J2118+0033 & 319.7380 & 0.5439 & 4.8582 & 2.2707 & 319.7118 & 0.5598 & 0.8107 & 0.2686 & 97.6065 & 0.0000 & 5.4214 & 1 & 1 & 1\\
 PSZ-SN3~J2119+0120 & 319.9540 & 1.3562 & 3.9086 & 4.4458 & 319.9744 & 1.3381 & 0.3680 & 0.1237 & 31.1533 & 0.0955 & 2.8914 & 1 & 1 & 1\\
 PSZ-SN3~J0514-1951 & 78.6623 & -19.9231 & 4.1396 & 4.9319 & 78.6868 & -19.8534 & 0.8936 & 0.1357 & 48.7245 & 0.0156 & 3.3206 & 1 & 1 & 1\\
 PSZ-SN3~J0502-1813 & 75.7216 & -18.1993 & 3.4844 & 5.6044 & 75.6171 & -18.2214 & 1.0885 & 0.5749 & 60.3850 & 0.1406 & 5.5214 & 1 & 1 & 1\\
 PSZ-SN3~J0548-2530 & 87.1631 & -25.4926 & 6.3700 & 5.4270 & 87.1715 & -25.5048 & 0.1593 & 0.0311 & 36.2221 & 0.0431 & 1.6597 & 1 & 1 & 1\\
 PSZ-SN3~J0528-2942 & 82.0841 & -29.7250 & 5.7007 & 3.8145 & 82.0811 & -29.7101 & 0.2386 & 0.1583 & 40.3784 & 0.0691 & 4.1431 & 1 & 1 & 1\\
 PSZ-SN3~J0538-2038 & 84.5847 & -20.6454 & 5.4451 & 4.9996 & 84.5856 & -20.6370 & 0.1010 & 0.0871 & 37.0128 & 0.0421 & 2.9113 & 1 & 1 & 1\\
 PSZ-SN3~J0520-2625 & 80.1255 & -26.4429 & 4.9942 & 1.7900 & 80.1125 & -26.4237 & 0.7537 & 0.2790 & 77.4486 & 0.0000 & 5.2686 & 1 & 1 & 1\\
 PSZ-SN3~J0516-2237 & 79.2379 & -22.6223 & 4.9108 & 4.2022 & 79.2386 & -22.6249 & 0.0390 & 0.2949 & 98.6374 & 0.0000 & 5.6110 & 1 & 1 & 1\\
 PSZ-SN3~J0529-2253 & 82.4507 & -22.8641 & 4.4810 & 5.8471 & 82.4575 & -22.8849 & 0.2227 & 0.1776 & 57.6785 & 0.0221 & 3.7982 & 1 & 1 & 1\\
 PSZ-SN3~J0516-2521 & 79.1042 & -25.3215 & 4.4406 & 6.2807 & 79.0631 & -25.3624 & 0.5279 & 0.2808 & 56.1890 & 0.0331 & 4.6955 & 1 & 1 & 1\\
 PSZ-SN3~J0519-2056 & 80.0775 & -21.0020 & 3.8548 & 5.1985 & 79.9872 & -20.9448 & 1.1760 & 0.3039 & 77.7563 & 0.0000 & 4.9633 & 1 & 1 & 1\\
 PSZ-SN3~J0540-2127 & 85.1922 & -21.4732 & 3.5656 & 5.4369 & 85.2095 & -21.4615 & 0.2195 & 0.5242 & 77.4497 & 0.0461 & 5.7749 & 1 & 1 & 1\\
 PSZ-SN3~J0521-2754 & 80.3559 & -27.9377 & 3.9751 & 3.6068 & 80.3628 & -27.9156 & 0.3821 & 0.3150 & 98.6217 & 0.0000 & 4.7847 & 1 & 1 & 1\\
 PSZ-SN3~J0545-2556 & 86.3885 & -25.9053 & 4.0783 & 5.5571 & 86.3658 & -25.9377 & 0.4132 & 0.0469 & 37.0815 & 0.0371 & 1.5132 & 1 & 1 & 1\\
 PSZ-SN3~J0533-2823 & 83.5642 & -28.3836 & 3.7296 & 5.2059 & 83.4791 & -28.3934 & 0.8706 & 0.1916 & 49.4947 & 0.0542 & 3.1730 & 1 & 1 & 1\\
 PSZ-SN3~J0530-2226 & 82.6648 & -22.4653 & 4.5761 & 3.3990 & 82.6643 & -22.4453 & 0.3528 & 0.1680 & 71.5937 & 0.0067 & 3.6625 & 1 & 1 & 1\\
 PSZ-SN3~J0546-2410 & 86.6542 & -24.1719 & 3.1268 & 5.3599 & 86.5557 & -24.1669 & 1.0076 & 0.3247 & 46.8889 & 0.0797 & 4.1055 & 1 & 1 & 1\\
 PSZ-SN3~J0530-2556 & 82.6693 & -25.7861 & 3.1967 & 5.1885 & 82.7240 & -25.9483 & 1.9600 & 0.1926 & 49.8588 & 0.0535 & 2.9369 & 1 & 1 & 1\\
 PSZ-SN3~J0605-3519 & 91.4696 & -35.3091 & 15.2181 & 2.3889 & 91.4492 & -35.3206 & 0.5087 & 0.5210 & 155.6621 & 0.0000 & 12.7754 & 1 & 1 & 1\\
 PSZ-SN3~J0553-3342 & 88.3485 & -33.7020 & 13.0358 & 1.6357 & 88.3436 & -33.7078 & 0.2613 & 0.4174 & 146.0235 & 0.0000 & 10.8541 & 1 & 1 & 1\\
    \hline
    \hline

\end{tabular}
\begin{tablenotes}
 \footnotesize
   \item[\emph{*}]{In the case of a second prominent optical counterpart (with $f_{\rm cont} < 0.3$) at a different $z_{\rm \MCMF}$, we provide an entry for that counterpart as well.}
  \item[\emph{$\dagger$}]{Masses derived in Section~\ref{sec:masses} are divided by 0.8 to correct for the estimated hydrostatic mass bias.}
 \end{tablenotes}
\end{threeparttable}
\end{table}
\end{landscape}

\section{Redshift comparisons}

\subsection{Spectroscopic redshifts}
\label{sec:specz_comp}
\begin{figure}
    \centering
    \includegraphics[width=\linewidth]{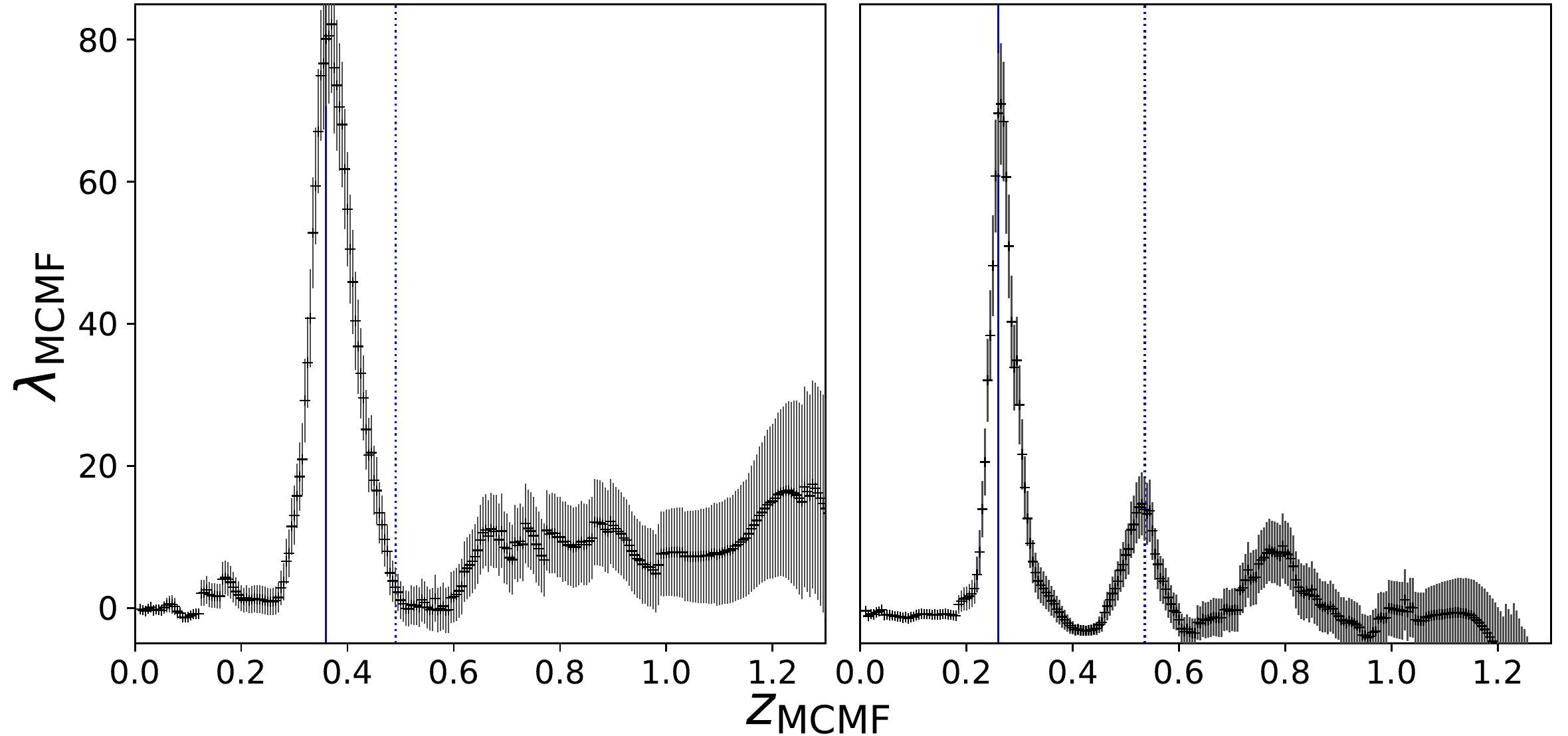}
    \vskip-0.1in
    \caption{Richness as a function of redshift for PSZ-SN3~J2145-0142 (left) and PSZ-SN3~J2347-0009 (right). The PSZ-MCMF redshift is shown with a blue continuous line, while the spec-$z$ is shown with a blue dotted line for both cases.}
    \label{fig:spec-z-comparison}
\end{figure}

As discussed in Section~\ref{sec:spec-z_comparison4}, the full cross-matched sample contains two sources that have no significant secondary peak and exhibit a large redshift offset with respect to $z_{\rm \MCMF}$. We inspect the DES images of these two clusters, namely PSZ-SN3~J2145-0142 ($z_{\rm \MCMF}=0.36$ and $f_{\rm cont} = 0.0$) and PSZ-SN3~J2347-0009 ($z_{\rm \MCMF}=0.26$ and $f_{\rm cont}=0.02$), where the separation between the spectroscopic and optical counterparts are $\sim$150 and $\sim$180 arcseconds, respectively, and find that in both cases the spectroscopic redshift points towards a different structure. In the case of PSZ-SN3~J2145-0142, the spectroscopic redshift seems to be associated with a single galaxy. Fig.~\ref{fig:spec-z-comparison} shows the richness as a function of redshift for both PSZ-SN3~J2145-0142 (left) and PSZ-SN3~J2347-0009 (right), with the spec-$z$ marked with blue dotted lines. In the case of PSZ-SN3~J2347-0009, the measured \fcont for the structure at $z\approx0.53$ is greater than our $f_{\rm cont}^{\rm max} = 0.3$ threshold, indicating that this is not a significant richness peak. 

\subsection{PSZ2 redshifts}
\label{sec:PSZ2_comp}

\begin{figure}
    \centering
    \includegraphics[width=\linewidth]{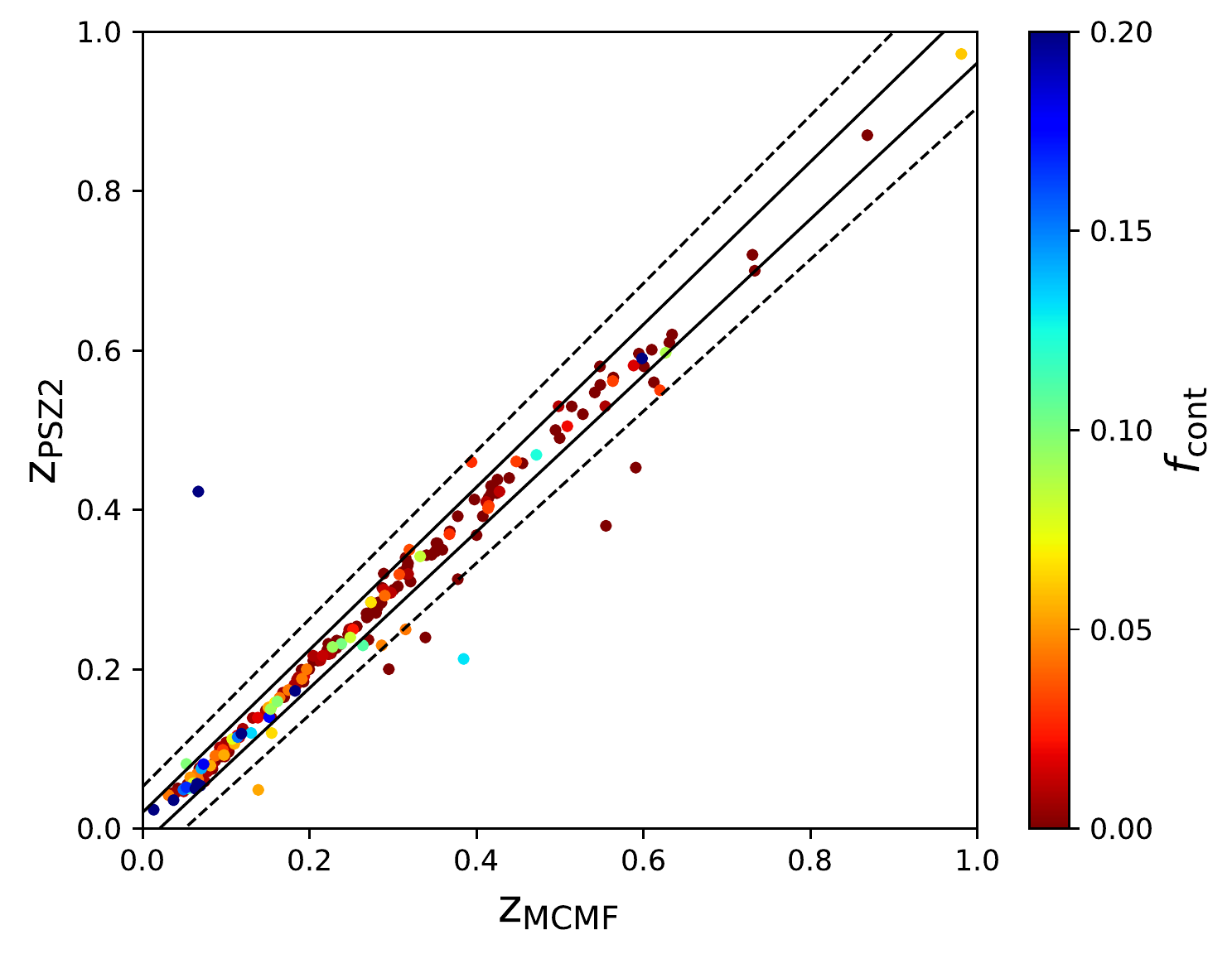}
    \vskip-0.15in
    \caption{Comparison of \MCMF photometric redshifts and those listed in the PSZ2 catalogue for the 216 matching clusters. Each cluster is color coded by the estimated $f_{\rm cont}$, saturated at $f_{\rm cont} = 0.2$. The solid and dashed lines enclose the areas where $|(z_{\rm MCMF} - z_{\rm PSZ2})\times(1+z_{\rm PSZ2})^{-1}| \leq 0.02, 0.05$, respectively.}
    \label{fig:redshift_comparison}
\end{figure}

In Fig.~\ref{fig:redshift_comparison} we show the comparison of PSZ2 redshifts to the \MCMF for the 216 matching systems. On the x-axis, we show the photometric redshift from \MCMF\hspace{-1 ex}, while redshifts from the PSZ2 catalogue are shown on the y-axis. Each source is color-coded according to their $f_{\rm cont}$ estimation. Continuous (dotted) lines show the enclosed area where $\delta z = |(z_{\rm MCMF} - z_{\rm PSZ2})\times(1+z_{\rm PSZ2})^{-1}| \leq 0.02$ (0.05). In case of multiple prominent redshift peaks with $f_{\rm cont} < 0.2$, we choose to plot only the redshift peak with the smaller $\delta z$ for each match.

Fig.~\ref{fig:redshift_comparison} shows that, although most of the estimated \MCMF redshifts have $\Delta z$ offsets at 2\% level or less in comparison to the PSZ2 catalogue, there are some clusters with a higher offset or with $f_{\rm cont} \geq 0.2$. Out of the 216 matching clusters, 207 have $f_{\rm cont} < 0.2$, and 197 (205) have a redshift offset, with respect to the first redshift peak, lower than 2\% (5\%). If we consider also structures with a second peak, we get 201 (209) matches with an offset lower than 2\% (5\%). To further study the reasons for these catalogue discrepancies, we separate between high $f_{\rm cont}$ ($>0.2$) and high $\delta z$ ($>0.05$).

First, out of the 9 clusters with $f_{\rm cont} \geq 0.2$, 8 have redshift offsets $\delta z < 0.02$, with 7 of them having $0.2 \lesssim f_{\rm cont} \lesssim 0.3$. DES images with artifacts such as missing bands can impact the \MCMF estimation of the photometric redshifts or the cluster centres. The \MCMF algorithm includes a masking of regions with artifacts when generating the galaxy density maps, thus avoiding the region entirely. Bright saturated stars can also bias the estimations of the richness and centers depending on where they are located. Thus, \MCMF also masks areas with bright saturated stars for the estimation of the different parameters. 

%As previously noted, 
Out of the 11 matches with $\delta z > 0.05$, 4 have a second significant richness peak that is in agreement with the reported redshift from the PSZ2 catalogue. Of the remaining 7, 1 has a masked area due to a bright star. For the others, the correct counterpart (and therefore redshift) is a matter of debate. 
For one of the systems, the \MCMF analysis finds a peak at the PSZ2 redshift, although the estimated $f_{\rm cont}$ is 0.31, indicating that this counterpart has a much higher probability of being contamination as compared to the primary richness peak with  $f_{\rm cont} = 0.05$.

\section{PSZ2 comparison examples}
\label{sec:PSZ2_comp_example}

\begin{figure}
    \centering
    \includegraphics[width=\linewidth]{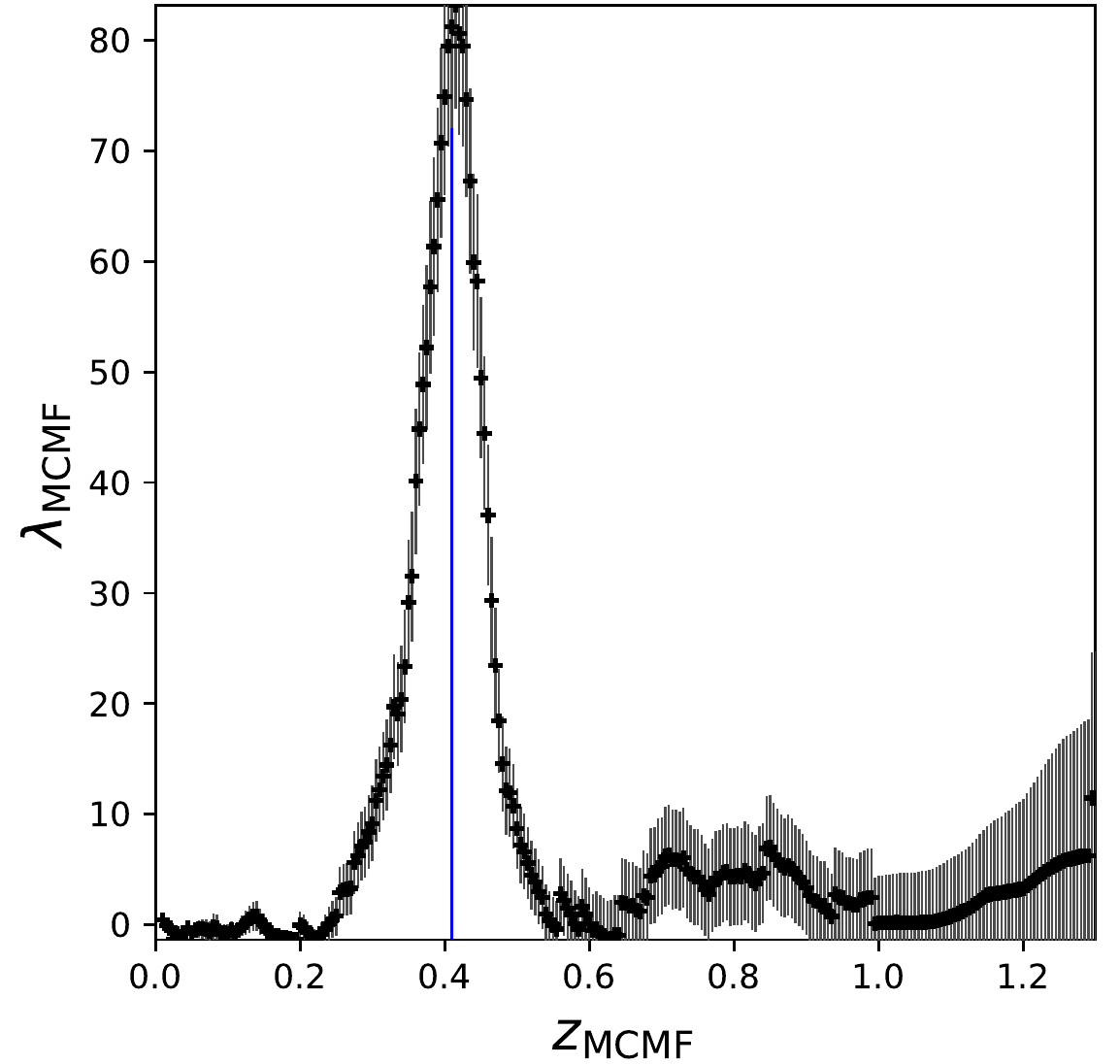}
    \includegraphics[width=\linewidth]{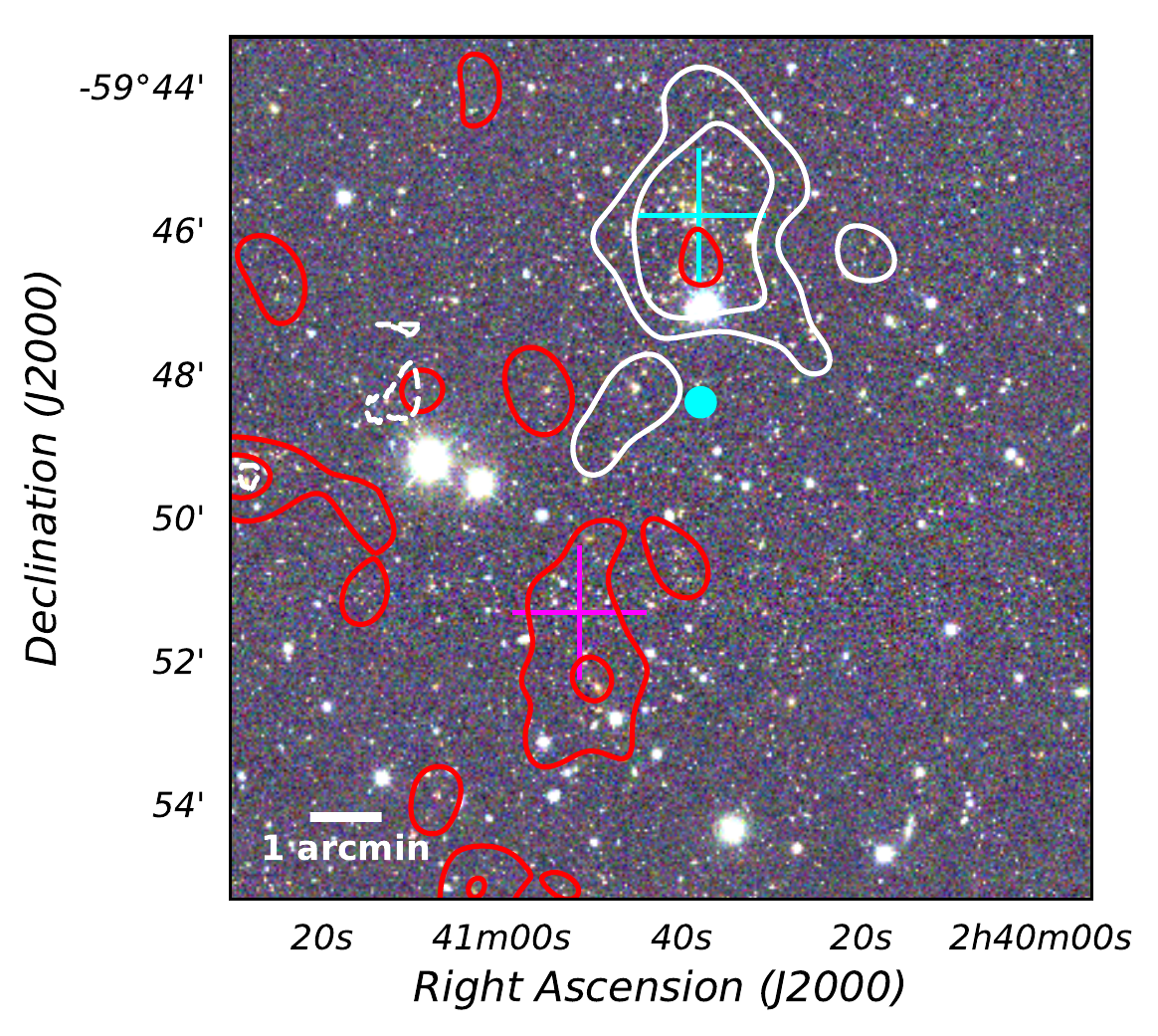}
    \vskip-0.15in
    \caption{\textit{top}: Richness as a function of redshift for the \Planck source PSZ-SN3~J0240-5945, with the best fit peak at $z_{\rm MCMF} = 0.41$. \textit{bottom}: pseudo-color image from DES \textit{g}, \textit{r}, \textit{i} bands near the \Planck candidate coordinates, which are marked with a cyan dot. Contours are from the galaxy density maps of the counterpart at $z_{\rm MCMF} = 0.41$ (white) and at $z_{\rm MCMF} = 0.605$ (red). The cyan cross marks the position of the optical counterpart found using \MCMF\hspace{-1 ex}, and the magenta cross marks the position of PSZ2~G280.76-52.30.}
    \label{fig:PSZ2_example_incorrect}
\end{figure}

There are two PSZ2 clusters for which we do not find a match (see Section~\ref{sec:mcmf_vs_psz2}) in our list of optical counterparts: PSZ2~G074.08-54.68 and PSZ2~G280.76-52.30. PSZ2~G074.08-54.68 is a cluster at $z_{\rm PSZ2} = 0.305$, with $M_{500}=5.40\times10^{14}$M$_\odot$ and S/N=6.1, which is within the DES footprint and that shows a prominent optical counterpart at (R.A., Dec.)= 347.04601, -1.92133, with the redshift coming from the REFLEX catalogue (ID: RXC J2308.3-0155). The area around this cluster is not masked due to bright stars or missing DES data. Nevertheless, this cluster is not in our \Planck SZE candidate catalogue. The PSZ2 cluster catalogue is a combination of three detection methods; PowellSnakes, MMF1 and MMF3, with the latter being the one used in this work. PSZ2~G074.08-54.68 is detected by the PowellSnakes algorithm, but not by MMF1 or MMF3. This could be due to the PSZ2 cluster being close to another cluster, PSZ2~G073.82-54.92, which might have been detected first, masking part (or all) of the flux of PSZ2~G074.08-54.68.

PSZ2~G280.76-52.30 is at $z_{\rm PSZ2} = 0.59$, with $M_{500}=4.88\times10^{14}$ M$_\odot$ and S/N=4.5, and it has the closest \Planck SZE position from our catalogue at 3.4~arcmin, with the optical position of that candidate having an offset of 5.8~arcmin to the PSZ2~G280.76-52.30 source.  Thus, it lies just outside our 3~arcmin matching radius. The PSZ2 redshift comes from the SPT catalogue, with the SPT ID of this cluster being SPT-CL~J0240-5952 \citep{Bocquet2019}. From the perspective of our analysis, the \Planck candidate (PSZ-SN3~J0240-5945) has $z_{\rm MCMF}=0.41$, with $\lambda_{\rm MCMF}=79$ and an estimated $f_{\rm cont}=0.03$. The second redshift peak that we find is at $z_{\rm MCMF}=0.605$, which is closer to the PSZ2 redshift. In Fig.~\ref{fig:PSZ2_example_incorrect} we show (top) the richness as a function of redshift, while below we show the $gri$ DES pseudo-color image. Overlayed are the density contours at $z_{\rm MCMF}=0.41$ (white) and $z_{\rm MCMF}=0.605$ (red). The cyan cross shows the optical position found by \MCMF\hspace{-1 ex}, and the magenta cross shows the position of PSZ2. For the peak at $z_{\rm MCMF}=0.605$ with $\lambda_{\rm MCMF}=50$, we estimate $f_{\rm cont}=0.28$, which means that we consider this to be a candidate with a second optical counterpart (requires $f_{\rm cont}<0.3$).  It is worth noting that, by using the same cross-match aperture, we find a match with the SPT-2500d catalogue \citep[][]{Bocquet2019}, SPT-CL~J0240-5946, whose reported redshift is $z_{\rm SPT}=0.4$.

\section{Further exploration of the \Planck candidate list contamination}\label{sec:Planckcontamination}

To investigate the difference between the observed contamination of $\sim$51\% and the 75\% contamination estimated from the \Planck sky simulations (see Section~\ref{sec:data_sz}) we compare the detection threshold $Y_{5R500}^{\rm min}=2\times10^{-4}$ arcmin$^2$ to the observed $Y_{5R500}$ distribution of our candidates. We will do this in three steps: First we will estimate an observed mass by means of the $\lambda_{\rm MCMF}$-$M_{500}$ relation derived in Section~\ref{sec:richness-mass_relation}. Secondly, we will determine the excess distribution of candidates with respect to the random lines-of-sight in different redshift ranges for the S/N$>$3 sample, which should give an estimate of the number of real clusters within this redshift range. Finally, we will use the derived parameters from the scaling relation and we will map from $\lambda_{\rm MCMF}$ to $Y_{5R500}$ on our excess clusters, using the $M_{500}-Y_{5R500}$ relation from  equation~(\ref{eq:y5r500}). With this, we can estimate the ratio of excess candidates with $Y_{5R500}\geq Y_{5R500}^{\rm min}$ with respect to the total number of excess candidates, which would give us an indication of how many real systems we expect to lose when applying this limiting value.

\begin{figure}
    \centering
    \includegraphics[width=\linewidth]{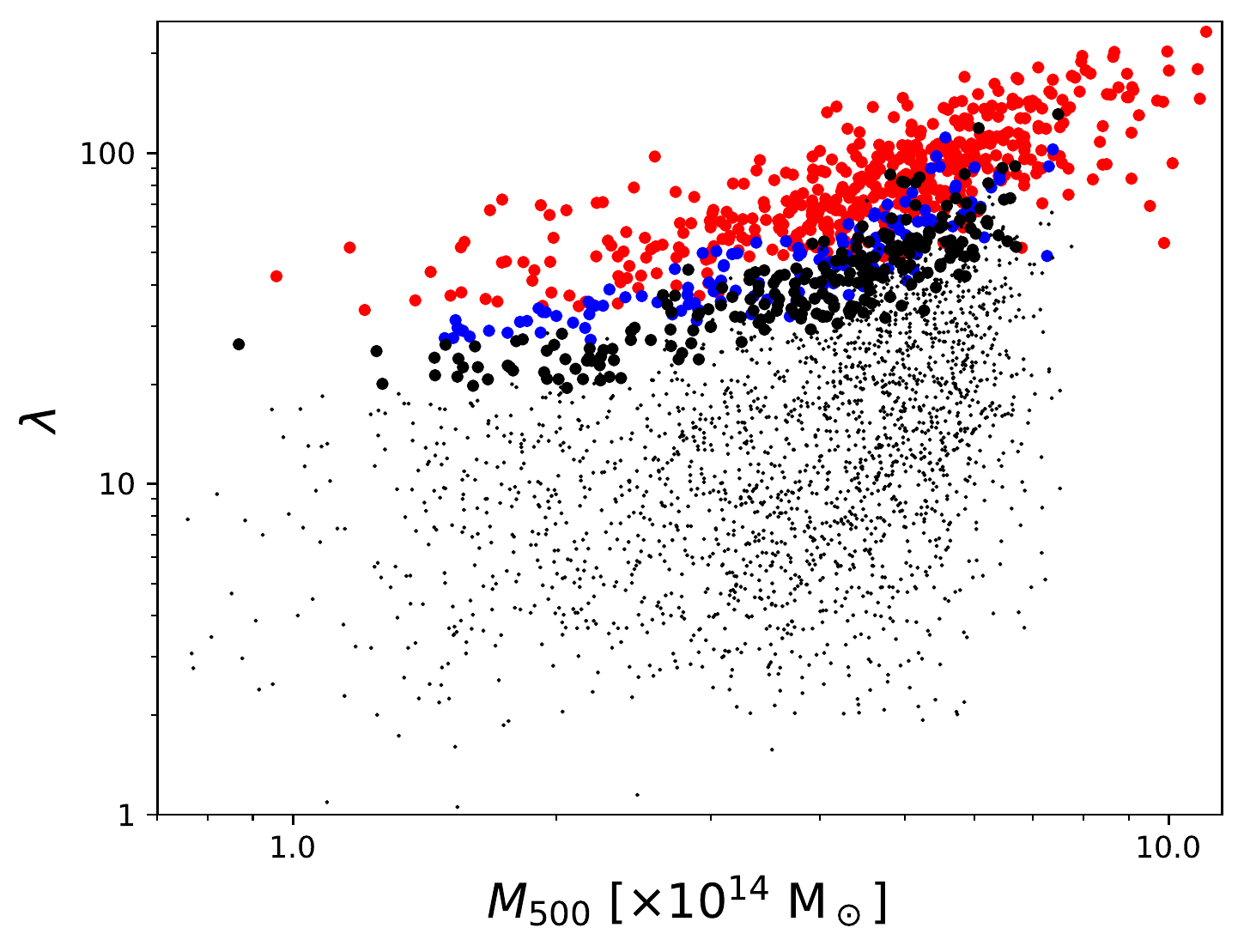}
    \vskip-0.15in
    \caption{Richness versus mass for the \Planck cluster candidates. Small black dots are candidates with $f_{\rm cont} \ge 0.2$. Bigger black, blue and red points represent counterparts with $0.1\le f_{\rm cont} < 0.2$, $0.05\le f_{\rm cont} < 0.1$ and $f_{\rm cont}<0.05$, respectively.
    }
    \label{fig:richness_vs_mass}
\end{figure}

\subsection{Richness--mass relation}\label{sec:richness-mass_relation}

Previous analyses have shown that the number of galaxies in a cluster (or richness) is approximately linearly proportional to the cluster mass \citep{Lin2004,Gladders07, Rozo2009, Rozo2009b, Klein2019},  with some intrinsic scatter \citep[$\sigma_{\rm int} \approx 25\% $, ][]{Rozo2014}. %
Fig.~\ref{fig:richness_vs_mass} shows how the derived masses behave with the estimated \MCMF richness of the candidates. Colors red, blue and black represent the different $f_{\rm cont}$ selection thresholds following Fig.~\ref{fig:f_cont}. Although at $f_{\rm cont} > 0.2$ the cloud of points does not seem follow any particular relation, the more reliable clusters with $f_{\rm cont} < 0.2$ exhibit a roughly linear trend at $M_{500} \gtrsim 2.5 \times 10^{14}$ M$_\odot$. The trend is stronger at lower $f_{\rm cont}$, where the contamination of the cluster sample is lowest.

We fit a $\lambda_{\rm MCMF}-M_{500}$ relation to our data but only for the high S/N sample (S/N$>$4.5 and $f_{\rm cont}^{\rm max} = 0.3$), which, assuming a catalogue contamination of $f_{\rm SZE-cont} \approx 8.5\%$ (Section~\ref{sec:data_sz}), means a purity of $\sim$97.4\%. For the fitting, we follow a similar procedure as the one described in \cite{Klein2021}, where the distribution of richnesses $\lambda_{\rm MCMF}$ is assumed to follow a log-normal distribution which depends on the mass $M_{500}$ and redshift $z$, so that 
\begin{equation}\label{eq:p_lambda}
    P(\ln \lambda | M_{500}, z) = \mathcal{N} ( \ln \lambda; \left<\ln \lambda\right>(M_{500},z), \sigma^2(M_{500}, z) ),
\end{equation}
\noindent
with mean
\begin{equation}\label{eq:p_lambda_mean}
    \left<\ln\lambda\right>(M_{500},z) = \ln\lambda_0 + \alpha_0 + \alpha_M \ln \left( \frac{M_{500}}{M_0} \right) + \alpha_z \ln \left( \frac{1+z}{1+z_0} \right)
\end{equation}
\noindent
and variance
\begin{equation}\label{eq:p_lambda_var}
    \sigma^2(M_{500}, z) = \exp ( \ln\zeta(z)-\left<\ln\lambda\right>) + \exp (s),
\end{equation}
\noindent
where $\lambda_0$, $M_0$ and $z_0$ are pivots and ($\alpha_0$, $\alpha_M$, $\alpha_z$ and $s$) are constrained by the likelihood analysis. For the pivots we use the median values of the richness, mass and redshift of the S/N$>$4.5 sample. The $\zeta(z)$ parameter corresponds to the richness correction factor used on \MCMF \citep[see  equation~7 from][]{Klein2019}. This first term on the variance captures the Poisson noise on the measured richness, while the second term represents the intrinsic variance within the cluster population. We refer the reader to Appendix A of \cite{Klein2021} for further details on the likelihood analysis. We find best fit values for the parameters of $\alpha_0 = -0.004 \pm 0.023 $, $\alpha_M = 0.961 \pm 0.071$, $\alpha_z = 0.095 \pm 0.252$ and $s = -2.151 \pm 0.101$.

\begin{figure}
    \centering
    \includegraphics[width=\linewidth]{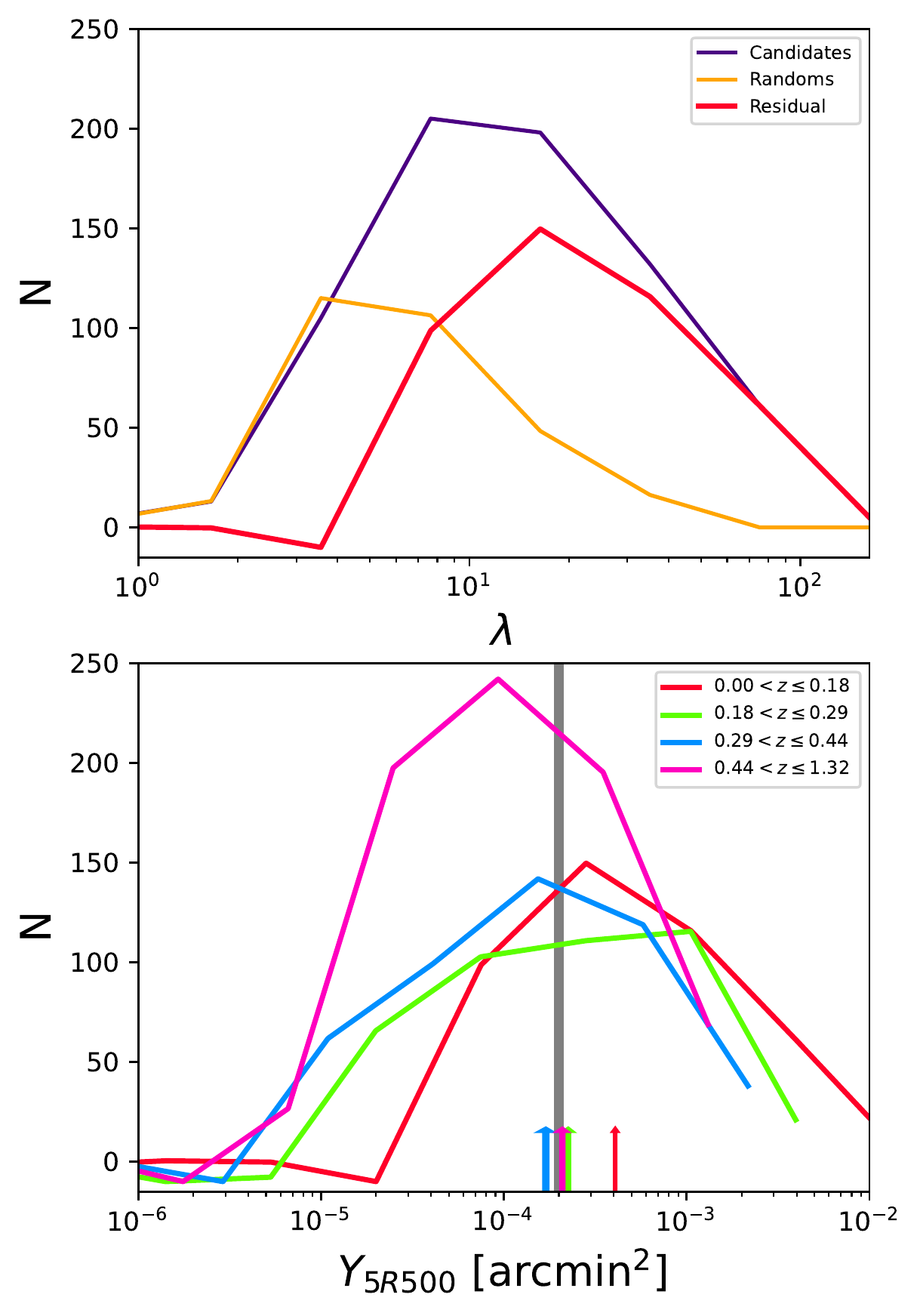}
    \vskip-0.15in
    \caption{\textit{top:} Richness distributions in the redshift range $0<z_{\rm MCMF}\leq 0.18$ of candidates, rescaled randoms and excess candidates shown as purple, orange and red lines, respectively. \textit{bottom:} $Y_{5R500}$ distributions of excess candidates for different redshift bins. $Y_{5R500}$ is determined from $\lambda_{\rm MCMF}$ using scaling relations. Colored arrows correspond to the richness derived $Y_{5R500}$ a candidate would need to have for us to consider it a real cluster in our sample, estimated using the median redshift of all the candidates for each redshift bin. The gray line marks the $Y_{5R500}^{\rm min}$ used to estimate the purity in Section~\ref{sec:data_sz}.}
    \label{fig:y5r500_dist}
\end{figure}

\subsection{On the difference between 51\% and 75\% initial contamination}

For the distribution of excess candidates, we use the PSZ-MCMF cluster catalog to define the redshift ranges using the 25\%, 50\% and 75\% percentiles, corresponding to ranges of $0<z\leq0.18$, $0.18<z\leq 0.29$, $0.29<z\leq 0.44$ and $0.44<z\leq1.32$.
For each of these ranges we look at the $\lambda_{\rm MCMF}$ distribution of \Planck candidates and that of the random lines-of-sight, rescaling the latter to fit the candidates distribution at low $\lambda_{\rm MCMF}$. We then subtract the number of scaled random lines-of-sight to the number of \Planck candidates for each $\lambda_{\rm MCMF}$ bin within a redshift range. We refer to this as the distribution of excess candidates, which maps the distribution of real clusters down to low $\lambda$ without accounting for catalogue purity, unlike when a \fcont threshold $f_{\rm cont}^{\rm max}$ is applied.

Finally we transform this $\lambda_{\rm MCMF}$ distribution into a $M_{500}$ distribution, and then into a $Y_{5R500}$ distribution. Fig.~\ref{fig:y5r500_dist} shows the different steps on the estimation of the excess and the final transformations. The top panel shows the distributions of candidates and scaled random lines-of-sight for the $0<z\leq0.18$ redshift range as purple and orange lines, respectively, with the excess candidates, labeled as ``residual'', shown in red. The bottom shows the distribution of the excess candidates for the different redshift bins in terms of $Y_{5R500}$. The vertical gray line marks the $Y_{5R500}^{\rm min}$. Depending on how we scale the randoms to fit the low richness regime of the candidates, the ratio of the total excess candidates (summed at all redshifts) to that of the total number of candidates varies between 55-65\%. Of those, $\sim$50\% are below the $Y_{5R500}^{\rm min}$ threshold, regardless of the normalisation, meaning that we would expect to lose half of the real sources by applying this threshold.

We note that we do not probe the $\lambda_{\rm MCMF}-M_{500}$ relation at $\lambda < 10$ (see Fig.~\ref{fig:richness_vs_mass}), which, depending on the redshift, could translate to $M_{500} < 10^{14}$ M$_\odot$. However, this does not affect our analysis because, as can be seen in Fig.~\ref{fig:f_cont}, the minimum $\lambda$ in our sample is $\lambda_{\rm min} \sim20$ at $z\approx0.03$, which corresponds to a mass $M_{500}>10^{14}$ M$_\odot$ using our scaling relation.

The different arrows on the bottom panel show, for each redshift range, the richness of a source with $f_{\rm cont}=0.2$, $\lambda_{\rm MCMF, min}$, translated into a $Y_{5R500, \rm min}$, color codded according to the redshift range. Each $\lambda_{\rm MCMF, min} (z)$ is estimated using the median redshift of each redshift range. These arrows can be interpreted as the selection thresholds that are applied when selecting candidates with $f_{\rm cont}<0.2$, showing good agreement with $Y_{5R500}^{\rm min}$ at all redshifts. This can also be seen in Fig.~\ref{fig:ntrue_ncand}, where at $f_{\rm cont}^{\rm max}=0.2$, the number of clusters over the number of candidates is $\sim$25\%, similar to the value expected using $Y_{5R500}^{\rm min}$ (Section~\ref{sec:data_sz}).

We note that groups and clusters corresponding to the difference between 51\% and 75\% have $0.2<f_{\rm cont}<0.65$. They correspond to small black dots in Fig.~\ref{fig:richness_vs_mass} and are thus subject to a strong selection bias. For these systems, the measured SZE signal is dominated by a positive noise fluctuation on top of an actual small SZ signal from the cluster. The conversion of the measured SZE signal to the mass thus provides strongly overestimated values. However, we are already excluding most of these systems in the final catalogue where we only add 11 S/N$>$4.5 clusters with $0.2\leq f_{\rm cont}<0.3$.

\noindent
{\it
$^{1}$ Faculty of Physics, Ludwig-Maximilians-Universit\"at, Scheinerstr. 1, 81679 Munich, Germany\\
$^{2}$ Excellence Cluster Origins, Boltzmannstr.\ 2, 85748 Garching, Germany\\
$^{3}$ Max Planck Institute for Extraterrestrial Physics, Giessenbachstrasse, 85748 Garching, Germany\\
$^{4}$ IRFU, CEA, Universit\'e Paris-Saclay, 91191 Gif-sur-Yvette, France\\
$^{5}$ AIM, CEA, CNRS, Universit\'e Paris-Saclay, Universit\'e Paris Diderot, Sorbonne Paris Cit\'e, 91191 Gif-sur-Yvette, France\\
$^{6}$ Observatorio Astron\'omico Nacional (OAN-IGN), C/ Alfonso XII 3, 28014, Madrid, Spain\\
$^{7}$ Cerro Tololo Inter-American Observatory, NSF's National Optical-Infrared Astronomy Research Laboratory, Casilla 603, La Serena, Chile\\
$^{8}$ Laborat\'orio Interinstitucional de e-Astronomia - LIneA, Rua Gal. Jos\'e Cristino 77, Rio de Janeiro, RJ - 20921-400, Brazil\\
$^{9}$ Department of Physics, University of Michigan, Ann Arbor, MI 48109, USA\\
$^{10}$ Institute of Cosmology and Gravitation, University of Portsmouth, Portsmouth, PO1 3FX, UK\\
$^{11}$ CNRS, UMR 7095, Institut d'Astrophysique de Paris, F-75014, Paris, France\\
$^{12}$ Sorbonne Universit\'es, UPMC Univ Paris 06, UMR 7095, Institut d'Astrophysique de Paris, F-75014, Paris, France\\
$^{13}$ Department of Physics \& Astronomy, University College London, Gower Street, London, WC1E 6BT, UK\\
$^{14}$ Kavli Institute for Particle Astrophysics \& Cosmology, P. O. Box 2450, Stanford University, Stanford, CA 94305, USA\\
$^{15}$ SLAC National Accelerator Laboratory, Menlo Park, CA 94025, USA\\
$^{16}$ Instituto de Astrofisica de Canarias, E-38205 La Laguna, Tenerife, Spain\\
$^{17}$ Universidad de La Laguna, Dpto. Astrofísica, E-38206 La Laguna, Tenerife, Spain\\
$^{18}$ Center for Astrophysical Surveys, National Center for Supercomputing Applications, 1205 West Clark St., Urbana, IL 61801, USA\\
$^{19}$ Department of Astronomy, University of Illinois at Urbana-Champaign, 1002 W. Green Street, Urbana, IL 61801, USA\\
$^{20}$ Institut de F\'{\i}sica d'Altes Energies (IFAE), The Barcelona Institute of Science and Technology, Campus UAB, 08193 Bellaterra (Barcelona) Spain\\
$^{21}$ Institut d'Estudis Espacials de Catalunya (IEEC), 08034 Barcelona, Spain\\
$^{22}$ Institute of Space Sciences (ICE, CSIC),  Campus UAB, Carrer de Can Magrans, s/n,  08193 Barcelona, Spain\\
$^{23}$ Astronomy Unit, Department of Physics, University of Trieste, via Tiepolo 11, I-34131 Trieste, Italy\\
$^{24}$ INAF-Osservatorio Astronomico di Trieste, via G. B. Tiepolo 11, I-34143 Trieste, Italy\\
$^{25}$ Institute for Fundamental Physics of the Universe, Via Beirut 2, 34014 Trieste, Italy\\
$^{26}$ Hamburger Sternwarte, Universit\"{a}t Hamburg, Gojenbergsweg 112, 21029 Hamburg, Germany\\
$^{27}$ Department of Physics, IIT Hyderabad, Kandi, Telangana 502285, India\\
$^{28}$ Fermi National Accelerator Laboratory, P. O. Box 500, Batavia, IL 60510, USA\\
$^{29}$ Jet Propulsion Laboratory, California Institute of Technology, 4800 Oak Grove Dr., Pasadena, CA 91109, USA\\
$^{30}$ Institute of Theoretical Astrophysics, University of Oslo. P.O. Box 1029 Blindern, NO-0315 Oslo, Norway\\
$^{31}$ Kavli Institute for Cosmological Physics, University of Chicago, Chicago, IL 60637, USA\\
$^{32}$ Instituto de Fisica Teorica UAM/CSIC, Universidad Autonoma de Madrid, 28049 Madrid, Spain\\
$^{33}$ Observat\'orio Nacional, Rua Gal. Jos\'e Cristino 77, Rio de Janeiro, RJ - 20921-400, Brazil\\
$^{34}$ School of Mathematics and Physics, University of Queensland,  Brisbane, QLD 4072, Australia\\
$^{35}$ Santa Cruz Institute for Particle Physics, Santa Cruz, CA 95064, USA\\
$^{36}$ Center for Cosmology and Astro-Particle Physics, The Ohio State University, Columbus, OH 43210, USA\\
$^{37}$ Department of Physics, The Ohio State University, Columbus, OH 43210, USA\\
$^{38}$ Center for Astrophysics $\vert$ Harvard \& Smithsonian, 60 Garden Street, Cambridge, MA 02138, USA\\
$^{39}$ Australian Astronomical Optics, Macquarie University, North Ryde, NSW 2113, Australia\\
$^{40}$ Lowell Observatory, 1400 Mars Hill Rd, Flagstaff, AZ 86001, USA\\
$^{41}$ Centre for Gravitational Astrophysics, College of Science, The Australian National University, ACT 2601, Australia\\
$^{42}$ The Research School of Astronomy and Astrophysics, Australian National University, ACT 2601, Australia\\
$^{43}$ Department of Astrophysical Sciences, Princeton University, Peyton Hall, Princeton, NJ 08544, USA\\
$^{44}$ Centro de Investigaciones Energ\'eticas, Medioambientales y Tecnol\'ogicas (CIEMAT), Madrid, Spain\\
$^{45}$ Instituci\'o Catalana de Recerca i Estudis Avan\c{c}ats, E-08010 Barcelona, Spain\\
$^{46}$ Department of Astronomy, University of California, Berkeley,  501 Campbell Hall, Berkeley, CA 94720, USA\\
$^{47}$ Institute of Astronomy, University of Cambridge, Madingley Road, Cambridge CB3 0HA, UK\\
$^{48}$ Department of Physics and Astronomy, University of Pennsylvania, Philadelphia, PA 19104, USA\\
$^{49}$ Department of Physics and Astronomy, Pevensey Building, University of Sussex, Brighton, BN1 9QH, UK\\
$^{50}$ School of Physics and Astronomy, University of Southampton,  Southampton, SO17 1BJ, UK\\
$^{51}$ Computer Science and Mathematics Division, Oak Ridge National Laboratory, Oak Ridge, TN 37831\\
$^{52}$ Lawrence Berkeley National Laboratory, 1 Cyclotron Road, Berkeley, CA 94720, USA\\
}

%%%%%%%%%%%%%%%%%%%%%%%%%%%%%%%%%%%%
% Don't change these lines
\bsp	% typesetting comment
\label{lastpage}
\end{document}